\documentclass{houches}
\usepackage{cite}
\usepackage{amssymb}
\usepackage{epsfig}
\newtheorem{HW}{Problem}[section]
\newenvironment{hw}{\begin{HW}}{\end{HW}}
\def\res#1{ { \hat {#1}}}
\def\bar#1{\overline #1}
\newcommand{\dsl}{\raise.15ex\hbox{/}\kern-.57em\partial}
\def\aslash{\raise.15ex\hbox{/}\mkern-14mu A}
\newcommand{\pslash}{\raise-.15ex\hbox{/}\mkern-9.5mu p}
\newcommand{\abs}[1]{\left| #1 \right|}
\newcommand{\vev}[1]{\left\langle #1 \right\rangle}
\newcommand{\veV}[1]{\langle #1 \rangle}
\newcommand{\Dslash}{\hbox{/\kern-.6000em D}}
\newcommand{\dslash}{\,\raise.15ex\hbox{/}\mkern-13.5mu D}

\newcommand{\lqcd}{\ensuremath{\Lambda_{\rm QCD}}}
\newcommand{\me}[3]{\ensuremath{\left\langle{#1}\vphantom{#2 #3}
\right|{#2}\left|\vphantom{#1 #2}{#3}\right\rangle}}
\newcommand{\mE}[3]{\ensuremath{\langle{#1}\vphantom{#2 #3}
|{#2} |\vphantom{#1 #2}{#3}\rangle}}

\newcommand{\ord}[1]{\ensuremath{\mathcal{O}\left(#1\right)}}
\newcommand{\bra}[1]{\left\langle #1 \right|}
\newcommand{\ket}[1]{\left| #1 \right\rangle}

\newcommand{\drawsquare}[2]{\hbox{%
\rule{#2pt}{#1pt}\hskip-#2pt
\rule{#1pt}{#2pt}\hskip-#1pt
\rule[#1pt]{#1pt}{#2pt}}\rule[#1pt]{#2pt}{#2pt}\hskip-#2pt
\rule{#2pt}{#1pt}}

\newcommand{\Yfund}{\raisebox{-.5pt}{\drawsquare{6.5}{0.4}}}
\newcommand{\Ysymm}{\raisebox{-.5pt}{\drawsquare{6.5}{0.4}}\hskip-0.4pt%
    \raisebox{-.5pt}{\drawsquare{6.5}{0.4}}}
\newcommand{\Ythrees}{\raisebox{-.5pt}{\drawsquare{6.5}{0.4}}\hskip-0.4pt%
     \raisebox{-.5pt}{\drawsquare{6.5}{0.4}}\hskip-0.4pt%
     \raisebox{-.5pt}{\drawsquare{6.5}{0.4}}}
\newcommand{\Yasymm}{\raisebox{-3.5pt}{\drawsquare{6.5}{0.4}}\hskip-6.9pt%
    \raisebox{3pt}{\drawsquare{6.5}{0.4}}}

\newcommand{\Yadjoint}{\raisebox{-3.5pt}{\drawsquare{6.5}{0.4}}\hskip-6.9pt%
    \raisebox{3pt}{\drawsquare{6.5}{0.4}}\hskip-0.4pt
    \raisebox{3pt}{\drawsquare{6.5}{0.4}}}
\begin{document}
\setcounter{chapter}{0}
\setcounter{page}{1}
\author[A.V. Manohar]{Aneesh V. Manohar\thanks{e-mail: amanohar@ucsd.edu}}
\address{Department of Physics 0319\\
University of California, San Diego\\
9500 Gilman Drive, La Jolla, CA 92093, USA}
\editor{F. David and R. Gupta}
\title{Probing the Standard Model of Particle Interactions}
\chapter{Large N QCD}
\section{Introduction}

Quantum chromodynamics, the theory of the strong interactions, is a non-Abelian
gauge theory based on the gauge group $SU(3)$. It was first pointed out by 
't~Hooft~\cite{thooft1,thooft2} that many features of QCD can be understood by
studying a gauge theory based on the gauge group $SU(N)$ in the limit $N
\rightarrow \infty$. One might think that letting $N \rightarrow \infty$ would
make the analysis more complicated because of the larger gauge group and
consequent increase in the number of dynamical degrees of freedom. One might
also think that $SU(N)$ gauge theory has very little to do with QCD because
$N=\infty$ is not close to $N=3$. However, we will soon see that $SU(N)$ gauge
theory simplifies in the $N \rightarrow \infty$ limit, that the true expansion
parameter is $1/N$, not $N$, and that the $1/N$ expansion is equivalent to a
semiclassical expansion for an effective theory of color singlet mesons and
baryons. Results for QCD can be obtained from the $N \rightarrow \infty$ limit
by expanding in $1/N=1/3$, and are in good agreement with experiment.

To decide whether $1/N$ is a small expansion parameter for QCD requires further
analysis. In QED, as Witten has remarked, the coupling constant $e=\sqrt{4 \pi
\alpha} = 0.30$, which is not very different from $1/3$. Anyone who has
actually computed radiative corrections in QED knows that the true expansion
parameter is not $e$, but is closer to $\alpha/4\pi \approx 10^{-3}$, which is
much smaller than $e$. By the end of these lectures you will see several
examples which show that the expansion parameter for QCD is $1/N=1/3$. While
not as small as the QED expansion parameter $\alpha/4 \pi$, $1/N$ is still a
useful expansion parameter for QCD. $1/N$ corrections are comparable in size to
flavor $SU(3)$ breaking corrections due to the strange quark mass, and
expanding in flavor $SU(3)$ breaking is well-known to be an extremely useful
expansion in QCD. Furthermore, we will find many examples where the $1/N$ term
vanishes, so that the first correction is of order $1/N^2$. In such cases, one
can make predictions at the 10\% level. This is a level of computational
accuracy in low-energy hadronic physics that is difficult to match using other
techniques.

In these lectures, I will concentrate on the large $N$ expansion for QCD, and
in particular, on trying to obtain QCD results that can be compared with
experimental data. Sections~\ref{sec:gn}--\ref{sec:qcd} and
sections~\ref{axialu1}, \ref{sec:baryons}--\ref{sec:bncount} of these lectures
are based on the treatments by Coleman~\cite{coleman}, and
Witten~\cite{witten-ep,witten-baryon}, respectively. Large $N$ expansions have
also been used to study other field theories, such as the $O(N)$ model, $CP^N$
model, etc. They provide insight into quantum field theory dynamics, and have
many applications in high energy physics and statistical mechanics. They have
been extensively used in recent years to study matrix models. These topics have
been discussed in previous Les Houches summer schools, and will not be repeated
here. A good reference on large $N$ methods is the compilation by Brezin and
Wadia~\cite{brezin}.

\section{The Gross-Neveu Model}\label{sec:gn}

The Gross-Neveu model~\cite{gross-neveu} is an interesting $1+1$ dimensional
field theory that can be studied using the $1/N$ expansion. The model is
asymptotically free with a spontaneously broken chiral symmetry, and so shares
some dynamical features with QCD. It will provide a useful warm-up exercise
before we tackle the much more difficult problem of large $N$ QCD.

The Gross-Neveu Lagrangian is
\begin{equation}\label{2.1}
L = \bar \psi\, i \dsl\, \psi + {\lambda \over 2} \left(\bar \psi \psi
\right)^2,
\end{equation}
where $\psi^a$, $a=1,\ldots,N$ are $N$ Dirac fields, and a sum on $N$ is
implicit in the notation, so that $\bar \psi \psi = \sum_a
\bar{{\psi^a}}\psi^a$, etc. In $1+1$ dimensions, Dirac fields are
two-component spinors, and have mass dimension $1/2$. $\lambda$ is a
dimensionless coupling constant. Equation~(\ref{2.1}) is invariant under an
$SU(N)$ flavor symmetry on the $\psi$'s,
\[
\psi^a(x) \rightarrow U^a{}_b\ \psi^b(x),
\]
where $U$ is an $SU(N)$ matrix, and also invariant under a discrete chiral
symmetry
\begin{equation}\label{2.2}
\psi \rightarrow \gamma_5 \psi,\quad \bar \psi \rightarrow -\bar \psi \gamma_5,
\quad \bar \psi \psi \rightarrow - \bar \psi \psi.
\end{equation}
Equation~(\ref{2.1}) is the most general possible Lagrangian invariant under
these symmetries with terms of dimension less than or equal to two, and so
describes a renormalizable field theory in $1+1$ dimensions. The discrete
chiral symmetry eq.~(\ref{2.2}) forbids a mass term, so the fermions are
massless at any finite order in perturbation theory, and no mass counterterm
is needed to regulate the ultraviolet divergences.

\begin{figure}[tbp]
\begin{center}
\epsfig{file=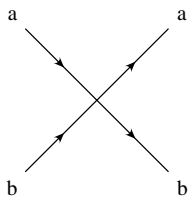,height=30mm}
\caption{The four-Fermi vertex of the Gross-Neveu model. $a,b$ are flavor
labels. \label{fig:1}}
\end{center}
\end{figure}

\begin{figure}[tbp]
\begin{center}
\epsfig{file=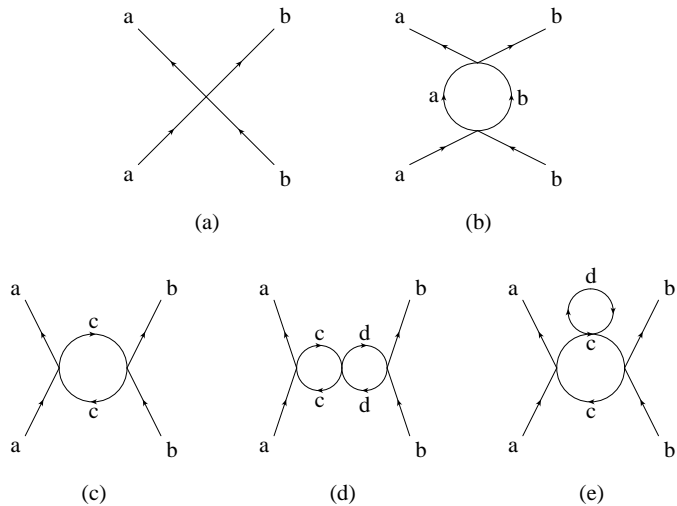,height=70mm}
\caption{Low-order diagrams for the scattering amplitude $a + \bar a
\rightarrow b + \bar b$, with $a \not = b$. Virtual flavors $c$ and $d$ are
summed over.\label{fig:2}}
\end{center}
\end{figure}

The basic interaction vertex is shown in fig.~\ref{fig:1}.
Consider the process $a + \bar a \rightarrow b + \bar b$, with $a \not = b$.
Some graphs contributing to this scattering amplitude are shown in
fig.~\ref{fig:2}. 
The leading order diagram fig.~\ref{fig:2}(a) is of order $\lambda$. The
one-loop correction fig.~\ref{fig:2}(b) is of order $\lambda^2$. The
intermediate fermion flavor $c$ in the one-loop correction fig.~\ref{fig:2}(c)
is arbitrary and must be summed, so the graph is order $\lambda^2 N$.
Similarly, figs.~\ref{fig:2}(d,e) are of order $\lambda^3 N^2$. One can clearly
see that the perturbation series does not have a well-defined limit as $N
\rightarrow \infty$, because the radiative corrections grow with powers of $N$.

One can obtain a well-defined large $N$ limit by rescaling the coupling
constant $\lambda$. Define $\lambda = g^2/N$, and take the limit $N \rightarrow
\infty$ with $g$ fixed, so that $\lambda$ is of order $1/N$. The graphs
considered in fig.~\ref{fig:2} then give a sensible expansion for the
scattering amplitude --- the lowest order term is $g^2/N$, and the correction
terms are $g^4/N^2$, $g^4/N$, $g^6/N$ and $g^6/N$ for figs.~\ref{fig:2}(b--e),
respectively. This shows that the perturbation series gives a scattering
amplitude of the form $(1/N) f(g^2,1/N)$, and has the large $N$ limit $(1/N)
f(g^2,0)$. In the large $N$ limit, diagrams figs.~\ref{fig:2}(a,c--e)
contribute to $f(g^2,0)$, but diagram fig.~\ref{fig:2}(b) is omitted. There is
some simplification of the diagrammatic expansion as $N \rightarrow \infty$,
but the limit is still highly non-trivial.

The Gross-Neveu Lagrangian is
\begin{equation}\label{2.3}
L = \bar \psi\, i \dsl\, \psi + {g^2 \over 2 N} \left(\bar \psi \psi
\right)^2,
\end{equation}
when written in terms of $g$. One way to understand the power of $1/N$ in the
interaction term is to note that $\bar \psi \psi/\sqrt{N}$ produces a flavor
singlet $\bar\psi \psi$ state with unit amplitude, since there is an implicit
sum over $N$ flavors in $\bar \psi \psi$. This is like in quantum mechanics,
where a state $\ket{\psi}$ which is the sum of $N$ orthonormal states with
equal amplitude, $\ket{\psi} = \alpha\left(\ket{1} + \ket{2} + \ldots +
\ket{N}\right)$, has normalization constant $N\abs{\alpha}^2=1$ to have unit
norm. Then $\left(\bar \psi \psi \right)^2$ should have a coefficient of order
$1/N$, so that $\bar \psi \psi$ scattering in the flavor singlet channel has an
amplitude of order unity. 

One can now study the perturbation series in $g$ for the Lagrangian
eq.~(\ref{2.3}) in the $N \rightarrow \infty$ limit. The diagrammatic 
expansion for the Gross-Neveu model simplifies in the large $N$ limit. For
example, we have seen that fig.~\ref{fig:2}(b) can be neglected. The
simplifications are sufficient to allow one to obtain exact results, though
this might not yet be apparent. To make the large $N$ analysis more
transparent, it is convenient to introduce an auxiliary field $\sigma$, and
write the Lagrangian eq.~(\ref{2.3}) as
\begin{equation}\label{2.4}
L = \bar \psi\, i \dsl\, \psi + \sigma \bar \psi \psi - {N \over 2 g^2}
\sigma^2.
\end{equation}
The Lagrangian is quadratic in $\sigma$, so integrating over $\sigma$ is
equivalent to minimizing the Lagrangian with respect to $\sigma$, which gives
\[
\sigma = {g^2 \over N} \bar \psi \psi.
\]
Substituting the answer back in eq.~(\ref{2.4}) gives the original form of the
Lagrangian, eq.~(\ref{2.3}).

\begin{figure}[tbp]
\begin{center}
\epsfig{file=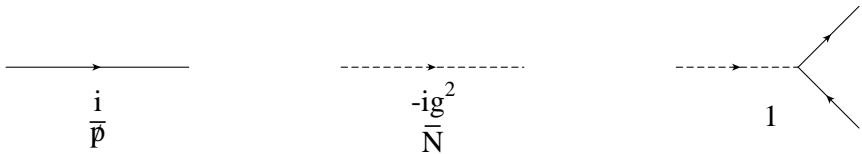,height=25mm}
\caption{Feynman rules for the Gross-Neveu Lagrangian eq.~(\ref{2.4}). The
$\psi$ field is represented by a solid line, and the auxiliary field $\sigma$
by a dashed line.\label{fig:3}}
\end{center}
\end{figure}

\begin{figure}[tbp]
\begin{center}
\epsfig{file=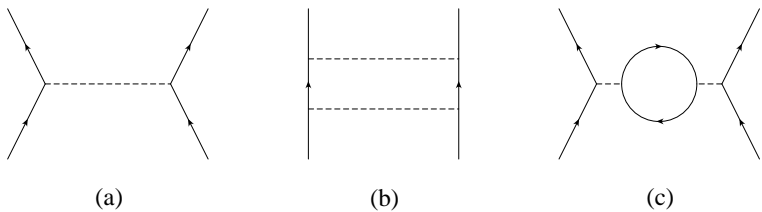,height=30mm}
\caption{The graphs of fig.~\ref{fig:2}(a,c) using the auxiliary field form of
the Lagrangian.\label{fig:4}}
\end{center}
\end{figure}

\begin{figure}[tbp]
\begin{center}
\epsfig{file=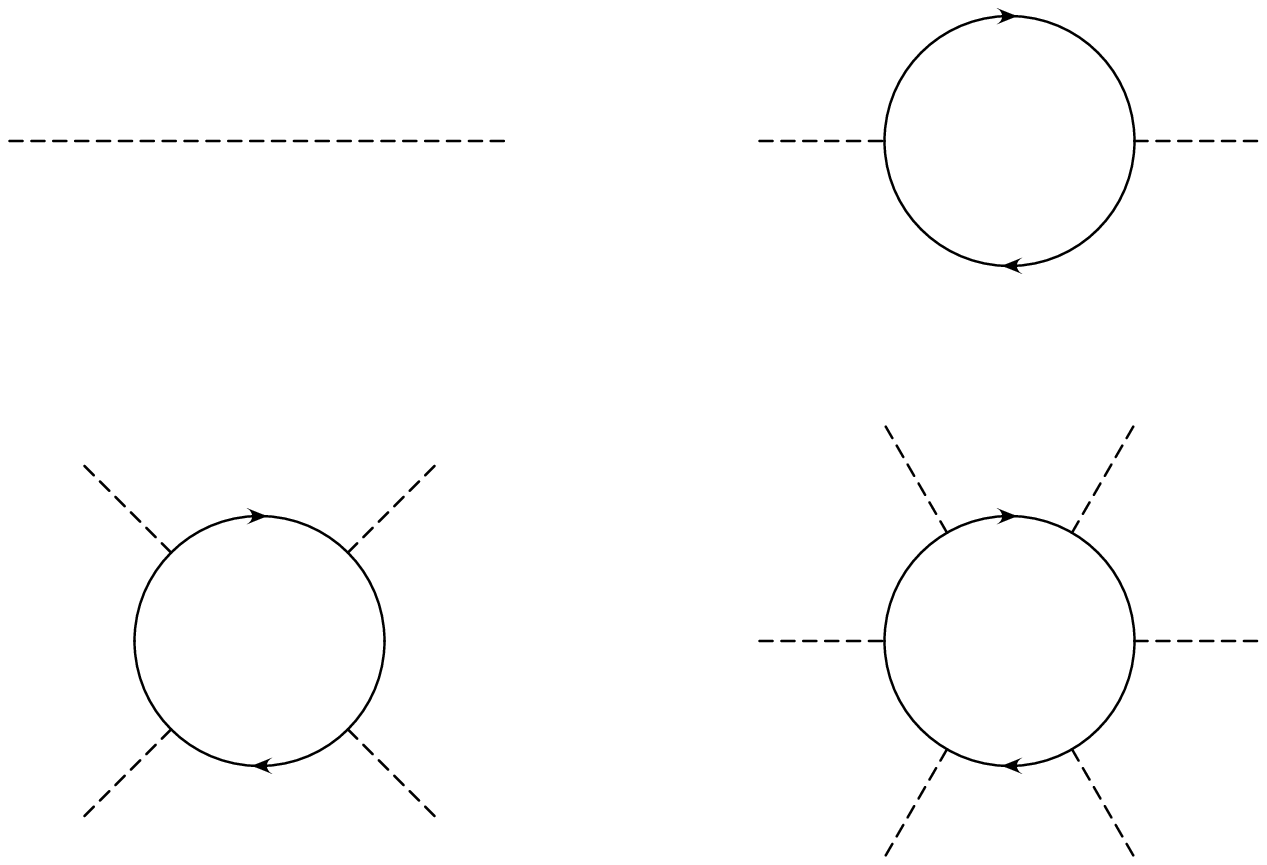,height=40mm}
\caption{Diagrammatic expansion for $L_{\rm eff}\left( \sigma \right)$.
\label{fig:5}}
\end{center}
\end{figure}

\begin{figure}[tbp]
\begin{center}
\epsfig{file=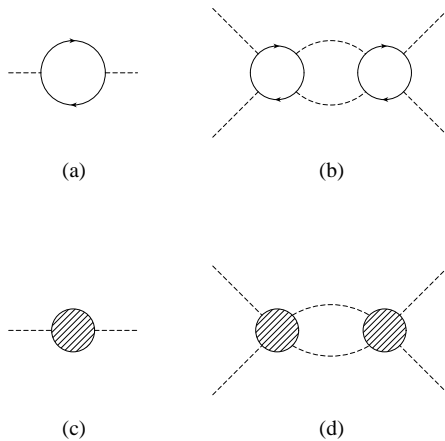,height=60mm}
\caption{Examples of $N$-counting. Graphs (a) and (b) are equivalent to (c) and
(d), where the blobs are interaction vertices in $N \tilde L_{\rm eff}$.
\label{fig:6}}
\end{center}
\end{figure}

The analysis of the large $N$ limit is simpler using the modified Lagrangian
eq.~(\ref{2.4}). The Feynman rules are given in fig.~\ref{fig:3}, 
and the scattering graphs of fig.~\ref{fig:2} are now represented as in
fig.~\ref{fig:4}. 
The auxiliary field representation allows one to obtain the $N$-counting rules
for diagrams. Consider the graphs of fig.~\ref{fig:4} with all external fermion
lines removed. The resulting graphs are generated by an effective action
$L_{\rm eff}(\sigma)$ which contains only external $\sigma$ lines. This
effective action is obtained by evaluating the fermion functional integral
using the Lagrangian eq.~(\ref{2.4}). The Lagrangian is quadratic in the
fermion fields, so $L_{\rm eff}(\sigma)$ is given exactly by the sum of
diagrams in fig.~\ref{fig:5}.
The first term is the tree-level inverse propagator $- {N /2 g^2} \sigma^2$,
and the remaining terms are the one-loop corrections. Each diagram in
fig.~\ref{fig:5} is of order $N$ --- the one-loop terms have $N$ fermions in a
closed loop, and the tree-level term is explicitly of order $N$. Thus the
effective action can be written as
\begin{equation}\label{2.5}
L_{\rm eff}(\sigma,g,N) = N \tilde L_{\rm eff}(\sigma,g).
\end{equation}
It is now straightforward to determine the power of $N$ in any Feynman graph
with only external $\sigma$ lines. Each term in the Lagrangian eq.~(\ref{2.5})
is of order $N$. Each interaction vertex has a factor of $N$, and each
propagator has a factor of $1/N$, since a vertex is a term in the Lagrangian,
and a propagator is the inverse of the quadratic terms in the Lagrangian. Thus
a diagram is proportional to $N^{V-I-E}$, where $V$ is the number of vertices,
$I$ is the number of internal $\sigma$ lines, and $E$ is the number of external
$\sigma$ lines. Factors of $1/N$ are included for external $\sigma$ propagators
since the physical scattering amplitudes have only external fermion fields, and
all the $\sigma$ lines are actually internal $\sigma$ lines in the full diagram
including $\psi$ propagators. An example is shown in fig.~\ref{fig:6}. Diagrams
(a) and (b) can be redrawn as (c) and (d), where the blob is an interaction
vertex in $N \tilde L_{\rm eff}$. The $N$-counting formula then shows that
fig.~\ref{fig:6}(c) is of order $N^{1-0-2}=1/N$, and fig.~\ref{fig:6}(d) is of
order $N^{2-2-4}=1/N^4$. These are also the $N$-counting rules for the original
diagrams figs.~\ref{fig:6}(a) and (b), respectively.
For any Feynman graph, one has the identity
\begin{equation}\label{2.6}
V-I+L=1,
\end{equation}
so that a Feynman diagram is of order 
\begin{equation}\label{2.6a}
N^{1-L-E}.
\end{equation}
The minus signs in front of $L$ and $E$ in eq.~(\ref{2.6a}) are important,
since they imply that additional loops or external lines bring an additional
suppression of powers of $1/N$, rather than enhancements by powers of $N$. 
This proves that the theory has a sensible $1/N$ expansion. Since the minimum
number of external $\sigma$ lines is at least two, the maximum power of $N$ is
$-1$.

The effective Lagrangian $L_{\rm eff}$ can be computed from the bubble sum in
fig.~\ref{fig:5}. It will be computed here only in the limit where all
$\sigma$ lines carry zero momentum, where it reduces to the effective potential
$V(\sigma)$. The effective potential is given by the sum of diagrams in
fig.~\ref{fig:5},
\begin{equation}\label{2.7}
-iV = -i {N\over 2 g^2} \sigma^2 - N \sum_{r=1}^\infty {1\over 2 r} \Tr \int
{d^2 p \over (2\pi)^2} \left( - {\pslash \sigma \over p^2}\right)^{2r}.
\end{equation}
There are only even powers of $\sigma$ because of the discrete symmetry $\sigma
\rightarrow - \sigma$. The first term in eq.~(\ref{2.7}) is the tree-level
amplitude. The second term is the sum of loop graphs. The $-N$ is from the $N$
flavors of fermions in the loop, $1/2r$ is the symmetry factor for the graph,
and the term in parentheses is the product of the fermion propagator $i
\pslash/p^2$ and the vertex $i \sigma$. Performing the trace, using the
identity
\[
\sum_{r=1}^\infty {x^{2r} \over 2r} = -\frac{1}{2} \log \left(1-x^2\right),
\]
and analytically continuing to Euclidean space gives
\begin{equation}\label{2.8}
V = {N\over 2 g^2} \sigma^2 - N \int
{d^2 p \over (2\pi)^2} \log \left(1 +{\sigma^2 \over p^2}\right).
\end{equation}
Regulating the loop integral using dimensional regularization in the
$\overline{MS}$ scheme gives
\begin{equation}\label{2.9}
V = N \left[{\sigma^2\over 2 g^2} + {\sigma^2\over 4 \pi}\left(
\log {\sigma^2\over \mu^2} - 1 \right) \right].
\end{equation}
The effective potential $V(\sigma)$ satisfies the renormalization group
equation
\begin{equation}\label{2.10}
\left[\mu {\partial \over \partial \mu} + \beta\left( g \right ) 
{\partial \over \partial g} - \gamma_\sigma \left( g \right) \sigma {\partial
\over \partial \sigma} \right] V\left(\sigma \right)=0
\end{equation}
Substituting eq.~(\ref{2.9}) into eq.~(\ref{2.10}) gives
\begin{equation}\label{2.11}
\gamma_\sigma \left(g \right)=0,\qquad \beta \left( g \right) = -{g^3 \over 2
\pi}.
\end{equation}
These are the exact anomalous dimension and $\beta$-function \emph{to all
orders in $g$} in the $N\rightarrow \infty$ limit. The Gross-Neveu model is an
asymptotically free theory, since the $\beta$-function is negative.

The Gross-Neveu model also exhibits spontaneous symmetry breaking. The extrema
of the effective potential eq.~(\ref{2.9}) are at 
\[
\sigma=0,\qquad \sigma = \pm \mu e^{-\pi/g^2} \equiv \pm \sigma_0,
\]
at which $V$ has the values
\[
V\left(0\right)=0,\qquad V\left(\pm\sigma_0\right) = - N {\sigma_0^2 \over 4
\pi}<0,
\]
so that the global minima of the potential are $\sigma = \pm \sigma_0$.
The discrete symmetry eq.~(\ref{2.2}) is spontaneously broken, since
\[
\vev{\sigma} = {g^2\over N} \vev{\bar \psi \psi},
\]
and the two minima $\pm \sigma_0$ are mapped into each other under this broken
symmetry. The fermions get a mass $m=\sigma_0$, since the Yukawa coupling is
$\sigma \bar \psi \psi$, and there is no wavefunction renormalization of the
$\sigma$ field.

The key simplification of the large $N$ limit was that the diagrams of the
theory reduced to a subset, fig.~\ref{fig:5}, which could be summed exactly to
give an effective Lagrangian $L_{\rm eff}$. The large $N$ limit is the same as
the semiclassical limit for $L_{\rm eff}(\sigma)$. This is evident from the
overall factor of $N$ in $L_{\rm eff}(\sigma)$, eq.~(\ref{2.5}). The form of
the functional integral
\[
\int \mathcal{D}\sigma\ e^{i S_{\rm eff}/ \hbar}
=\int \mathcal{D}\sigma\ e^{i N \tilde S_{\rm eff}/\hbar}
\]
shows that an expansion in $\hbar$ is equivalent to an expansion in $1/N$.

The Gross-Neveu Lagrangian eq.~(\ref{2.3}) can be written as
\begin{equation}\label{2.14}
L = N \bar \Psi\, i \dsl\, \Psi + N g^2 \left(\bar \Psi \Psi
\right)^2,
\end{equation}
using rescaled fermion fields $\psi = \sqrt{N} \Psi$. This also has an overall
factor of $N$, so one might naively think that the large $N$ limit is the same
as the semiclassical limit for the Lagrangian~(\ref{2.14}). This is incorrect,
because the terms in the Lagrangian have hidden $N$ dependence, because there
are $N$ flavors of $\Psi$, and Feynman diagrams have factors of $N$ from the
flavor index sums. The effective action $S_{\rm eff} \left( \sigma \right)$
has no hidden factors of $N$, since $\sigma$ is a single component flavor
singlet field. In this case, the overall factor of $N$ does imply that the
large $N$ and semiclassical limits are the same. The effective action $S_{\rm
eff}\left( \sigma \right)$ for the composite field $\sigma$ is obtained by
adding tree and loop graphs in the original Gross-Neveu theory, and so contains
quantum corrections in the Gross-Neveu Lagrangian eq.~(\ref{2.3}). The large
$N$ limit of the Gross-Neveu model is thus equivalent to the semiclassical
expansion of an effective theory of flavor singlet $\sigma$ fields
(``mesons''). The $\sigma$ effective action includes quantum corrections in the
Gross-Neveu model, so the large $N$ limit is not the same as the semiclassical
limit of the original Gross-Neveu model. A similar result holds for QCD. We
will see that the large $N$ limit of QCD is the same as the semiclassical limit
of an effective theory of color singlet mesons and baryons.

\begin{hw}[Unitarity Bound]\sl
\begin{itemize}
\end{itemize}
\noindent Show that the $a + \bar a \rightarrow b + \bar b$ amplitude must be
of order $1/N$ (or smaller) to avoid violating unitarity in the large $N$
limit.
\end{hw}

\begin{hw}\sl
\begin{itemize}
\end{itemize}
\noindent Prove eq.~(\ref{2.6}).
\end{hw}

\begin{hw}[Effective Potential in the $\overline{{\bf MS}}$ Scheme]\sl 
\begin{itemize}
\end{itemize}
\noindent Evaluate the effective potential eq.~(\ref{2.8}) by analytically
continuing the momentum integral to $2-2\epsilon$ dimensions. You can apply the
familiar rules for Feynman graphs in $D$ dimensions by using the identity
\[
\left. {\partial \over \partial \alpha}\right|_{\alpha=0} 
\left(1 +{\sigma^2 \over p^2}\right)^\alpha = 
\log \left(1 +{\sigma^2 \over p^2}\right).
\]
\end{hw}

\begin{hw}[1/N Corrections to V]\sl
\begin{itemize}
\end{itemize}
\noindent The effective potential for $\sigma$ contains two loop corrections,
such as fig.~\ref{fig:7}(a). In the method of calculation outlined above, where
one first computes the fermion functional integral, these terms are obtained by
computing loop graphs using $L_{\rm eff}$, as in fig.~\ref{fig:7}(b). These are
suppressed by $1/N$, and were neglected above. Include these (unknown) order
$g^2/N$ terms in $V$, and repeat the derivation of eq.~(\ref{2.11}). Assume
that $\gamma_\sigma\left( g \right)$ starts at order $g^2$, and $\beta\left( g
\right)$ starts at order $g^3$. Show that one obtains
\begin{eqnarray*}
\gamma_\sigma\left(g \right) &=& {1\over N}\mathcal{O}\left(g^2\right),\\
\beta\left(g \right) &=& -{g^3\over2\pi}-g \gamma_\sigma\left(g \right) + 
{1\over N}\mathcal{O}\left(g^5\right).\\
\end{eqnarray*}
\begin{figure}[tbp]
\begin{center}
\epsfig{file=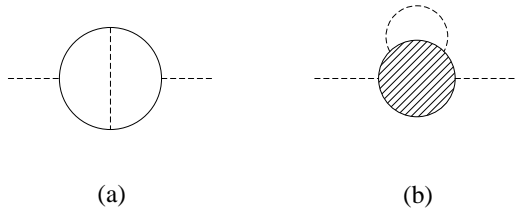,height=30mm}
\caption{A two loop correction to the effective potential $V(\sigma)$. The
correction in the theory with both $\psi$ and $\sigma$ fields is given by (a).
The same diagram in the theory obtained by integrating out $\psi$ is given by
(b).\label{fig:7}}
\end{center}
\end{figure}
\end{hw}

\section{QCD}\label{sec:qcd}

\subsection{$N$-Counting Rules for Diagrams}

The analysis of the $N$-counting rules for QCD is more complicated than that
for the Gross-Neveu model studied in the previous section. The main reason for
this is that gluons transform under the adjoint representation of the gauge
group, rather than the fundamental representation. In the Gross-Neveu model,
the dynamics could be rewritten in terms of a singlet field $\sigma = \bar \psi
\psi$. In QCD, one can construct an infinite number of gauge singlets, e.g.\ 
$\Tr F_{\mu \nu}^2$, $\Tr F_{\mu \nu}^3$, $\ldots$, $\Tr F_{\mu \nu}^N$, from
the gluon field-strength tensor $F_{\mu \nu}$.

The theory we will study is an $SU(N)$ gauge theory with $N_F$ flavors of
fermions (quarks) in the fundamental representation of $SU(N)$. The gauge field
is an $N \times N$ traceless hermitian matrix, $A_\mu=A_\mu^A T^A$, and the
covariant derivative is
\[
D_\mu = \partial_\mu + i {g\over\sqrt N} A_\mu.
\]
The matrices $T^A$ are normalized so that
\[
\Tr T^A T^B = \frac{1}{2} \delta^{AB}.
\]
The coupling constant has been chosen to be $g/\sqrt N$, rather than $g$,
because this will lead to a theory with a sensible (and non-trivial) large $N$
limit. The field strength is
\[
F_{\mu\nu}=\partial_\mu A_\nu - \partial_\nu A_\mu + i {g \over \sqrt N}\left[
A_\mu , A_\nu \right],
\]
and the Lagrangian is
\begin{equation}\label{3.1}
L = - \frac{1}{2}\Tr F_{\mu \nu} F^{\mu \nu} + \sum_{k=1}^{N_F}
\bar \psi_k \left( i\, \Dslash - m_k \right) \psi_k.
\end{equation}
The large $N$ limit will be taken with the number of flavors $N_F$ fixed. It is
also possible to consider other limits, such as $N \rightarrow \infty$ with
$N_F/N$ held fixed~\cite{veneziano}.

One way to understand the $g/\sqrt{N}$ scaling of the coupling constant is to
look at the QCD $\beta$-function,
\begin{equation}\label{3.2}
\mu {d g \over d \mu} = - b_0 {g^3 \over 16 \pi^2} +
\mathcal{O}\left(g^5\right),\qquad b_0=\frac{11}{3}N - \frac{2}{3}N_F,
\end{equation}
using the conventionally normalized coupling constant. This equation does not
have a sensible large $N$ limit since $b_0$ is order $N$. Replacing $g$ by 
$g/\sqrt{N}$ in eq.~(\ref{3.2}) gives
\[
\mu {d g \over d \mu} = - \left( \frac{11}{3} - \frac{2}{3}\frac{N_F}{N}
\right) {g^3 \over 16 \pi^2} +
\mathcal{O}\left(g^5\right).
\]
The $\beta$-function equation now has a well-defined limit as $N\rightarrow
\infty$. The $N_F$ term is suppressed by $1/N$, and we will soon see that all
fermion loop effects are $1/N$ suppressed. The scale parameter of the strong
interactions, \lqcd, is held fixed as $N \rightarrow \infty$, since $N$ drops
out of the equation for the running of $g$. Thus the large $N$ limit for QCD
with the coupling constant scaling like $1/\sqrt{N}$ is equivalent to taking
the limit $N \rightarrow \infty$ holding the string tension, or a meson mass
such as the $\rho$ mass, fixed.

\begin{figure}[tbp]
\begin{center}
\epsfig{file=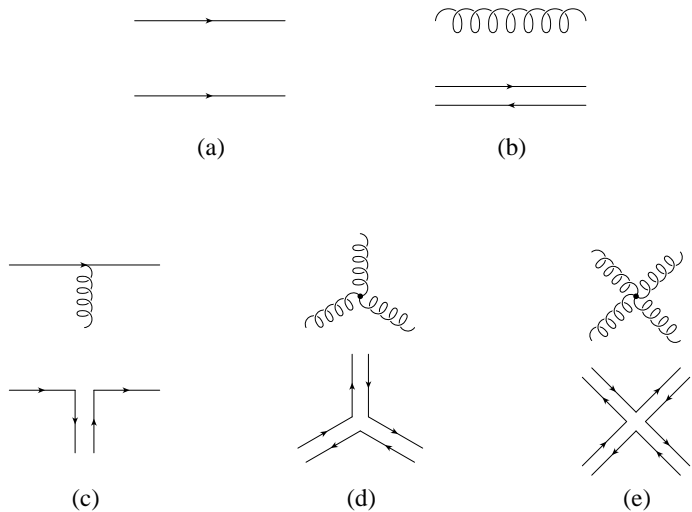,height=70mm}
\caption{'t~Hooft's double line notation. The lower diagram shows each QCD
propagator or interaction vertex in the double line notation.\label{fig:8}}
\end{center}
\end{figure}

To analyze the $N$-counting rules for QCD, one needs a simple way to count the
powers of $N$ in a given Feynman diagram. This can be done with the help of a
trick due originally to 't~Hooft. The quark propagator is
\begin{equation}\label{3.4}
\vev{\psi^a \left(x \right) \bar{{\psi^b}}\left(y \right)} = \delta^{ab}
S\left( x-y \right).
\end{equation}
This is represented diagrammatically by a single line (fig.~\ref{fig:8}(a)),
and the color at the beginning of the line is the same as at the end of the
line, because of the $\delta^{ab}$ in eq.~(\ref{3.4}). The gluon propagator is
\[
\vev{A^A_\mu \left(x \right) A_\nu^B \left(y \right)} = \delta^{AB}
D_{\mu\nu}\left( x-y \right),
\]
where $A$ and $B$ are indices in the adjoint representation. Instead of
treating a gluon as a field with a single adjoint index, it is preferable to
treat it as an $N\times N$ matrix with two indices in the $N$ and $\overline
N$ representations, $\left( A_\mu \right)^a{}_b= A_\mu^A \left( T^A \right)^ a
{}_b$. The gluon propagator can be rewritten as
\[
\vev{ A_{\mu b}^a \left(x \right) A_{\nu d}^c 
\left( y \right)} = D_{\mu\nu}\left( x-y \right)\left( \frac{1}{2}\delta^a_d
\,\delta^c_b - \frac{1}{2N} \delta^a_b\, \delta^c_d \right),
\]
where the identity
\begin{equation}\label{3.7}
\left(T^A \right)^a{}_b \left(T^A\right)^c{}_d = \frac{1}{2}\delta^a_d
\, \delta^c_b - \frac{1}{2N} \delta^a_b \, \delta^c_d \qquad \left(SU(N)\right)
\end{equation}
has been used. The corresponding identity for $U(N)$ is
\begin{equation}\label{3.8}
\left(T^A \right)^a{}_b \left(T^A\right)^c{}_d = \frac{1}{2}\delta^a_d
\, \delta^c_b \qquad \left(U(N)\right),
\end{equation}
where the $U(1)$ generator has the same normalization as the $SU(N)$
generators. It is convenient to use the $U(N)$ identity eq.~(\ref{3.8}) instead
of the $SU(N)$ identity eq.~(\ref{3.7}) for analyzing the $N$-dependence of
Feynman diagrams. The correct $SU(N)$ propagator is given by including an
additional $U(1)$ ghost field that cancels the extra $U(1)$ gauge boson in
$U(N)$. The effects of the $U(1)$ gauge boson are $1/N^2$ suppressed, as we
will see later. In most applications, the difference between $U(N)$ and 
$SU(N)$ will turn out to be unimportant. The reason is that, typically, what
one can prove is that a certain amplitude first occurs at some order in $1/N$.
For such computations, the numerical size of the $1/N$ corrections is
irrelevant.

The $U(N)$ gluon propagator can then be represented using 't~Hooft's double
line notation, as in fig.~\ref{fig:8}(b). The color lines represent the $N$ and
$\overline N$ indices $a$ and $b$ on $A_\mu^a{}_b$, and color is conserved
during propagation because of the $\delta$-function structure in
eq.~(\ref{3.8}). The gauge-fermion vertex $\bar \psi_a\, {\aslash}^a{}_b
\psi^b$ is shown in fig.~\ref{fig:8}(c). The double line notation provides a
simple way to keep track of the color index contractions in a Feynman graph.

\begin{figure}[tbp]
\begin{center}
\epsfig{file=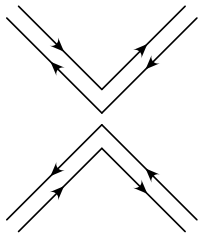,height=30mm}
\caption{Color flow for a four-gluon interaction which has two color traces.
\label{fig:9}}
\end{center}
\end{figure}

The three and four gauge boson vertices arise from the $\Tr F_{\mu\nu} 
F^{\mu\nu}$ gluon kinetic energy. Each kinetic energy term is a single trace
over color. The three-gluon vertex arises from terms such as
\begin{equation}\label{3.15}
\Tr A_\mu A_\nu \partial_\mu A_\nu = A_\mu^a{}_b A_\nu^b{}_c 
\partial_\mu A_\nu^c{}_a,
\end{equation}
where the color indices have been shown explicitly. It is represented using the
double line notation as fig.~\ref{fig:8}(d), and the four gluon vertex as in
fig.~\ref{fig:8}(e). It is important that all the interactions arise from a
single color trace --- otherwise one could have color flow in a four-gluon
vertex as in fig.~\ref{fig:9}, where the diagram can be broken up into two
disconnected color flows.

\begin{figure}[tbp]
\begin{center}
\epsfig{file=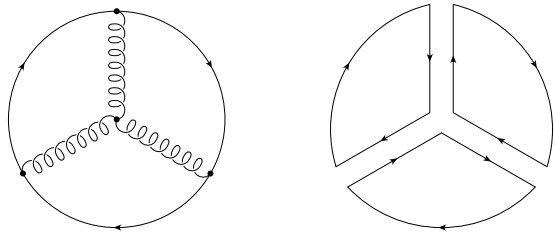,height=30mm}
\caption{A Feynman graph, and one of its partner double line graphs,
representing a particular color index contraction. This is an example of a
planar diagram.\label{fig:10}}
\end{center}
\end{figure}

\begin{figure}[tbp]
\begin{center}
\epsfig{file=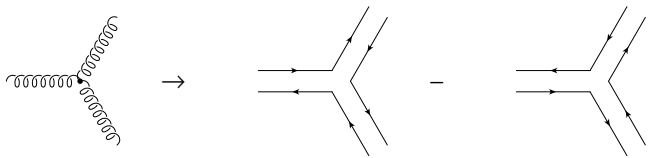,height=20mm}
\caption{The complete three-gluon vertex in double line notation.
\label{fig:20}}
\end{center}
\end{figure}

Every Feynman graph in the original theory can then be written as a sum of
double line graphs. Each double line graph gives a particular color index 
contraction of the original diagram. An example of a Feynman graph and one of 
its double line partners is shown in fig.~\ref{fig:10}.
One Feynman graph can give rise to several double line graphs. For example, the
three-gluon vertex is given by eq.~(\ref{3.15}), plus a term with $A_\mu
\leftrightarrow A_\nu$. The complete three-gluon vertex in the double line
notation is represented in fig.~\ref{fig:20}.

One can think of each double line graph as a surface obtained by gluing
polygons together at the double lines. Since each line has an arrow on it, and
double lines have oppositely directed arrow, one can only construct orientable
polygons in an $SU(N)$ theory. For $SO(N)$, the fundamental representation is a
real representation, and the lines do not have arrows. In this case, it is
possible to construct non-orientable surfaces such as Klein bottles.

To compute the $N$-dependence requires counting powers of $N$ from sums over
closed color index loops, as well as factors of $1/\sqrt{N}$ from the explicit
$N$ dependence in the coupling constants. It is convenient to use a rescaled
Lagrangian to simplify the derivation of the $N$-counting rules. Define
rescaled gauge fields $g A /\sqrt{N} \rightarrow \res{A} $ so that the
covariant derivative is $D_\mu = \partial_\mu + i \res A_\mu$, and rescaled
fermion fields $\psi \rightarrow \sqrt N \res \psi$ so that the Lagrangian
becomes
\begin{equation}\label{3.9}
L = N\left[- \frac{1}{2g^2}\Tr \res F_{\mu \nu} \res F^{\mu \nu} +
\sum_{k=1}^{N_F} \bar \res \psi_k \left( i\, \Dslash - m_k \right) \res \psi_k
\right] .
\end{equation}
The Lagrangian has an overall $N$, but the theory does not reduce to a
classical theory of quarks and gluons in the $N \rightarrow \infty$ limit,
because the number of components of $\res \psi$ and $\res A_\mu$ grows with
$N$. 

One can read off the powers of $N$ in any Feynman graph from eq.~(\ref{3.9}). 
Every vertex has a factor of $N$, and every propagator has a factor of $1/N$.
In addition, every color index loop gives a factor of $N$, since it represents
a sum over $N$ colors. In the double line notation where Feynman graphs
correspond to polygons glued to form surfaces, each color index loop is the
edge of a polygon, and is the face of the surface. Thus one finds that a
connected vacuum graph (i.e.\ with no external legs) is of order
\begin{equation}\label{3.10}
N^{V-E+F} = N^\chi,
\end{equation}
where $V$ is the number of vertices, $E$ is the number of edges, $F$ is the
number of faces, and $\chi \equiv V-E+F$ is a topological invariant known as
the Euler character. For a connected orientable surface
\begin{equation}\label{3.11}
\chi = 2 - 2 h - b,
\end{equation}
where $h$ is the number of handles, and $b$ is the number of boundaries (or
holes). For a sphere $h=0$, $b=0$, $\chi=2$; for a torus, $h=1$, $b=0$,
$\chi=0$.

A quark is represented by a single line, and so a closed quark loop is a
boundary. Thus every closed quark loop brings a $1/N$ suppression. The maximum
power of $N$ is two, from graphs with $h=b=0$. These are connected graphs with
no closed quark loops, with the topology of a sphere. Remove one polygon from
the sphere, so that one obtains a sphere with one hole. This can be flattened
into a diagram drawn on a flat sheet of paper, with the hole as the outermost
edge. One can then glue back the removed polygon by thinking of it as the
paper exterior to the diagram with infinity identified. Thus the order $N^2$
graphs are planar diagrams with only gluons, that is, they can be drawn on the
surface of a sheet of paper without having a gluon ``jump'' over another. All
points where gluon lines cross have to be interaction vertices.

A diagram with $c$ connected pieces can be of order $N^{2c}$, and so grows with
the number of connected pieces. This is not surprising; the sum of all diagrams
is the exponential of the connected diagrams. The connected diagrams are of
order $N^2$, corresponding to a vacuum energy of order $N^2$, which is to be
expected since there are $\mathcal{O}\left(N^2\right)$ gluon degrees of
freedom. Expanding the exponential gives arbitrary high powers of $N$. From
now on, we will restrict $N$-counting to the connected diagrams. One can obtain
the $N$ dependence of a disconnected diagram by multiplying together the $N$
dependence of all the connected pieces.

\begin{figure}[tbp]
\begin{center}
\epsfig{file=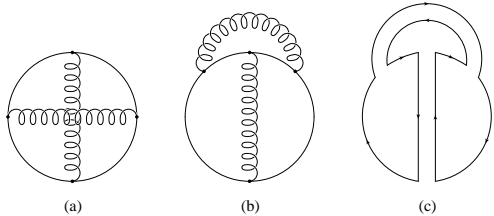,height=30mm}
\caption{An example of a non-planar diagram. Diagram (a) can be redrawn as (b),
but it still non-planar since the quark loop must form the outermost edge of
the diagram. The double line version of the graph is shown in (c).
\label{fig:11}}
\end{center}
\end{figure}

We will often be interested in correlation functions that depend on properties
of the quarks, such as masses. The leading graphs that depend on quarks must
have at least one quark line, and are order $N$, with $h=0$ and $b=1$. One
might expect the quark contribution to the vacuum energy to be of order $N$,
since there are $N$ quarks of each flavor. The order $N$ quark diagrams have
the topology of a sphere with one hole, with the quark loop forming the edge of
the hole. One can then flatten them out into a planar diagram as for gluons. In
this case, the order $N$ diagrams are written as planar diagrams with a single
quark loop which forms the outermost edge of the diagram. An example of a
planar quark diagram is fig.~\ref{fig:10}. Figure~\ref{fig:10} is of order $N$,
since $h=0$ and $b=1$. This can also be seen using the original $N$-counting
rules of the Lagrangian eq.~(\ref{3.1}). Each vertex has a factor $g/\sqrt{N}$,
and each closed index loop brings a factor of $N$, so the graph is of order
$(1/\sqrt{N})^4 \times N^3 = N$. Figure~\ref{fig:11}(a) is not a planar
diagram, even though it can be drawn as fig.~\ref{fig:11}(b), because the
diagram must be planar when drawn with the quark line as the outermost edge.
Figure~\ref{fig:11}(a) is of order $1/N$, since the vertex factors give 
$(1/\sqrt{N})^4$ and the color index sum gives a factor of $N$. Note that for a
given number of quark loops (boundaries $b$), the expansion is in powers of
$1/N^2$, rather than $1/N$, because of the $-2$ in front of $h$ in
eq.~(\ref{3.11}). 

Large N diagrams for QCD look like two-dimensional surfaces. For example, the
leading diagram in the pure-glue sector has the topology of a sphere, and the
leading diagram in the quark sector is a surface with the quark as the
outermost edge. One can imagine all possible planar gluon exchanges as filling
out the surface into a two-dimensional world-sheet. It has been conjectured
that this is the way in which large N QCD might be connected with string
theory, with planar diagrams representing the leading order string theory
diagrams. The topological counting rules for the $1/N$ suppression factors in
QCD are the same as that for the string coupling constant in the string loop
expansion. The connection between large N QCD and string theory has never been
made precise. Two major obstacles are that QCD is neither supersymmetric nor
conformally invariant. One result in this direction is that the partition
function for $SU(N)$ Yang-Mills theory (i.e.\ no quarks) in two dimensions was
shown to agree with the partition function of a string
theory~\cite{GrossTaylor} by explicit calculation. The connection was possible
because Yang-Mills theory in two-dimensions is a free field theory, and the
partition function only depends on the topology of the background spacetime. To
reproduce the correct $N$-factors, it was necessary to use a modified string
theory, in which folds are suppressed.

\subsubsection{$U(1)$ Ghosts}

\begin{figure}[tbp]
\begin{center}
\epsfig{file=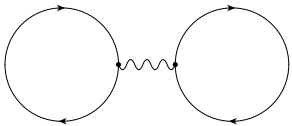,height=15mm}
\caption{A connected diagram with one $U(1)$ ghost gauge boson exchange.
\label{fig:21}}
\end{center}
\end{figure}

The corrections due to the difference between using $SU(N)$ and $U(N)$ are
straightforward to analyze. As mentioned earlier, the $SU(N)$ theory can be
thought of as a $U(N)$ theory with an additional $U(1)$ ghost gauge field to
cancel the extra $U(1)$ gauge field in $U(N)$. The $U(1)$ ghost field does not
couple to the $U(N)$ gauge bosons, since the $U(1)$ generator commutes with all
the $U(N)$ generators. We only need to consider exchanges of the $U(1)$ ghost
field between quark lines. The additional powers of $N$ due to the $U(1)$ ghost
are most simply counted using the original form of the Lagrangian
eq.~(\ref{3.1}), rather than the rescaled version eq.~(\ref{3.9}). Consider a
connected diagram, with some gluon lines and ghost $U(1)$ lines. The $U(1)$
ghost does not change the color structure of the diagram, so the $N$ dependence
from the color index loops, etc.\ is obtained using the counting rules
discussed above, for the diagram with the $U(1)$ ghosts erased. In addition,
one gets a factor of $1/N$ from each $U(1)$ ghost propagator (see
eq.~(\ref{3.7})), and a factor of $1/N$ from the two coupling constants at the
two ends of each gauge boson line. Thus each $U(1)$ ghost brings a $1/N^2$
suppression. The only subtlety in this argument is that $U(1)$ exchange can
turn an otherwise disconnected diagram into a connected diagram. Thus the
leading diagram with one $U(1)$ ghost has two quark loops, as in
fig.~\ref{fig:21}, and is order $N \times N \times 1/N^2 = \ord 1$. This is
only $1/N$ suppressed relative to the leading connected quark diagrams, which
are order $N$. However, the effect of the $U(1)$ ghost in fig.~\ref{fig:21} is
to precisely cancel the corresponding graph with a $U(N)$ boson, since
fig.~\ref{fig:21} with $SU(N)$ boson exchange vanishes, because $\Tr T^A=0$.
The net result is that the difference between $SU(N)$ and $U(N)$ is order
$1/N^2$.

\begin{hw} \sl
\begin{itemize}
\end{itemize}
\noindent Prove eqs.~(\ref{3.7}), (\ref{3.8}).
\end{hw}

\subsection{The 't Hooft Model}

The 't Hooft model is large $N$ QCD in $1+1$ dimensions. This theory was solved
by 't Hooft~\cite{thooft1,thooft2} to obtain the exact meson spectrum. There
is an extensive discussion of this model in Coleman's lectures~\cite{coleman}.
I will not repeat the solution of the model here. Instead, I will summarize why
the model is solvable, and show some of the numerical solutions.

In $1+1$ dimensions, the gauge coupling constant $g$ has dimensions of a mass,
and becomes relevant in the infrared. The theory is confining, and has a linear
potential at large distances. Even QED confines in $1+1$ dimensions, and has a
linear potential. It is convenient to use light-cone coordinates
\[
x^{\pm} = \frac{1}{\sqrt{2}}\left(x^0\pm x^1\right),
\]
with metric
\[
ds^2 = 2 dx^+ dx^-,\qquad g_{\mu\nu}=
\left(\begin{array}{cc}
0 & 1 \\ 1 & 0 
\end{array}\right),
\]
and light-cone gauge $A^+=A_-=0$. The field-strength tensor has a single
non-vanishing component,
\[
F_{+-} = \partial_+ A_- - \partial_- A_+ + i{g \over \sqrt{N}}\left[
A_+,A_-\right] = - \partial_- A_+ ,
\]
so the theory becomes effectively Abelian. This is the first important
simplification. The second is planarity, which allows one to solve the theory
exactly.
 
\begin{figure}[tbp]
\begin{center}
\epsfig{file=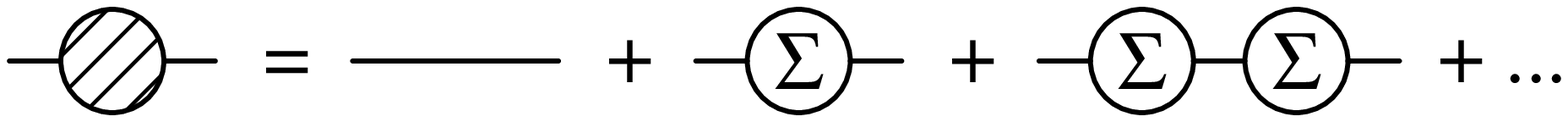,height=15mm}
\caption{The quark propagator in terms of the one-particle irreducible 
self-energy $\Sigma$.\label{fig:12}}
\end{center}
\end{figure}

\begin{figure}[tbp]
\begin{center}
\epsfig{file=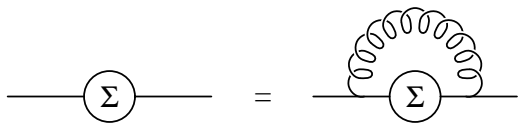,height=20mm}
\caption{The equation for $\Sigma$. \label{fig:13}}
\end{center}
\end{figure}

Consider the quark propagator, which can be written in terms of the
one-particle irreducible piece $\Sigma$, as in fig.~\ref{fig:12}. 
The equation for $\Sigma$ is given graphically in fig.~\ref{fig:13}. Planarity
is crucial for the derivation of this relation. We earlier derived the 
$N$-counting rules only for vacuum graphs. The results can easily be extended
to the fermion two-point function, which is obtained by differentiating a
vacuum graph once with respect to a fermion bilinear source, i.e., by cutting
the closed quark loop at one point. Planarity of the vacuum graphs then implies
that the quark propagator is given by planar graphs with all gluons on one side
of the quark line. The first gluon emitted by the quark must be the last gluon
absorbed, because the diagram is planar, and there are no gluon
self-interactions. This immediately leads to fig.~\ref{fig:13}.

\begin{figure}[tbp]
\begin{center}
\epsfig{file=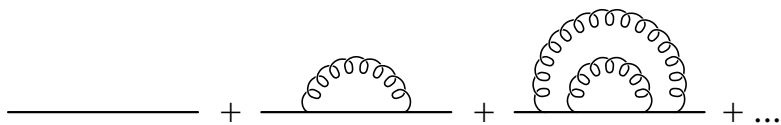,height=20mm}
\caption{Rainbow sum for the exact quark propagator.\label{fig:50}}
\end{center}
\end{figure}

The equation for $\Sigma$ in fig.~\ref{fig:13} can be iterated, to give the
solution fig.~\ref{fig:50} for the quark propagator. 
The diagrams in fig.~\ref{fig:50} are known as rainbow diagrams, and the
rainbow approximation is exact in the 't~Hooft model. The analytical solution
of fig.~\ref{fig:13} is straightforward~\cite{coleman}. The solution is that
the quark propagator is given by the free-quark form $1/\left( \pslash-M
\right)$, with the renormalized quark mass $M$ given in terms of the bare quark
mass by
\[
M^2=m^2-{g^2\over 2 \pi}.
\]
Note that there is still a pole in the quark propagator, even though the theory
is confining, and there are no quark states in the spectrum of the theory. Also
note that the pole in the quark propagator can be tachyonic if $m^2 <
g^2/2\pi$. Nevertheless, the theory has a sensible meson spectrum for all
values of $m^2 \ge 0$.

\begin{figure}[tbp]
\begin{center}
\epsfig{file=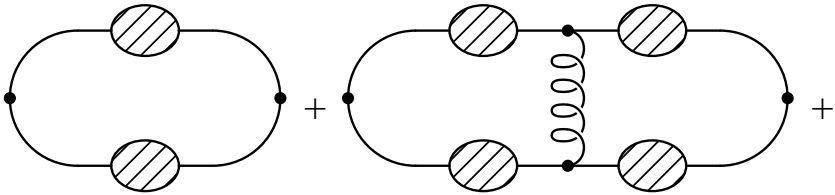,height=25mm}
\caption{Graphical expansion for the meson propagator. The shaded blobs are
full quark propagators. \label{fig:14}}
\end{center}
\end{figure}

\begin{figure}[tbp]
\begin{center}
\epsfig{file=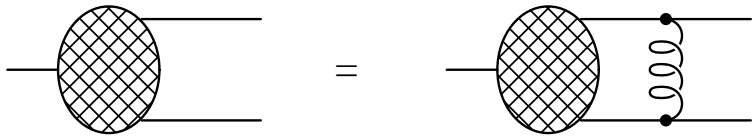,height=20mm}
\caption{The Bethe-Salpeter equation for the meson wavefunction. The
cross-hatched blob is the meson wavefunction $\phi(x)$. \label{fig:15}}
\end{center}
\end{figure}

The meson propagator has the exact graphical expansion in fig.~\ref{fig:14}, 
and the ladder graph approximation is exact in the large $N$ limit. This again
follows trivially from the planar diagram structure of the theory and the
absence of gluon self interactions: gluons are not allowed to cross in any
graph. The Bethe-Salpeter equation for the meson wavefunction, known as the 
't~Hooft equation, is shown graphically in fig.~\ref{fig:15}, 
and follows from the meson propagator fig.~\ref{fig:14}. Let $P$ be the total
momentum of the meson, $q$ be the momentum of the quark (the quark is
well-defined, since there is no pair creation in the large $N$ limit),
$x=q_-/P_-$, and $\phi(x)$ the amplitude to find the quark with this light-cone
momentum fraction. A simple calculation leads to the 't Hooft
equation~\cite{thooft1,thooft2,coleman} for the meson wavefunction
\[
\mu^2 \phi(x) = \left[{M_1^2\over x} + {M_2^2 \over 1-x} \right] \phi(x) -
{g^2 \over 2 \pi} \int_0^1 dy\ P\left( {1 \over (x-y)^2} \right) \phi(y),
\]
where $P$ denotes the principal value, $\mu$ is the meson mass, $M_1$ is the
renormalized quark mass, and $M_2$ is the renormalized antiquark mass.

\begin{figure}[tbp]
\begin{center}
\epsfig{file=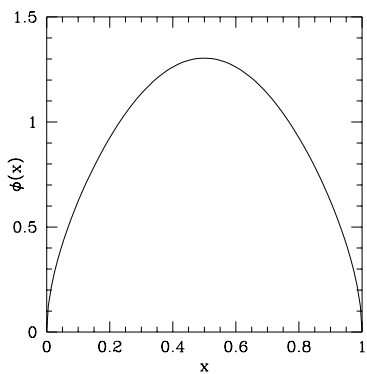,height=50mm}
\caption{The ground state wavefunction of the 't~Hooft model with $m_1=m_2=1$.
\label{fig:16}}
\end{center}
\end{figure}

\begin{figure}[tbp]
\begin{center}
\epsfig{file=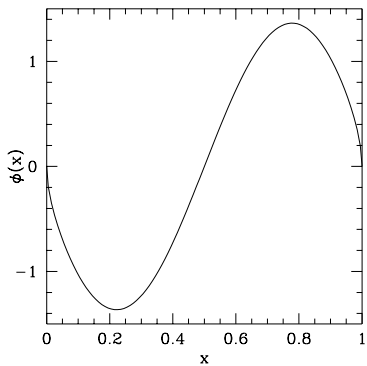,height=50mm}
\caption{The wavefunction for the first excited state of the 't~Hooft model 
with $m_1=m_2=1$. \label{fig:17}}
\end{center}
\end{figure}

\begin{figure}[tbp]
\begin{center}
\epsfig{file=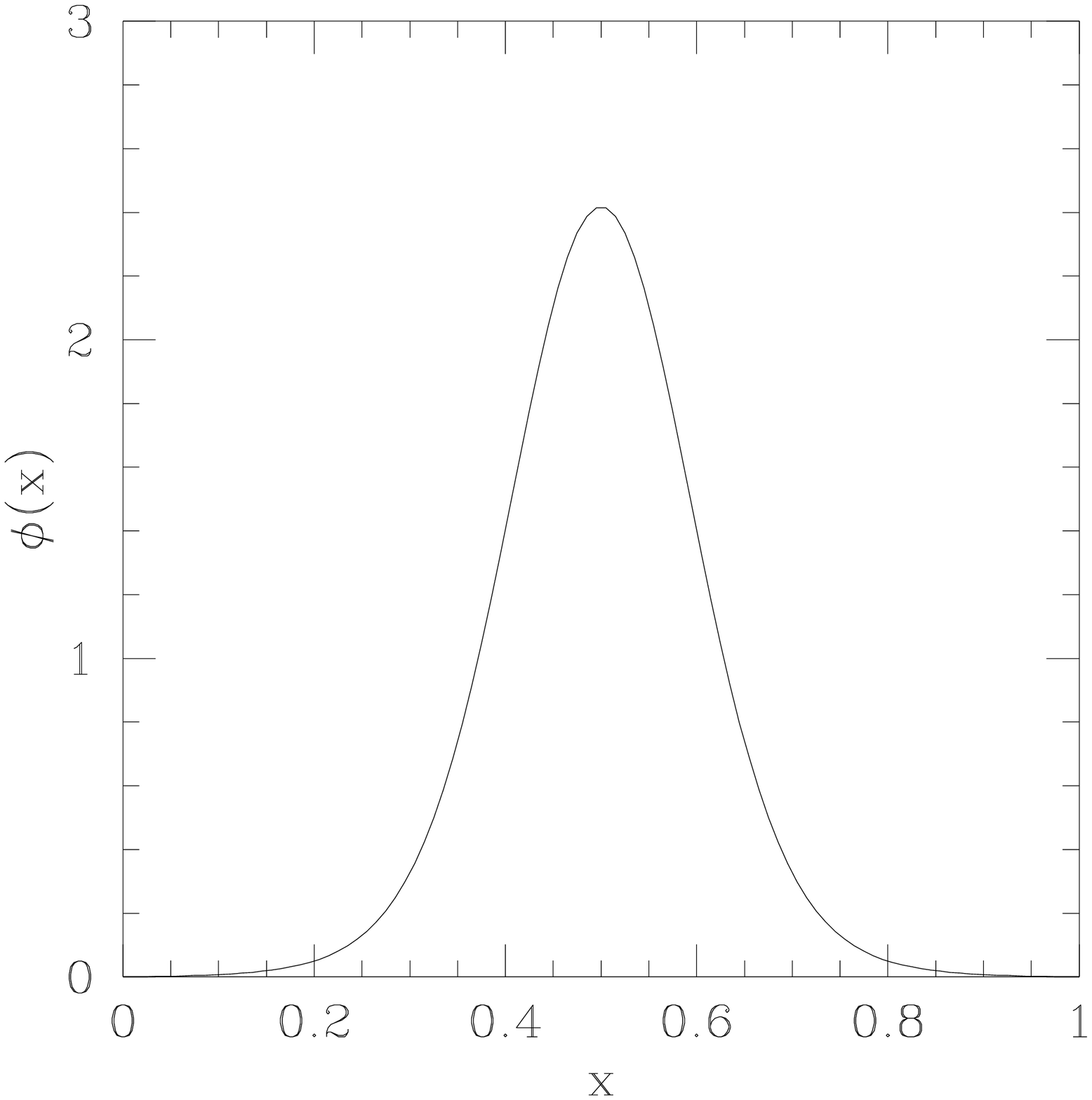,height=50mm}
\caption{The ground state wavefunction of the 't~Hooft model with $m_1=m_2=10$.
\label{fig:18}}
\end{center}
\end{figure}

\begin{figure}[tbp]
\begin{center}
\epsfig{file=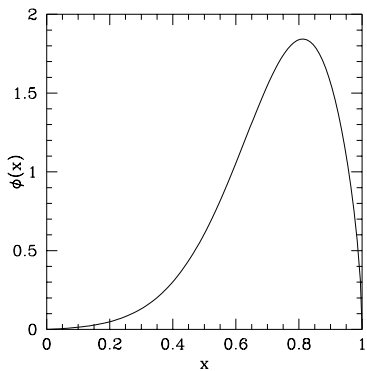,height=50mm}
\caption{The ground state wavefunction of the 't~Hooft model with $m_1=5$ and
$m_2=1$. \label{fig:19}}
\end{center}
\end{figure}

This equation can be solved numerically using a Multhopp
transform~\cite{multhopp}. The ground state wavefunction $\phi(x)$ for
$m_1=m_2=1$ (renormalized masses in units of $g/\sqrt{2\pi}$) is shown in
fig.~\ref{fig:16}. The meson mass is $\mu=2.7$, which is larger than the sum of
the two quark masses. The wavefunction of the first excited state is shown in
fig.~\ref{fig:17}, and corresponds to a meson with mass $\mu=4.16$. One can
show that for large excitation number $n$, the meson mass is linear in $n$. The
ground state wavefunction for $m_1=m_2=10$ is shown in fig.~\ref{fig:18}; the
meson mass is $\mu=20.55$. Clearly, increasing the quark mass narrows the
momentum spread in the wavefunction, and also decreases $\mu - m_1 - m_2$, the
strong interaction contribution to the mass. For unequal masses, the momentum
distribution is asymmetric, and the heavy quark carries most of the momentum.
Figure~\ref{fig:19} shows the ground state wavefunction for $m_1=5$, $m_2=1$,
with a meson mass $\mu=6.72$. For light quarks, $m\rightarrow 0$, the meson
wavefunction is not affected much by the quark mass. As one might expect, the
structure of the wavefunction is governed by the scale $g$ rather than $m$. The
wavefunctions of the 't~Hooft model show many of the features one would expect
for the wavefunctions of mesons in QCD in $3+1$ dimensions.

\subsection{$N$-Counting Rules for Correlation Functions}

\begin{figure}[tbp]
\begin{center}
\epsfig{file=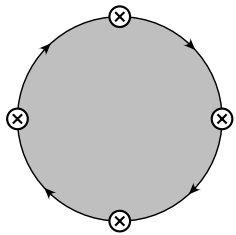,height=30mm}
\caption{Leading contribution to a quark correlation function. The
shaded region represents planar gluons, and $\otimes$ are insertions of the
fermion bilinears $\res O_i$. \label{fig:22}}
\end{center}
\end{figure}

We have discussed the $N$-counting rules for connected vacuum diagrams. These
can be used to derive $N$-counting rules for gluon and quark correlation
functions. The correlators we will study are vacuum expectation values of
products of gauge invariant quark and gluon operators. The operators need not
be local; all we require is that they cannot be split into pieces which are
separately color singlets. Allowed operators are
\[
\bar{\res \psi} \res \psi,\ \res F_{\mu\nu} \res F^{\mu\nu},
\ \bar{\res \psi} D^\mu \res \psi,\
\bar{\res \psi} \res F^{\mu\nu} \res \psi,\ 
\bar{\res \psi}(y) P e^{-i \int_x^y \res A_\mu dz^\mu} \res \psi(x),
\]
but not
\[
\left( \bar {\res \psi} \res \psi \right)^2.
\]
Operators involving quark fields must be bilinear in the quarks. Let $\res O_i$
denote a generic operator written in terms of rescaled fields, and add the
source term $N J_i \res O_i$ to the rescaled Lagrangian eq.~(\ref{3.9}). The
entire Lagrangian still has an overall factor of $N$, so the $N$-counting rule
eq.~(\ref{3.10}) still holds. Connected correlation functions are then obtained
by differentiating $W(J)$, the sum of connected vacuum graphs, with respect to
the sources
\[
\vev{\res O_1 \res O_2 \ldots \res O_r}_C = 
{1 \over i N} {\partial \over \partial J_1}
\ldots {1 \over i N} {\partial \over \partial J_r} W(J).
\]
The order $N^2$ contribution to $W(J)$ is from graphs with only gluon lines.
This can contribute to the correlation function $\veV{\res O_1 \res O_2 \ldots
\res O_r}_C$, provided none of the operators contain any quark fields. Thus
pure-gluon $r$-point correlation functions are of order $N^{2-r}$. The diagrams
that contribute to these are planar diagrams with insertions of $\res O_i$. The
first contribution to $W(J)$ that involves quarks is of order $N$. Thus
$r$-point correlation functions that involve quark fields are of order
$N^{1-r}$. The diagrams that contribute to them are planar diagrams with a
single quark loop with the fermion bilinears inserted on the quark line, as
shown in fig.~(\ref{fig:22}).

The $N$-counting rules were derived using the rescaled Lagrangian
eq.~(\ref{3.9}). One can obtain $N$-counting rules for correlation functions
with fields normalized as in the original Lagrangian eq.~(\ref{3.1}) by using
the replacement
\begin{equation}\label{resfield}
\psi = \sqrt{N} \res \psi,\ 
A = {\sqrt{N}\over g} \res A.
\end{equation}
In particular, quark bilinears are related by a factor of $N$, $O_i = N \res
O_i$, e.g. $\bar \psi \psi = N \bar \res \psi \res \psi$.

The $N$-counting rules for correlation functions can be used to derive the
$N$-counting rules for meson and glueball scattering amplitudes. We will use
the notation $\res G_i$ to denote a gauge invariant pure-gluonic operator, and
$\res H_i$ to denote a gauge invariant operator bilinear in the quark fields,
where the operators are written in terms of the rescaled fields. Gluon
operators $\res G_i$ can create glueballs, and quark bilinears $\res H_i$ can
create mesons. The two point function $\veV{\res G_1 \res G_2}_c$ is of order
unity, so $\res G_i$ creates glueball states with unit amplitude. The $r$-point
function $\veV{\res G_1 \ldots \res G_r}_c$ is of order $N^{2-r}$. Thus
$r$-glueball interaction vertices are of order $N^{2-r}$, and \emph{each
additional glueball gives a $1/N$ suppression}. The meson two point correlation
function $\veV{\res H_1 \res H_2}_c \sim 1/N$. Thus $\sqrt N \res H_i$ creates
a meson with unit amplitude. The $r$-point function $\veV{\sqrt N \res H_1
\ldots \sqrt N \res H_r}_c$ is of order $N^{1-r/2}$. Thus $r$-meson interaction
vertices are of order $N^{1-r/2}$, and \emph{each additional meson gives a
$1/\sqrt N$ suppression}. Finally, one can look at mixed glue-ball meson
correlators, $\veV{\res G_1 \ldots \res G_r \res H_1 \ldots \res H_s}_c$,
which are of order $N^{1-r-s}$. Thus an interaction vertex involving $r$
glueballs and $s$ mesons is of order $N^{1-r-s/2}$, so that $\veV{\res G \sqrt
N \res H}$ is of order $1/\sqrt N$ --- meson-glueball mixing is of order
$1/\sqrt N$, and vanishes in the $N \rightarrow \infty$ limit.

One important result that we will need later is that the pion decay constant
$f_\pi$ is of order $\sqrt N$. The matrix element $\mE{0}{\bar q \gamma^\mu
\gamma_5 T^a q}{\pi^b(p)}=if_\pi p^\mu \delta^{ab}$, where the quark fields are
normalized as in the original Lagrangian eq.~(\ref{3.1}). The $N$-dependence of
the matrix element can be obtained from $\veV{\res H_1 \sqrt N \res H_2}\sim
\mathcal{O}(1/\sqrt N)$, where the first bilinear is the axial current, and
the second bilinear produces a pion from the vacuum with amplitude of order
unity. One needs to multiply $\res H_1$ by an additional factor of $N$ to
convert the axial current from rescaled fields to the original quark fields,
so that
\[
f_\pi \propto \sqrt N.
\]

The $N$-counting rules imply that one has a weakly interacting theory of mesons
and glueballs with a coupling constant $1/\sqrt N$. As in any weakly
interacting theory, one can expand in the coupling constant $1/\sqrt N$. The
leading order graphs are tree-graphs, and the leading order singularities are
poles. At one-loop (i.e.\ $1/N$), one gets 2 particle cuts, at two loops,
three-particle cuts, and so on. QCD, a strongly interacting theory of quarks
and gluons, has been rewritten as a weakly interacting theory of hadrons.  The
leading (in $N$) interactions bind the quarks and gluons into color singlet
hadrons. The residual interactions between these hadrons are $1/N$ suppressed.
The $1/N$ expansion is also equivalent to the semiclassical expansion for the
meson theory. These results will also hold for baryons.

The spectrum of the theory contains an infinite number of narrow glueball and
meson resonances. The resonances are narrow, because their decay widths vanish
as $N \rightarrow \infty$, since all decay vertices are proportional to powers
of the weak coupling constant $1/\sqrt N$ and hadron masses (i.e.\ phase space
factors) do not grow with $N$. There must be an infinite number of resonances
to reproduce the logarithmic running of QCD correlation functions. A meson
two-point correlation function can be written as a sum over resonances,
\[
\int d^4 x e^{i q \cdot x} \vev{Q(x) Q(0)}_C = \sum_i {Z_i \over q^2-m_i^2},
\]
since single meson exchange dominates in the large $N$ limit. The left hand
side has logarithms of $q^2$, which can only be reproduced by the right hand
side if there are an infinite number of terms in the sum.

\subsection{The Master Field}

The $N$-counting rules imply that the functional integral measure is
concentrated on a single gauge orbit, and that fluctuations vanish in the $N
\rightarrow \infty$ limit. For example, if $\res G$ is a gauge invariant
operator made of gluon fields, $\veV{\res G} \sim \mathcal{O}(N)$, and its
variance is
\[
\left( \Delta \res G \right)^2 \sim \vev{\res G^2}-\vev{\res G}^2 = 
\vev{\res G^2}_c \sim 
\mathcal{O}(1).
\]
Thus $\Delta \res G / \res G \sim \mathcal{O}(1/N) \rightarrow 0$. It is easy
to see that all gauge invariant observables have no fluctuations in the $N
\rightarrow \infty$ limit. The functional integral measure must then be
concentrated on a single gauge orbit, known as the master orbit, represented by
a set of gauge equivalent vector potentials $A_\mu(x)$. It is expected that one
can find a point on this orbit where the vector potential is given by constant
matrices $A_\mu$. This is the master field, and all correlation functions are
given by computing them in this single field configuration. A lot of
information can be encoded in the master field, since it is an $\infty \times
\infty$ matrix. The master field has recently been determined for QCD in $1+1$ 
dimensions~\cite{GG,douglas}.

\begin{figure}[tbp]
\begin{center}
\epsfig{file=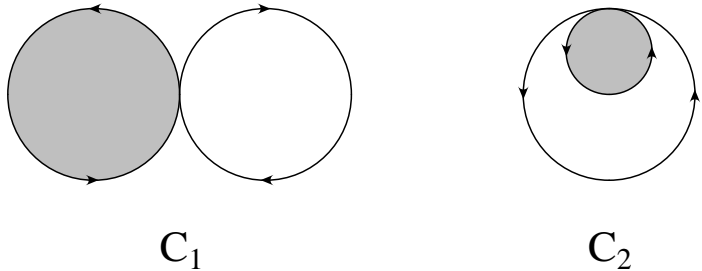,height=40mm}
\caption{Examples of Wilson loops. $A_1$ and $A_2$ are the unshaded and shaded
areas, respectively. \label{fig:23}}
\end{center}
\end{figure}

There are some other interesting results for large $N$ QCD in $1+1$ dimensions
which have been obtained recently. The basic gauge invariant observables in a
gauge theory are the Wilson loops. Kazakov and Kostov obtained the exact
expression for all Wilson loops~\cite{KK}. The Wilson loop for a close curve
$C$ is the expectation value of the trace of the path-ordered exponential
\begin{equation}\label{wloop}
W(C) = \vev{ \Tr P e^{-i g \int A_\mu\ dx^\mu} },
\end{equation}
and is of order $N$. For a simple closed curve, the Wilson loop for $1+1$
dimensional large $N$ QCD satisfies an exact area law,
\[
W(C) = N\, e^{-g^2 A/2},
\]
where $A$ is the area enclosed by $C$. More interesting examples are
self-intersecting curves, such as those in fig.~\ref{fig:23}, 
with Wilson
loops
\[
W(C_1) = N\, e^{-g^2 \left(A_1 + A_2\right)/2},\qquad 
W(C_2) = N\, e^{-g^2 \left(A_1 + 2 A_2\right)/2}\left[1-g^2 A_2 \right],
\]
where $A_1$ and $A_2$ are the unshaded and shaded areas, respectively. The
master field for 2D QCD has been shown to reproduce the Kazakov-Kostov results
for the Wilson loops.

\begin{hw}\sl
\begin{itemize}
\end{itemize}
\noindent In a non-Abelian theory, the field strength tensor $F_{\mu\nu}$ does
not determine all the gauge invariants. In $1+1$ dimensions for gauge group
$SU(2)$, construct two different vector potentials which both produce the same
constant field strength tensor $F_{12} =f \tau_3$, where $f \not=0$ is a
constant, and yet give different values for Wilson loops.
\end{hw}

\begin{hw}\sl
\begin{itemize}
\end{itemize}
\noindent Define the matrix
\[
U_R = P e^{-i g \int A_\mu\ dx^\mu}
\]
where $A_\mu=A_\mu^A T^A$ with $T^A$ in representation $R$ of $SU(N)$. The
Wilson loop in the fundamental representation (denoted by $F$), 
eq.~(\ref{wloop}), is
\[
W_F(C) = \vev{ \Tr U_F }
\]
where $\vev{\cdot}$ represents a functional integral over all gauge field
configurations. Show that the Wilson loop in the adjoint representation
is
\[
W_{\rm adj}(C) = \vev { \Tr U_{\rm adj} } = \vev {\Tr U_F \Tr U^\dagger_F} - 1.
\]
This equation holds in any number of dimensions, and does not depend on taking
the large $N$ limit. The Wilson loop $W_F(C)$ is expected to have an area law,
\[
W_F(C) \sim N\, \exp(-\lambda\ {\rm area(C)}),
\]
where $\lambda$ is an unknown constant.
Show that in the large $N$ limit, one expects~\cite{greensite}
\[
W_{\rm adj}(C) \sim N^2\, \exp(-2\ \lambda\ {\rm area(C)})+ \exp(-\lambda'\
{\rm perimeter(C))},
\]
where $\lambda'$ is another unknown constant.
What are the implications of this formula for confinement/screening of adjoint
quarks?
\end{hw}

\section{Meson Phenomenology}\label{sec:mesons}

\subsection{Zweig's Rule}

\begin{figure}[tbp]
\begin{center}
\epsfig{file=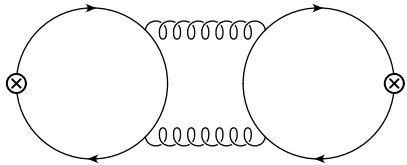,height=20mm}
\caption{ An annihilation graph, which violates Zweig's
rule and is $1/N$ suppressed.
\label{fig:24}}
\end{center}
\end{figure}

Meson correlation functions at leading order in $1/N$ are given by diagrams
with a single quark loop. Annihilation graphs, such as those in
fig.~\ref{fig:24}, 
have two quark loops and are suppressed by one power of $N$. The suppression of
annihilation graphs is known as Zweig's rule. One consequence of this is that
mesons occur in nonets for three light quark flavors. There are nine possible
$\bar q q$ states, which are divided into flavor octets $\bar q\, T^A\, q$,
where $T^A$ is a $SU(3)$ {\it flavor} matrix, and a singlet $\bar q q$.
Usually, the singlet and octet mesons are not related, because the singlet
meson can mix with gluons. In the large $N$ limit, this mixing is suppressed.
The singlet and octet meson couplings are related by treating them as members
of a $U(N)$ multiplet, with the $U(1)$ and $SU(N)$ couplings having the same
normalization. Examples of this are: in the large $N$ limit, the vector meson
nonet of $\left\{\rho, \omega, \phi, K^*\right\}$ all have the same mass, and
the decay constants of the $\pi$ and $\eta'$ are equal, $f_\pi=f_{\eta'}$.

\subsection{Exotics}

\begin{figure}[tbp]
\begin{center}
\epsfig{file=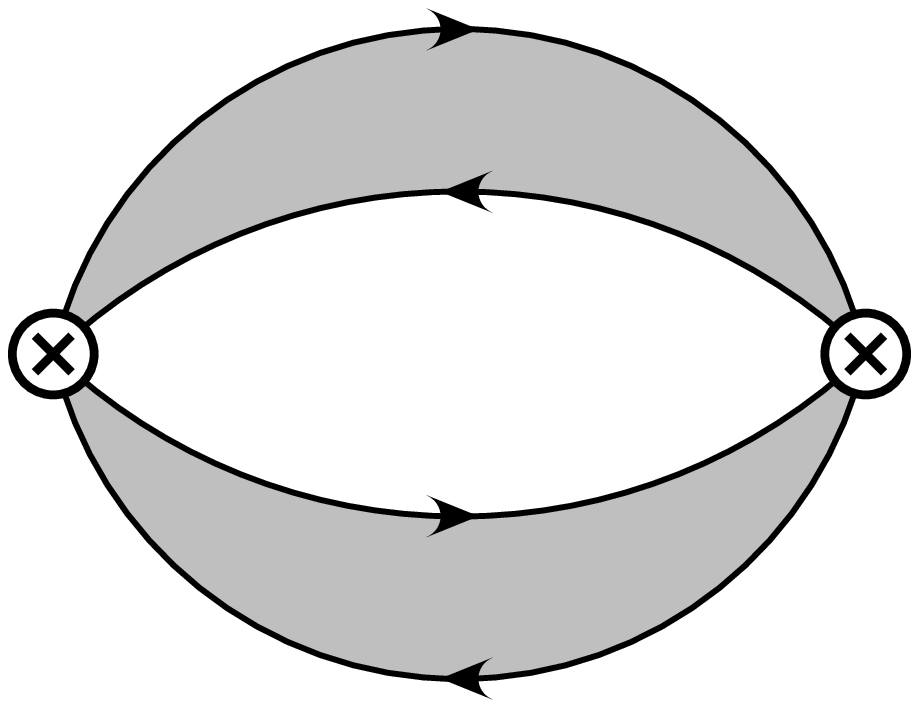,height=25mm}
\caption{The leading order graphs for the $\vev{\left(\bar q q \right)^2
\left(\bar q q \right)^2}$ correlation function. The shaded regions represent
planar gluons. \label{fig:25}}
\end{center}
\end{figure}

At leading order in $1/N$, there are no exotic states. This does not mean that
there are no exotic states in QCD, only that their binding is subleading in the
$1/N$ expansion. As an example, consider the propagator $\veV{\left(\bar q q
\right)^2 \left(\bar q q \right)^2}$ for a four-quark state $\bar q q \bar q
q$. It is easy to see that at leading order, the graphs which contribute have
the form shown in fig.~\ref{fig:25},
i.e.\ the correlation function is $\vev{\bar q q\ \bar q q }^2$, the square
of the $\bar q q$ correlation function. This is the correlation function for
two non-interacting mesons, and so $\left(\bar q q\right)^2$ creates two
mesons, rather than a four-quark state.

\subsection{Chiral Perturbation Theory}\label{sec:chiral}

The $U(N_F)_L \times U(N_F)_R$ chiral symmetry symmetry of QCD is spontaneously
broken to a diagonal $U(N_F)_V$ vector symmetry, resulting in (pseudo-)
Goldstone bosons.\footnote{The axial anomaly is $\ord{1/N}$. See
section~\ref{axialu1}.} This form for the breaking can be proved in the large
$N$ limit~\cite{coleman-witten}. The low-energy interactions of the
pseudo-Goldstone bosons of QCD, the $\pi$, $K$, $\eta$, and $\eta'$, can be
described in terms of an effective Lagrangian known as the chiral Lagrangian.
One can imagine computing the chiral Lagrangian by evaluating the QCD
functional integral with sources for the pseudo-Goldstone bosons. The source
terms are fermion bilinears. In the large $N$ limit, we have seen that the
leading order diagrams that contribute to correlation functions of fermion
bilinears are of order $N$, and contain a single quark loop, as in
fig.~\ref{fig:22}. This implies that the leading order terms in the chiral
Lagrangian are of order $N$. It is also clear from the structure of
fig.~\ref{fig:22} that the leading order terms can be written as a single
flavor trace, since the outgoing quark flavor at one vertex is the incoming
flavor at the next vertex. Similarly, diagrams with two quark loops have two
flavor traces, and are of order unity, and in general, those with $r$ quark
loops have $r$ traces, and are of order $N^{1-r}$.

The chiral Lagrangian is written in terms of a unitary matrix
\begin{equation}\label{umatrix}
U = e^{2 i \Pi/f_\pi},
\end{equation}
where $f_\pi\approx 93$~MeV is the pion decay constant, and
\begin{equation}
\Pi = {1\over \sqrt 2}
\left(\begin{array}{ccc}
{\pi^0 \over \sqrt 2} + {\eta \over \sqrt 6} +{\eta' \over \sqrt 3} &
\pi^+ & K^+ \\[0.75em]
\pi^- & -{\pi^0 \over \sqrt 2} + {\eta \over \sqrt 6} +{\eta' \over \sqrt 3} &
K^0 \\[0.75em]
K^- & {\bar K}^0 & -{2\eta \over \sqrt 6} +{\eta' \over \sqrt 3} \\
\end{array}\right),\label{umatrix1}
\end{equation}
is the matrix of pseudo-Goldstone bosons. The $\eta'$ has been included, since
it is related to the octet pseudo-Goldstone bosons in the large $N$ limit, by
Zweig's rule. The order $p^2$ terms in the chiral Lagrangian are
\begin{equation}\label{3.17}
L^{(2)} = {f^2_\pi \over 4} \Tr D^\mu U D_\mu U^{-1} + 
{f^2_\pi \over 4} B \Tr \left( m^\dagger U + m U^{-1} \right),
\end{equation}
where $m$ is the quark mass matrix in the QCD Lagrangian. The first term is
order $N$, since $f_\pi \propto \sqrt N$. The second term in eq.~(\ref{3.17})
also has a single trace and is of order $N$, so $B$ is of order unity. The $U$
field has an expansion in powers of $\pi/f_\pi$. Thus each additional meson
field has a factor of $1/f_\pi\propto 1/\sqrt N$, which gives the required
$1/\sqrt N$ suppression for mesons derived earlier. The effective Lagrangian
eq.~(\ref{3.17}) has an overall factor of $N$, and the $U$ matrix is $N$
independent, so the $1/N$ expansion is equivalent to a semiclassical expansion.
Graphs computed using the chiral Lagrangian have a $1/N$ suppression for each
loop.

The order $p^4$ terms in the chiral Lagrangian are conventionally written
as~\cite{GL}
\begin{eqnarray}
&&L^{(4)}
= L_1 \left[ \Tr D_\mu U^{-1} D^\mu U \right]^2
+L_2 \Tr D_\mu U^{-1} D_\nu U \Tr D^\mu U^{-1} D^\nu U \nonumber \\
&&+L_3 \Tr D_\mu U^{-1} D^\mu U D_\nu U^{-1} D^\nu U \nonumber \\
&&+L_4 \Tr D_\mu U^{-1} D^\mu U \Tr \left( U^{-1} m + m^\dagger U
\right)\nonumber \\
&&+L_5 \Tr D_\mu U^{-1} D^\mu U \left( U^{-1} m + m^\dagger U \right)
+L_6 \left[ \Tr \left( U^{-1} m + m^\dagger U \right) \right]^2
\nonumber \\
&&+L_7 \left[ \Tr \left( U^{-1} m - m^\dagger U \right) \right]^2
+L_8 \Tr \left( m^\dagger U m^\dagger U + U^{-1} m U^{-1} m
\right) \nonumber \\
&&-iL_9 \Tr \left( F^{\mu \nu}_R D_\mu U D_\nu U^{-1} +
F^{\mu \nu}_L D_\mu U^{-1} D_\nu U \right) \nonumber \\
&&+L_{10} \Tr U^\dagger F^{\mu \nu}_R U^{-1} F_{L\mu \nu},
\label{3.18}
\end{eqnarray}
where $F_{L,R}$ are the field-strength tensors of the (flavor) $U(3)_L$ and
$U(3)_R$ gauge fields. The terms with a single trace, $L_3$, $L_5$, $L_8$,
$L_9$ and $L_{10}$ should be of order $N$, and those with two traces, $L_1$,
$L_2$, $L_4$, $L_6$ and $L_7$ should be of order one. This is not correct,
because of one subtlety. There is an identity
\begin{equation}\label{3.19}
\Tr A B A B = - 2 \Tr A^2 B^2 + \frac 1 2 \Tr A^2 \Tr B^2 + \left( \Tr A B
\right)^2,
\end{equation}
which is valid for arbitrary traceless $3 \times 3$ matrices $A$ and $B$. 
Using
$A=D_\mu U U^{-1}$ and $B=D_\nu U U^{-1}$ in eq.~(\ref{3.19})
gives the relation
\begin{eqnarray}
&&\Tr D_\mu U D_\nu U^{-1} D^\mu U D^\nu U^{-1} = -2 \Tr D_\mu U D^\mu U^{-1} 
D_\nu U D^\nu U^{-1} \nonumber \\
&&+ \frac 1 2 \Tr D_\mu U D^\mu U^{-1} 
\Tr D_\nu U D^\nu U^{-1} + \Tr D_\mu U D_\nu U^{-1} \Tr D^\mu U D^\nu U^{-1}.
\nonumber \\
\label{3.20}
\end{eqnarray}
It is important to remember that this relation is special to three flavors, and
would not hold for an arbitrary number of flavors. The operator $\Tr D_\mu U
D_\nu U^{-1} D^\mu U D^\nu U^{-1}$ is a single trace operator and can occur in
the Lagrangian with a coefficient $c$ which is of order $N$. Eliminating the
operator using the identity eq.~(\ref{3.20}) gives the contributions $\delta
L_1=c/2$, $\delta L_2=c$ and $\delta L_3=-2c$ to $L_{1-3}$. $L_3$ was already
of order $N$, and so remains order $N$. $L_1$ and $L_2$ are now of order $N$,
because of the $c$ term, but $2L_1-L_2$ is of order unity. Thus one finds the
$N$-counting rules
\begin{eqnarray*}
\mathcal{O}(N): & \qquad & L_1, L_2, L_3, L_5, L_8, L_9, L_{10} \\
\mathcal{O}(1): & \qquad & 2 L_1 - L_2, L_4, L_6, L_7 
\end{eqnarray*}
These are the $N$-counting rules given in ref.~\cite{GL}, with the exception of
$L_7$, which is taken from ref.~\cite{rafael}. In ref.~\cite{GL},
$L_7$ was argued to be of order $N^2$. We will return
to $L_7$ in section~\ref{sec:resonances}, after discussing the $\eta'$. The
experimental values for the $L$'s are given in Table~\ref{tab:1}. The terms of
order $N$ are systematically larger than those of order unity.
\begin{table}[tbp]
\caption{Experimental values for the coefficients of the order $p^4$ terms in
the chiral Lagrangian, eq.~(\ref{3.18}). Values are from ref.~\cite{Pich} 
\label{tab:1}}
\begin{tabular}{lr@{\hspace{0.2em}}c@{\hspace{0.2em}}lc}
\hline
$L_i$ & \multispan{3} Value & Order \\
\hline
$2 L_1 - L_2 $ & $-0.6$ & $\pm$ & $0.5$ & $1$ \\
$L_4$ & $-0.3$ & $\pm$ & $0.5$ & $1$ \\
$L_6$ & $-0.2$ & $\pm$ & $0.3$ & $1$ \\
$L_7$ & $-0.4$ & $\pm$ & $0.2$ & $1$ \\
$L_2$ & $1.4$ & $\pm$ & $0.3$ & $N$ \\
$L_3$ & $-3.5$ & $\pm$ & $1.1$ & $N$ \\
$L_5$ & $1.4$ & $\pm$ & $0.5$ & $N$ \\
$L_8$ & $0.9$ & $\pm$ & $0.3$ & $N$ \\
$L_9$ & $6.9$ & $\pm$ & $0.7$ & $N$ \\
$L_{10}$ & $-5.5$ & $\pm$ & $0.7$ & $N$ \\
\hline
\end{tabular} 
\end{table}

Higher derivative terms in the chiral Lagrangian are suppressed by powers of
the chiral symmetry breaking scale $\Lambda_\chi \sim
1$~GeV~\cite{weinberg,georgi-am}. In the large $N$ limit, $\Lambda_\chi$ is of
order unity, and so stays at around $1$~GeV. Loop graphs in the chiral
Lagrangian are proportional to $1/(4 \pi f_\pi)^2$ and are of order $1/N$. Thus
in the large $N$ limit, the chiral Lagrangian can be used at tree level, and
loop effects are suppressed by powers of $1/N$.

\subsection{Non-leptonic $K$ Decay}

Weak decays of hadrons can be computed in terms of an effective low energy
Lagrangian, since hadron masses are much smaller than the $W$ mass.
Semileptonic weak decays, such as $K \rightarrow \pi \ell \nu$, can be computed
using the weak Lagrangian
\[ 
L = - {4 G_F\over \sqrt 2}\ \bar u \gamma^\mu P_L s\,
\bar \ell \gamma_\mu P_L \nu,
\]
where $P_L=(1-\gamma_5)/2$ is the left-handed projection operator. To all
orders in the strong interactions (and neglecting electromagnetic corrections),
the matrix element for the decay $H_i \rightarrow H_f \ell \bar \nu$ can be
written as
\[
- i{4 G_F\over \sqrt 2} \me{H_f}{\bar u \gamma^\mu P_L s}{H_i}
 \me{\ell \bar \nu}{\bar \ell \gamma_\mu P_L \nu}{0},
\]
i.e.\ it factorizes into the product of a hadronic matrix element, and a
leptonic matrix element. The leptonic part can be computed explicitly using
free Dirac spinors. The hadronic part is the matrix element of a current, and
for $K$ decays, it can be computed in terms of the decay constant $f_K$. The
decay constant is determined from the measured $K \rightarrow \mu \bar \nu$
decay rate, and can then be used to predict other decay rates such as $K
\rightarrow \pi e \bar \nu$, etc.

\begin{figure}[tbp]
\begin{center}
\epsfig{file=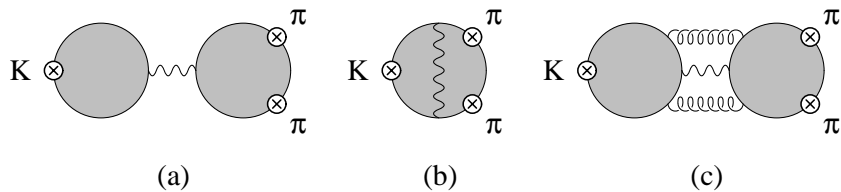,height=30mm}
\caption{Diagrams for $K \rightarrow \pi \pi$. There are related diagrams with
$K \leftrightarrow \pi$. The shaded regions represent planar gluons. The wavy
line is the $W$ boson. The $\otimes$ are insertions of the axial current.
Diagram (a) is of order $N^2$. The factorization violation contributions (b)
and (c) are order $N$ and order unity, respectively. \label{fig:51}}
\end{center}
\end{figure}

Non-leptonic $K$ decay amplitudes are more difficult to compute, since they
depend on the hadronic matrix elements of four-quark operators. To analyze
these using the large $N$ limit, it is convenient to first look at the weak
amplitudes directly in terms of $W$ exchange, rather than using the effective
weak Lagrangian. The $K \rightarrow \pi \pi$ amplitude in the large $N$ limit,
and to lowest order in the electroweak interactions, is given by diagrams with
a single $W$ boson. The $W$ boson is color neutral, so the $N$-counting for a
diagram with a $W$ boson is the same as that for the diagram with the $W$ boson
removed. The diagram must contain a quark loop, since the $K$ and $\pi$ mesons
contain quarks. The leading order diagrams are planar diagrams with two quark
loops (fig.~\ref{fig:51}(a)). 
These are disconnected diagrams when the $W$ boson is removed, so each quark
loop subgraph is of order $N$, and the overall graph is of order $N^2$. The $K
\rightarrow \pi \pi$ amplitude is order $\sqrt N$, since each current produces
a meson with amplitude $\sqrt N$. This should be compared with a three-meson
amplitude in the strong interactions, which is $1/\sqrt N$. Weak interaction
perturbation theory is an expansion in $G_F \lqcd^2 N$. Formally, this diverges
if one takes $N \rightarrow \infty$ with $G_F$ fixed. This does not mean that
one should add lots of $W$'s to strong interaction processes to get an
amplitude that grows with $N$. In the end, the results are going to be applied
to $N=3$, with $G_F$ set to its experimental value. One can first use
perturbation theory in $G_F$ to write the weak decay amplitudes in terms of
hadronic matrix elements, and then apply the $1/N$ expansion to evaluate the
purely strong interaction matrix elements.

There are no gluon exchanges between the two quark loops, so the leading order
$K \rightarrow \pi \pi$ amplitude fig.~\ref{fig:51}(a) has clearly factorized
into the product of $\me{0} {j^\mu_W} {K} \times \me{\pi\pi} {j_{\mu W}} {0}$
plus terms with $\pi \leftrightarrow K$, where $j^\mu_W$ is the weak current.
Factorization is exact in the large $N$ limit. Corrections to the factorization
approximation, such as figs.~\ref{fig:51}(b,c) are suppressed by $1/N$ and
$1/N^2$, respectively. The $K \rightarrow \pi \pi$ amplitudes can be computed
in terms of $f_K$ in the factorization approximation, since each hadronic
matrix element is that of a current. In particular, the ratio $A_{1/2}/A_{3/2}$
of the $\Delta I=1/2$ and $\Delta I=3/2$ amplitudes is equal to $\sqrt 2$. One
easy way to compute this is to note that the amplitude for $K^0 \rightarrow
\pi^0 \pi^0$ from fig.~\ref{fig:51}(a) vanishes, since all the mesons are
neutral, and the $W$ boson is charged. 

Experimentally, $A_{1/2}/A_{3/2}\approx 21$, which is the famous $\Delta I=1/2$
rule in non-leptonic $K$ decays. There is no $\Delta I=1/2$ enhancement at
leading order in the $1/N$ expansion~\cite{ihalf,cfg}. At first sight, this is
a disaster. However, it is important to keep in mind that non-leptonic weak
decays are a multiscale problem, and involve both $M_W$ and $\lqcd$.
Renormalization group scaling of the effective weak Lagrangian from $M_W$ to a
low scale $\mu\sim 1$~GeV produces an enhancement of the $\Delta I=1/2$
amplitude. Formally, this enhancement is $1/N \times \log M_W/\mu$. $N=3$ and
$\log M_W/\mu \sim 4$, so one should sum all powers $1/N \times \log M_W/\mu$.
This is done by using the renormalization group to scale the weak Hamiltonian
down to some low scale $\mu$ of order the the hadronic scale. Matrix elements
of the low-energy weak Hamiltonian do not contain any large logarithms, and
should have $1/N$ corrections of canonical size. This has been examined in
detail, and is claimed to produce a satisfactory understanding of the $\Delta
I=1/2$ rule~\cite{bbg}. The analysis is involved and will not be repeated here.
A simpler example is considered in the next section.

\begin{hw}\sl
\begin{itemize}
\end{itemize}
\noindent The $K$ decay amplitudes are 
\begin{eqnarray*} \mathcal{A}\left(K^+ \rightarrow
\pi^+ \pi^0 \right) &=& {3 \over 2} A_{3/2} \\ 
\mathcal{A}\left(K^0 \rightarrow \pi^+ \pi^- \right) &=& 
A_{1/2} + {1\over \sqrt 2} A_{3/2} \\
\mathcal{A}\left(K^0 \rightarrow \pi^0 \pi^0 \right) &=& 
A_{1/2} -  \sqrt 2 A_{3/2}, 
\end{eqnarray*} 
neglecting final state interaction phases. Compute the amplitudes $K^+
\rightarrow \pi^+ \pi^0$, $K^0 \rightarrow \pi^+ \pi^-$, and $K^0
\rightarrow \pi^0 \pi^0$ in the factorization approximation, and use these to
obtain $A_{3/2}$ and $A_{1/2}$. Compute the decay widths for $K^+$ and $K_S^0$
decay, and compare with experiment to obtain $A_{3/2}$ and $A_{1/2}$. Note that
$K_S^0 \not = K^0$.
\end{hw}

\subsection{$K-\bar K$ mixing}

\begin{figure}[tbp]
\begin{center}
\epsfig{file=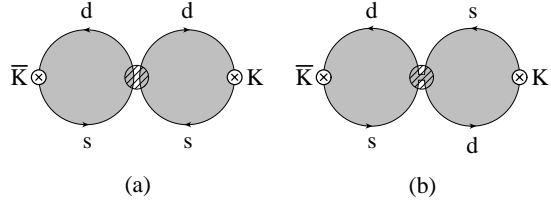,height=30mm}
\caption{$K - \bar K$ mixing diagrams. The blob represents the four-quark
operator $\bar d \gamma^\mu P_L s \, \bar d \gamma^\mu P_L s$, and the shaded
regions represent planar gluons. The operator vertex has been made
``transparent'' so that the color flow along the fermion lines is visible.
Diagram (a) is of order $N^2$, and Diagram (b), is of order $N$.
\label{fig:52}}
\end{center}
\end{figure}

The $K-\bar K$ mixing amplitude is of second order in the weak interactions.
The mixing amplitude is given by the matrix element of the $\Delta S=2$
Lagrangian between $K$ and $\bar K$. The $\Delta S=2$ Lagrangian is
\begin{equation}
L = {G_F^2 \over 16 \pi^2} \eta(\mu)\ 
\bar d \gamma^\mu P_L s \, \bar d \gamma^\mu P_L s,
\label{eq:4.10}
\end{equation}
where $\eta$ has dimension two, can be computed in perturbation theory, and
includes renormalization group scaling from $M_W$ down to some hadronic scale
$\mu$. The $\mu$ dependence of $\eta$ is cancelled by the anomalous dimension
of the four-quark operator. It is conventional to write the $K-\bar K$ matrix
element of the four-quark operator as
\[
\me{\bar K} {\bar d \gamma^\mu P_L s\, \bar d \gamma^\mu P_L s}{K} = \frac 4 3
f_K^2 M_K^2 B_K(\mu),
\]
where $B_K(\mu)$ parameterizes the hadronic matrix element of the four-quark
operator renormalized at $\mu$. In the large $N$ limit, the matrix element is
given by fig.~\ref{fig:52}(a), 
and factorizes into the matrix element of two currents,
\[
\me{\bar K} {\bar d \gamma^\mu P_L s\, \bar d \gamma^\mu P_L s}{K} = 2
\me{\bar K} {\bar d \gamma^\mu P_L s}{0} 
\times\me{0} {\bar d \gamma^\mu P_L s}{K}.
\]
The factor of two arises because there are two ways of combining the weak
currents with the mesons. Corrections to the factorization approximation
are of order $1/N$ from fig.~\ref{fig:52}(b). The matrix element $\me {0}{\bar
d \gamma^\mu P_L s}{K}=f_K p^\mu/\sqrt 2$, so in the factorization
approximation, one finds $B_K=3/4+\mathcal{O}\left(1/N\right)$~\cite{wise}. The
same result holds for $B-\bar B$ mixing, 
$B_B=3/4+\mathcal{O}\left(1/N\right)$. There is no $\mu$ dependence in $B$ in
the large $N$ limit, because the anomalous dimension of the four-quark operator
is of order $1/N$. The $1/N$ corrections to $B_{K,B}$ do not contain the scale
$M_W$, since renormalization group scaling has already been used to obtain the
effective interaction eq.~(\ref{eq:4.10}) at a low scale. One does not expect
enhanced $1/N$ corrections in $B_{K,B}$. Recent lattice results give $B_K(2\
{\rm GeV}) = 0.62 \pm 0.02 \pm 0.02$~\cite{gupta} and $B_K(2\ {\rm GeV}) =
0.628 \pm 0.042$~\cite{jlqcd}, which are consistent with
$3/4+\mathcal{O}\left(1/N\right)$.

\subsection{Axial $U(1)$ and the $\eta'$ Mass}\label{axialu1}

The $U(1)_A$ flavor symmetry is broken by anomalies, so the $\eta'$ is not a
Goldstone boson. The anomaly graph involves a quark loop, and is suppressed in
the large $N$ limit. This allows us to study the $\eta'$ as a pseudo-Goldstone
boson, with $1/N$ as a symmetry breaking parameter.

The QCD Lagrangian including the $\theta$ term is
\begin{equation}\label{3.25}
L = - \frac{1}{2}\Tr F_{\mu \nu} F^{\mu \nu} + {g^2 \over 8 \pi^2}{\theta\over
N} \Tr F_{\mu \nu} \tilde F^{\mu \nu} + \bar \psi \left( i\, \Dslash - m
\right) \psi,
\end{equation}
where the usual coupling constant $g^2$ has been replaced by $g^2/N$. $\theta$
is a periodic variable with periodicity $2\pi$. It is clear from the form of
eq.~(\ref{3.25}) that quantities will depend on the combination $\theta / N$,
since that is the parameter combination which occurs in the QCD Lagrangian.

There is no $\theta$-dependence of any physical quantity in perturbation
theory, which makes the analysis of the $U(1)_A$ sector tricky. The vacuum
energy $E$ is of order $N^2$. If one assumes that there is no $1/N$ 
suppression of the $\theta$ dependence, it must have the form
\[
E = N^2 F(\theta/N),
\]
where $F$ is some function with periodicity $2\pi/N$. In the dilute instanton
gas approximation, the $\theta$ dependence of quantities has the form
\[
e^{-8 \pi^2 N/g^2} e^{i \theta} = \left(e^{-8 \pi^2/g^2} e^{i \theta /N
}\right)^N
\]
in the one-instanton sector. This is exponentially suppressed in $N$, and one
might think that all $\theta$ dependence is exponentially small in $N$. This
conclusion is believed to be incorrect. The dilute instanton gas approximation
is not valid because of infrared divergences.

There is one important result about the $\theta$ dependence of QCD when
fermions are included --- if any fermion is massless, all $\theta$ dependence
vanishes. This leads to the following puzzle: fermion loop contributions to the
vacuum energy are order $N$, so how can they cancel the order $N^2$ vacuum
energy of the pure glue theory? To study this, look at the second derivative of
the vacuum energy with respect to $\theta$~\cite{veneziano-ep,witten-ep},
\begin{equation}\label{3.26}
{d^2E\over d\theta^2} = \left({g^2 \over 8 \pi^2 N}\right)^2
\int d^4 x \ \me{0}{T \left( \Tr F \tilde F(x) \Tr F \tilde F(0) \right)}{0},
\end{equation}
which is $\mathcal{O}(1)$. Define
\[
U(k) = \int d^4 x\ e^{ik\cdot x}
\me{0}{T \left( \Tr F \tilde F(x) \Tr F \tilde F(0) \right)}{0},
\]
so that
\[
{d^2E\over d\theta^2} = \left({g^2 \over 8 \pi^2 N}\right)^2 U(0).
\]
One can write the two-point correlation function as a sum over intermediate
single-particle states,
\[
U(k) = \sum_{\rm glueballs} {N^2 a_n^2 \over k^2 - m_n^2} + 
\sum_{\rm mesons} {N b_n^2 \over k^2 - M_n^2},
\]
where we have used the $N$-counting rules $\mE{0}{\Tr F \tilde F}{{\rm
glueball}} \sim N$, $\mE{0} {\Tr F \tilde F}{{\rm meson}} \sim \sqrt N$. 
Multiparticle states can be neglected in the large $N$ limit. Here $m_n$ and
$M_n$ are glueball and meson masses, and $N a_n$ and $\sqrt N b_n$ are the
amplitudes for $\Tr \tilde F F$ to create a glueball or meson from the vacuum.
In the pure-glue theory, only the first term is present, so $U(0) \sim N^2$ and
$d^2E/d\theta^2 \sim 1$. In the theory with quarks and gluons, the second term
is also present. The only way that the second term can cancel the first is if
one meson has a mass of order $1/\sqrt N$. This can cancel the first term at
$k=0$, which is what is needed to cancel the $\theta$ dependence of the vacuum
energy, but does not cancel the first term at arbitrary values of $k$. It is
believed that the $\eta'$ mass is of order $1/\sqrt N$, and produces the
required cancellation.\footnote{One might wonder how two terms of the same sign
can cancel each other.  The resolution is that there is an equal time
commutator that must be added to eq.~(\ref{3.26}).  See the appendix of
ref.~\cite{witten-ep} for more details.} With this assumption, one finds that
\[
U(0)_{\rm no\ quarks} = N { b_{\eta'}^2 \over M^2_{\eta'}}.
\]
The matrix element $\sqrt N b_{\eta'} g^2/8\pi^2 = (g^2/8\pi^2) \mE{0}{\Tr F
\tilde F}{\eta'}$, which can be written as
\[
(g^2/8\pi^2)\me{0}{\Tr F \tilde F}{\eta'} = {N \over 2 N_F} \me{0}{
\partial_\mu J^\mu_5}{\eta'} = {N \over 2 F} f_{\eta'} M^2_{\eta'}
\]
 using the anomaly equation
\[
\partial_\mu J^\mu_5 = N_F {g^2 \over 4 \pi^2 N} \Tr F \tilde F.
\]
This gives the Veneziano-Witten formula for the $\eta'$ mass,
\begin{equation}\label{3.30}
M^2_{\eta'} = {2 N_F\over f_\pi} \left({ d^2E_{\rm no\ quarks} \over d\theta^2}
\right)_{\theta=0}.
\end{equation}
The $\eta'$ is a Goldstone boson in the large $N$ limit, and $M^2_{\eta'}$ is
linear in the symmetry breaking parameter $1/N$. In general, one can show that
the $\eta'$ dependence of a zero-momentum amplitude in the theory with quarks
can be obtained from the $\theta$ dependence in the theory without quarks, by
the replacement 
\begin{equation}\label{3.29}
\theta \rightarrow \theta + \sqrt{2 N_F} \eta'/f_\pi.
\end{equation}
The $\eta'$ mass is $d^2E/d\eta'^2=M^2_{\eta'}$, which reduces to
eq.~(\ref{3.30}) using eq.~(\ref{3.29}).

The form eq.~(\ref{3.29}) can also be obtained using the $U(1)_A$ Ward
identity. Under an axial $U(1)$ transformation, $\psi_L \rightarrow e^{i
\alpha} \psi_L$, $\psi_R \rightarrow e^{- i \alpha} \psi_R$, and $\theta
\rightarrow \theta - 2 N_F \alpha$. The $\eta'$ is the Goldstone boson of the
$U(1)_A$ symmetry, and transforms as $\eta' \rightarrow \eta' + f_\pi \alpha$,
so the $U(1)_A$ invariant combination is eq.~(\ref{3.29}).

In chiral perturbation theory, the $U$ matrix can be extended to include the
$\eta'$,
\[
U \rightarrow U e^{2 i \eta'/f_{\eta'}\sqrt{2 N_F}},
\]
where $f_\pi=f_{\eta'}$ at leading order in $1/N$.\footnote{This has already
been done in the $U$ of eq.~(\ref{umatrix1}).} Then the linear combination in
eq.~(\ref{3.29}) is
\[
\theta - i \log \det U.
\]
One can obtain the zero-momentum $\eta'$ couplings by using this linear
combination for all the $\theta$ dependence in the effective Lagrangian. There
are also momentum-dependent $\eta'$ interactions which are not related to
$\theta$-dependence, and cannot be determined by this method.

In the large $N$ and chiral limits, the $\pi,K,\eta,\eta'$ nonet is massless.
Including non-zero quark masses and $1/N$ corrections gives mass to the mesons.
For simplicity, consider $m_u=m_d=0$, $m_s\not=0$. The neutral mesons in the
nonet can be chosen to be $\bar u u$, $\bar d d$ and $\bar s s$, rather than
the $\pi^0$, $\eta$ and $\eta'$. Since $m_u=m_d=0$, there is an exact $U(2)
\times U(2)$ chiral symmetry in the large $N$ limit, so the meson mass matrix
for the $\bar u u$, $\bar d d$ and $\bar s s$ mesons must have the form
\begin{equation}\label{3.31}
M^2 = \left(\begin{array}{ccc}
0 & 0 & 0 \\
0 & 0 & 0 \\
0 & 0 & C
\end{array}\right),
\end{equation}
where $C$ is some function of $m_s$. There can be no off-diagonal terms, since
Zweig's rule is exact in the large $N$ limit. The $1/N$ correction to the mass
matrix has the form
\begin{equation}\label{3.32}
M^2 = {a\over N} \left(\begin{array}{ccc}
1 & 1 & 1 \\
1 & 1 & 1 \\
1 & 1 & 1
\end{array}\right),
\end{equation}
(in the limit of equal quark masses), since the amplitude for $q_i\bar q_i
\rightarrow q_f \bar q_f$ does not depend on the flavors $i,f$ of the initial
and final quarks. The $1/N$ factor is explicit, so that $a$, which represents
the strength of the annihilation graphs, is of order one. The quark mass and
$1/N$ are both treated as small, so that effects of order $m_q/N$ have been
neglected. The complete mass matrix is the sum of
eqs.~(\ref{3.31})+(\ref{3.32}). It has one zero eigenvalue (the $\pi^0$) since
chiral $SU(2) \times SU(2)$ is still an exact symmetry. The two non-zero
eigenvalues give the ratio~\cite{georgi}
\begin{equation}
{M^2_\eta \over M^2_{\eta'}} = {3 + R - \sqrt{9-2R+R^2} \over 
3 + R + \sqrt{9-2R+R^2} }, \qquad R = CN/a.
\label{eq:4.21}
\end{equation}
Irrespective of the value of $R$, one finds
\begin{equation}
{M^2_\eta \over M^2_{\eta'}} \le {3 -\sqrt 3 \over 3 + \sqrt 3} = 0.27.
\label{eq:4.22}
\end{equation}
The experimental value for the ratio is $0.33$, which exceeds the bound, but
not by much. (Remember that our expansion parameter is about $1/3$.) The bound
was derived neglecting $m_{u,d}$, and keeping the lowest order term in $1/N$.
Including light quark masses, and adding the $1/N$ correction gives a bound
that is consistent with the experimental value~\cite{peris}.

The $\eta'$ mass $M^2_{\eta'}$ is a function of the quark masses $m_q$ and
$1/N$. In the limit $N \rightarrow \infty$, it is of order $m_q$, and in the
chiral limit $m_q \rightarrow 0$, it is of order $1/N$, so it is sometimes said
that the large $N$ and chiral limits do not commute. The origin of this
non-commutativity is clear; the mass matrix is the sum of two terms 
eqs.~(\ref{3.31})+(\ref{3.32}), and the eigenvalues depend on the ratio $z=N
m_q/\Lambda_\chi$, where $\Lambda_\chi$ is a typical hadronic scale which has
been used to make $z$ dimensionless. The large $N$ limit is $z \rightarrow
\infty$, and the chiral limit is $z \rightarrow 0$, and $M^2_{\eta'}$ has
different forms in these two limits. However, what is relevant for applying
large $N$ and chiral symmetry is that $1/N$ and $m_q/\Lambda_\chi$ are both
small; their relative size $z$ is irrelevant.

Chiral perturbation theory is an expansion in derivatives over some chiral
symmetry breaking scale $\Lambda_\chi$, which is $\mathcal{O}(1)$ in the large
$N$ limit. The $\eta'$ is light in the large $N$ and chiral limits,
irrespective of the relative size of $1/N$ and $m_q$. Since it is light, the
$\eta'$ should be included as an explicit degree of freedom in the large $N$
chiral Lagrangian, to avoid an inconsistent expansion. The  $U(N_F) \times
U(N_F)$ chiral Lagrangian in the large $N$ limit to order $p^4$ has been worked
out in ref.~\cite{hlpt} 

\begin{hw}[The $\eta'$ Mass]\sl
\begin{itemize}
\end{itemize}
\begin{enumerate}
\noindent Derive eqs.~(\ref{eq:4.21}) and ~(\ref{eq:4.22}).
\end{enumerate}
\end{hw}

\subsection{Resonances and $1/N$}\label{sec:resonances}

In section~\ref{sec:chiral}, we derived the $N$ dependence of an effective
Lagrangian for mesons. The leading order terms in the Lagrangian are of order
$N$, and have a single trace over flavor. The first correction is of order
unity, and has two flavor traces, etc. In addition, every meson field carries a
suppression factor of $1/\sqrt N$. The Lagrangian can be represented
schematically as
\begin{equation}\label{3.1.1}
L = N \Tr X + \Tr Y_1 \Tr Y_2 + {1\over N} \Tr Z_1 \Tr Z_2 \Tr Z_3 + \ldots,
\end{equation}
where $X$, $Y_i$ and $Z_i$ are functions of $M/\sqrt N$, where $M$ is a meson
field. Here $\Tr Y_1 \Tr Y_2$ is an abbreviation for the sum of all possible
terms written as the product of two traces, etc. For example, in the chiral
Lagrangian, $\Tr X$ represents terms such as $\Tr D_\mu U^{-1} D^\mu U$, where
$U = \exp(2 i \pi /f)$, with $f$ of order $\sqrt N$. Here we will consider a
more general low energy Lagrangian, that includes the Goldstone bosons as well
as additional meson fields, and study the form of terms induced by integrating
out heavy meson fields.

In the large $N$ limit, mesons form nonets. It is convenient to represent a
meson nonet (such as the $\rho$, $\omega$, $\phi$ and $K^*$) by a $3 \times 3$
matrix $M$. The one-meson couplings can be obtained from the Lagrangian
eq.~(\ref{3.1.1}), by retaining the terms which contain one power of $M/\sqrt
N$, and schematically have the form
\begin{eqnarray}
L^{(1)} = &&\sqrt N \Tr A M + {1\over \sqrt N}
\Tr B M \Tr C \nonumber \\
&&+ {1\over N^{3/2}} \Tr D M \Tr E_1 \Tr E_2 + \ldots.
\label{3.1.2}
\end{eqnarray}
The terms induced by integrating out the meson multiplet $M$ at tree level can
be obtained from eq.~(\ref{3.1.2}). The meson propagator (at zero momentum) is 
\begin{equation}\label{3.1.3}
\Delta^{ab}={1\over m_8^2} \delta^{ab} + \delta_{1/m^2} \delta^{a9}
\delta^{b9} = \left\{ \begin{array}{cl} 
{1\over m_8^2} \delta^{ab} & a,b=1,\ldots,8 \\[10pt]
{1\over m_9^2} \delta^{ab} & a,b=9 \\
\end{array}\right.
\end{equation}
where $m_8$ is the mass of the octet mesons,
\[
\delta_{1/m^2} = {1\over m_9^2} - {1\over m_8^2},
\]
is related to the mass difference of the the octet and singlet mesons, and the
first $\delta$ function in eq.~(\ref{3.1.3}) is over $a,b=1,\ldots,9$. Writing
$M$ as $M^a T^a$ ($a=1$--9), and using the identity eq.~(\ref{3.8}) with $N$
replaced by $N_F=3$, shows that the terms induced by meson exchange at lowest
order in the derivative expansion are
\begin{eqnarray}
&&\left[\sqrt N \Tr A T^a + {1\over \sqrt N}
\Tr B T^a \Tr C + {1\over N^{3/2}} \Tr D T^a \Tr E_1 \Tr E_2 + \ldots \right]
\nonumber \\
&&\qquad \qquad \times\ \Delta^{ab}\ \times \nonumber \\
&&\left[ \sqrt N \Tr A T^b + {1\over \sqrt N}
\Tr B T^b \Tr C + {1\over N^{3/2}} \Tr D T^b \Tr E_1 \Tr E_2 + \ldots \right]
 \nonumber \\
&&=
{1\over 2 m_8^2} \Bigl[ N \Tr A A + 2 \Tr A B \Tr C + \nonumber \\
&& \qquad \qquad {1\over N}\left( \Tr B B
\Tr C \Tr C + 2 \Tr A D \Tr E_1 \Tr E_2 \right) + \ldots \Bigr] \nonumber \\
&&+
{1\over 6} \delta_{1/m^2} \Bigl[ N \Tr A \Tr A + 2 \Tr A \Tr B \Tr C + 
\nonumber \\
&& \qquad
{1\over N} \left( \Tr B \Tr B
\Tr C \Tr C + 2 \Tr A \Tr D \Tr E_1 \Tr E_2 \right) + \ldots \Bigr].
\label{3.1.5}
\end{eqnarray}
For a meson multiplet other than the Goldstone bosons, $m_8^2 \sim \ord 1$ and
$\delta_{1/m^2} \sim \ord{1/N}$. Using this in eq.~(\ref{3.1.5}) shows that the
terms induced by integrating out a heavy meson nonet have the same $N$-counting
as those in the original Lagrangian eq.~(\ref{3.1.1}), as one might expect.

The singlet meson plays an important role in reproducing the correct 
$N$-counting, as there are non-trivial cancellations between singlet and octet
meson exchange. It is inconsistent to include the octet mesons but not the
singlet meson in the large $N$ limit. Neglecting the singlet meson is
equivalent to letting $m_9 \rightarrow \infty$, so that $\delta_{1/m^2} =
-1/m_8^2$ and is of order one. In this case, the $\delta_{1/m^2}$ terms in
eq.~(\ref{3.1.5}) violates the $N$-counting rules by one power of $N$.

It is also inconsistent to use large $N$ counting rules, and treat the $\eta'$
as heavy. Integrating out the $\eta'$ is equivalent to retaining only the
singlet meson contributions in eq.~(\ref{3.1.5}), which can be done by letting
$m_8 \rightarrow \infty$. In this case $\delta_{1/m^2} =
1/m_9^2=1/m_{\eta'}^2$. The $\eta'$ exchange terms violate $N$-counting by two
powers of $N$, if one uses $m_{\eta'}^2 \sim \ord{1/ N}$. It is inconsistent to
integrate out the $\eta'$, and at the same time assume that $m^2_{\eta'} \sim
\ord{1/N}$, since a light particle is being integrated out of the effective
Lagrangian. This is the origin of $L_7 \sim \ord{N^2}$. Retaining the $\eta'$
in the effective Lagrangian gives $L_7 \sim \ord{1}$~\cite{rafael}.

\section{Baryons}\label{sec:baryons}

Baryons are color singlet hadrons made up of quarks. The $SU(N)$ invariant
$\epsilon$-symbol has $N$ indices, so a baryon is an $N$-quark state,
\[
\epsilon_{i_1 \cdots i_N} q^{i_1} \cdots q^{i_N}.
\]
A baryon can be thought of as containing $N$ quarks, one of each color, since
all the indices on the $\epsilon$-symbol must be different for it to be
non-zero. Quarks obey Fermi statistics, and the $\epsilon$-symbol is
antisymmetric in color, so the baryon must be completely symmetric in the other
quantum numbers such as spin and flavor.

The number of quarks in a baryon grows with $N$, so one might think that large
$N$ baryons have little to do with baryons for $N=3$. However, we will soon see
that for baryons, as for mesons, the expansion parameter is $1/N$, and that one
can compute baryonic properties in a systematic semiclassical expansion in
$1/N$. The results are in good agreement with the experimental data, and
provide information on the spin-flavor structure of baryons. We will be able to
compute baryon properties such as masses, magnetic moments and axial couplings.
The $1/N$ expansion provides some deep connections between QCD and two popular
models, the quark model, and the Skyrme model, which provide a good
phenomenological description of baryonic properties.

\subsection{$N$-Counting Rules for Baryons}\label{sec:bncount}

\begin{figure}[tbp]
\begin{center}
\epsfig{file=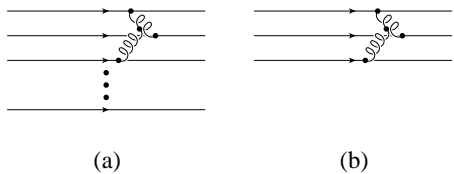,height=25mm}
\caption{A baryon interaction and the corresponding connected component.
\label{fig:28}}
\end{center}
\end{figure}

\begin{figure}[tbp]
\begin{center}
\epsfig{file=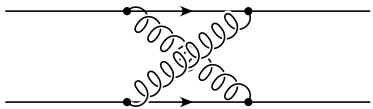,height=15mm}
\caption{An example of a ``planar'' two-body baryon graph.
\label{fig:56}}
\end{center}
\end{figure}

The $N$-counting rules for baryon graphs can be derived using our previous
results for meson graphs. Draw the incoming baryon as $N$-quarks with colors
arranged in order, $1 \cdots N$. The colors of the outgoing quark lines are
then a permutation of $1 \cdots N$. It is convenient to derive the $N$-counting
rules for connected graphs. For this purpose, the incoming and outgoing quark
lines are to be treated as ending on independent vertices, so that the
connected piece of fig.~\ref{fig:28}(a) is fig.~\ref{fig:28}(b). 
A connected piece that contains $n$ quark lines will be referred to as an
$n$-body interaction. The colors on the outgoing quarks in an $n$-body
interaction are a permutation of the colors on the incoming quarks, and the
colors are distinct. Each outgoing line can be identified with an incoming line
of the same color in a unique way. One can now relate connected graphs for
baryons interactions with planar diagrams with a single quark loop. The leading
in $N$ diagrams for the $n$-body interaction are given by taking a planar
diagram with a single quark loop, cutting the loop in $n$ places, and setting
the color on each cut line to equal the color of one of the incoming (or
outgoing) quarks. For example, the interaction in fig.~\ref{fig:28}(b) is
given by cutting fig.~\ref{fig:10} once at each of the three fermion lines.
Planar meson diagrams contain a single closed quark loop as the outer edge of the
diagram. Baryon $n$-body graphs obtained from cutting the quark loop require
that one twist the quark lines to orient them with their arrows pointing in
the same direction, and do not necessarily look planar when drawn on a sheet of
paper. For example, fig.~\ref{fig:56} is a ``planar'' diagram for
a two-body interaction. Baryon graphs in the double-line notation can
have color index lines crossing each other due to the fermion line twists.

The relationship between meson and baryon graphs immediately gives us the 
$N$-counting rules for an $n$-body interaction in baryons: an $n$-body
interaction is of order $N^{1-n}$, since planar quark diagrams are of order
$N$, and $n$ index sums over quark colors have been eliminated by cutting $n$
fermion lines. Baryons contain $N$ quarks, so $n$-body interactions are equally
important for any $n$. $n$-body interactions are of order $N^{1-n}$, but there
are $\mathcal{O}(N^n)$ ways of choosing $n$-quarks from a $N$-quark baryon.
Thus the net effect of $n$-body interactions is of order $N$.

\begin{figure}[tbp]
\begin{center}
\epsfig{file=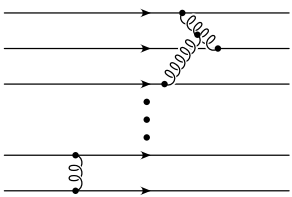,height=25mm}
\caption{An example of a disconnected baryon interaction. \label{fig:29}}
\end{center}
\end{figure}

Diagrams with two disconnected pieces, such as fig.~\ref{fig:29},
have a net effect of order $N^2$, those with three disconnected pieces are of
order $N^3$, and so on. This is easy to understand. The baryon mass $M_B$ is of
order $N$, since it contains $N$ quarks. The diagrams are an expansion of the
baryon propagator, and sum to give
\[
e^{- i M_B t} = 1 - i M_B t - {M_B^2 t^2 \over 2} + \ldots\ .
\]
Diagrams with a single connected component produce the order $t$ term (the
interaction can take place at any time) and are order $N \sim M_B$, those with
two connected components give the order $t^2$ term (each connected component
can take place at any time) and are order $N^2 \sim M_B^2$, etc. Baryon
interactions in the large $N$ limit are best studied in terms of connected
diagrams, and the diagrammatic methods are the same as used in many-body
theory.

Interactions of quarks in a baryon can be described by a non-relativistic
Hamiltonian if the quarks are very heavy. The Hamiltonian has the form
\begin{eqnarray}
H = &&N m + \sum_i {p_i^2 \over 2 m} + {1\over N}\sum_{i\not=j}V \left( x_i-x_j 
\right) \nonumber \\
&&+ {1\over N^2} \sum_{i\not=j\not=k}V \left( x_i-x_j ,x_i-x_k
\right) + \ldots. \label{4.2}
\end{eqnarray}
Each term contributes $\mathcal{O}(N)$ to the total energy. The interaction
terms in the Hamiltonian eq.~(\ref{4.2}) are the sum of many small
contributions, so fluctuations are small, and each quark can be considered to
move in an average background potential. Consequently, the Hartree
approximation is exact in the large $N$ limit. The ground state wavefunction
can be written as
\[
\psi_0\left(x_1,\ldots,x_N\right) = \prod_{i=1}^N \phi_0\left(x_i\right),
\]
where $x_i$ are the positions of the quarks.  The spatial wavefunction
$\phi_0\left(x\right)$ is $N$-independent, so the baryon size is fixed in the
$N \rightarrow \infty$ limit. The first excited state wavefunction is
\begin{equation}\label{excited}
\psi_1\left(x_1,\ldots,x_N\right) = {1\over \sqrt N} \sum_{k=1}^N
\phi_i\left(x_k\right) \prod_{i=1,i\not=k}^N \phi_0\left(x_i\right).
\end{equation}
Further details about this approach can be found in
refs.~\cite{coleman,witten-baryon}.

\begin{figure}[tbp]
\begin{center}
\epsfig{file=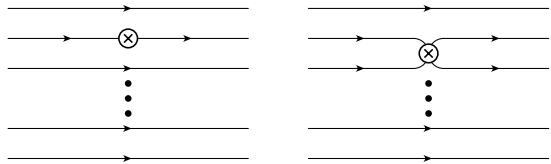,height=20mm}
\caption{Baryon matrix elements of a one-body operator such as $\bar q q$, and
a two-body operator such as $\bar q q \, \bar q q$. \label{fig:281}}
\end{center}
\end{figure}

The $N$-counting rules can be extended to baryon matrix elements of color 
singlet operators. Consider a one-body operator, such as $\bar q q$. The 
baryon matrix element $\me{B}{\bar q q}{B}$ has $N$ terms, since the operator
can be inserted on any of the quark lines (see fig.~\ref{fig:281}(a)).
The baryon matrix element is therefore $\le \mathcal{O}(N)$. One obtains an
inequality because there can be cancellations between the $N$ possible
insertions. These cancellations will be crucial in unraveling the structure of
baryons. Similarly, a two-body operator such as $\bar q q\, \bar q q$ has
matrix element $\le \mathcal{O}(N^2)$, since there are $\mathcal{O}(N^2)$ ways
of inserting the operator in a baryon (see fig.~\ref{fig:281}(b)). In general,
an $n$-body operator has matrix elements $\le \mathcal{O}(N^n)$.

\begin{figure}[tbp]
\begin{center}
\epsfig{file=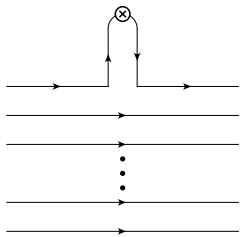,height=30mm}
\caption{Diagrams for baryon-meson couplings. \label{fig:291}}
\end{center}
\end{figure}

\begin{figure}[tbp]
\begin{center}
\epsfig{file=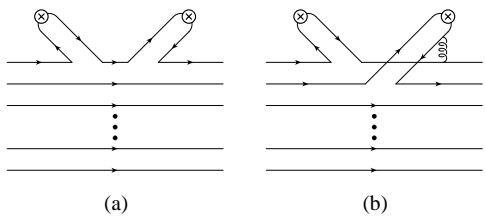,height=30mm}
\caption{Diagrams for baryon-meson scattering.
\label{fig:30}}
\end{center}
\end{figure}

The baryon-meson coupling constant is $\le \sqrt N$. This can be seen from
fig.~\ref{fig:291}, 
which shows the matrix element of a fermion bilinear in a baryon. There are $N$
possible insertions of the fermion bilinear, so the matrix element is order
$N$. The amplitude for a fermion bilinear to create a meson is the $Z$-factor,
which is order $\sqrt N$, so the baryon-meson coupling constant is of order
$N/Z = \sqrt N$. The baryon-meson scattering amplitude is $\le \mathcal{O}(1)$.
Two contributions to the scattering amplitude are shown in 
fig.~\ref{fig:30}. 
Figure~\ref{fig:30}(a) has $N$ possible insertions of the fermion bilinear,
and two meson $1/Z$-factors of $1/\sqrt N$ each, so the net amplitude is $\le
\ord 1$. The two bilinears must be inserted on the same quark line to 
conserve energy --- the incoming meson injects energy into the quark line,
which must be removed by the outgoing meson to give back the original baryon.
If the bilinears are inserted on different quark lines, as in
fig.~\ref{fig:30}(b), an additional gluon is needed to transfer energy
between the two quark lines. The number of ways of choosing two quarks is
$N^2$, the meson $1/Z$-factors are $1/\sqrt N$ each, and the two gluon
couplings give $1/\sqrt N$ each, so the total amplitude is again $\le
\mathcal{O}(1)$. We have seen that the amplitudes $\rm baryon \rightarrow
baryon + meson$ is of order $\sqrt N$, and $\rm baryon + meson \rightarrow 
baryon + meson$ is of order unity. One can similarly show that $\rm baryon +
meson \rightarrow baryon + 2\ mesons$ is of order $1/\sqrt N$, etc. As in
purely mesonic amplitudes, each additional meson gives a factor of $1/\sqrt
N$ suppression.

\begin{figure}[tbp]
\begin{center}
\epsfig{file=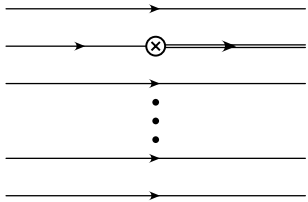,height=25mm}
\caption{Diagrams for heavy meson + baryon $\rightarrow$ heavy baryon. The
heavy quark is represented by a double line. \label{fig:301}}
\end{center}
\end{figure}

One can also look at the transition amplitudes for ground state baryon $+$
meson $\rightarrow $ excited baryon, or equivalently, for $B + M_{Q}
\rightarrow B_{Q}$, where $M_{Q}$ and $B_{Q}$ are mesons and baryons
containing a single heavy quark. In both processes, one of the quarks in the
final state is different from the others. In the transition amplitude diagram,
the meson amplitude must be inserted on the quark line that is different, as
in fig.~\ref{fig:301}; 
the meson operator either adds energy to the quark or converts it from a light
quark to a heavy quark. Thus the combinatorial factor is unity, instead of
$N$. The baryon with one excited (or heavy quark) has a wavefunction of the 
form eq.~(\ref{excited}) in which one sums over the $N$ possible quarks which 
can be different, and multiplies by a normalization factor of $1/\sqrt N$.
This produces an additional factor of $N \times 1/\sqrt N$, so the $B
\rightarrow M + B^{*}$ amplitude is of order $\sqrt N$. This is $1/\sqrt N$
suppressed relative to the corresponding amplitude between ground state
baryons.

\begin{figure}[tbp]
\begin{center}
\epsfig{file=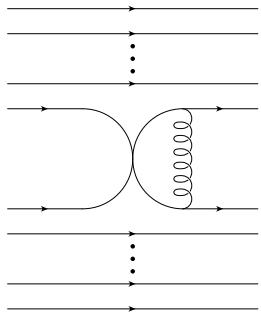,height=40mm}
\caption{Baryon-baryon scattering. \label{fig:31}}
\end{center}
\end{figure}

Baryon-baryon scattering amplitudes at fixed velocity are of order $N$. It is
important to study baryon scattering at fixed velocity, rather than fixed
momentum, because the baryon mass is of order $N$. Working at fixed velocity
avoids kinematic enhancements near threshold. The baryon-baryon scattering
amplitude from diagrams such as fig.~\ref{fig:31} 
has a combinatorial factor $N^2$ for the choice of two quarks, and $(1/\sqrt
N)^2$ for the two gluon couplings, for an overall factor of $N$. One could also
consider a similar diagram without the exchanged gluon. Then the two quarks
exchanged must have the same color, so the combinatorial factor is $N$. The net
result is that the baryon-baryon scattering amplitude at fixed velocity is of
order $N$. Baryon-baryon scattering can be described by classical trajectories
in the large $N$ limit, since the particle masses are order $N$, and the
scattering amplitude is also of order $N$.

The processes considered so far all have an $N$ dependence that is some power
of $N$. There are also processes that are exponentially small in $N$, such as
the cross-section for $e^+ e^- \rightarrow B \bar B$. The amplitude to create a
quark pair from the vacuum is some number $a<1$. The baryon has $N$ quarks, so
the amplitude to create a $B \bar B$ pair is of order $a^N$, and is
exponentially suppressed in $N$.

An important observation due to Witten is that all the $N$-counting rules
mentioned above are the same as in a field theory with coupling constant
$1/\sqrt N$, where the mesons are fundamental fields and the baryon is a
soliton.

\subsection{The Non-Relativistic Quark Model}

The non-relativistic quark model treats the baryon as made of three 
non-relativistic quarks bound by a potential. The precise details of the 
potential will not be important for these lectures. All we need assume is that
the ground state baryon is described by all three quarks in the same spatial
wavefunction $\phi(x)$. The wavefunction must then be completely symmetric in
spin and flavor. In the case of three flavors, there are six possible quark
states $u \uparrow$, $u\downarrow$, $d \uparrow$, $d\downarrow$, $s \uparrow$,
$s\downarrow$. The potential is assumed to be spin and flavor independent, so
the non-relativistic quark model has an $SU(6)$ spin-flavor symmetry under
which these six states transform as the fundamental representation. The ground
state baryons transform as the completely symmetric product of three ${\bf
6}$'s of $SU(6)$, which is the $\bf 56$ dimensional representation.

\begin{figure}[tbp]
\begin{center}
\setlength{\unitlength}{6mm}
\begin{picture}(6,5.1962)(-3,-2.5981)
\put(-3,2.5981){\makebox(0,0){$\Delta^-$}}
\put(-1,2.5981){\makebox(0,0){$\Delta^0$}}
\put(1,2.5981){\makebox(0,0){$\Delta^+$}}
\put(3,2.5981){\makebox(0,0){$\Delta^{++}$}}
\put(-2,0.866){\makebox(0,0){$\Sigma^{*-}$}}
\put(0,0.866){\makebox(0,0){$\Sigma^{*0}$}}
\put(2,0.866){\makebox(0,0){$\Sigma^{*+}$}}
\put(-1,-0.866){\makebox(0,0){$\Xi^{*-}$}}
\put(1,-0.866){\makebox(0,0){$\Xi^{*0}$}}
\put(0,-2.5981){\makebox(0,0){$\Omega^{-}$}}
\end{picture}
\caption{Flavor $SU(3)$ weight diagram for the decuplet baryons. The horizontal
axis is $I_3$, and the vertical axis is hypercharge. \label{fig:32}}
\end{center}
\end{figure}
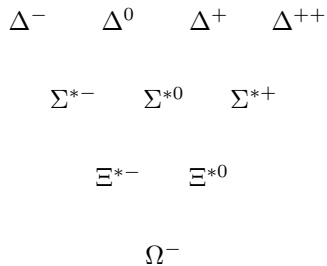

Three spin $1/2$'s added together can give spin $1/2$ or spin $3/2$, so the
baryon $\bf 56$ contains spin-1/2 and spin-3/2 baryons. The spin-$3/2$ 
wavefunction in the $m=3/2$ state is simple, it is $\uparrow \uparrow 
\uparrow$, and is completely symmetric. The flavor wavefunction must also be
completely symmetric. It can have the form $uuu$, $(uud+udu+duu)/\sqrt 3$,
etc. The spin-3/2 baryons are the decuplet baryons, shown in fig.~\ref{fig:32}.

\begin{figure}[tbp]
\begin{center}
\setlength{\unitlength}{6mm}
\begin{picture}(4,3.4642)(-2,-1.7321)
\put(-1,1.7321){\makebox(0,0){$n$}}
\put(1,1.7321){\makebox(0,0){$p$}}
\put(-2,0){\makebox(0,0){$\Sigma^-$}}
\put(0,0){\makebox(0,0){$\Lambda,\Sigma^0$}}
\put(2,0){\makebox(0,0){$\Sigma^+$}}
\put(-1,-1.7321){\makebox(0,0){$\Xi^-$}}
\put(1,-1.7321){\makebox(0,0){$\Xi^0$}}
\end{picture}
\caption{Flavor $SU(3)$ weight diagram for the baryon octet. The horizontal
axis is $I_3$, and the vertical axis is hypercharge. \label{fig:33}}
\end{center}
\end{figure}
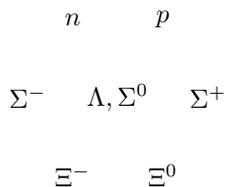

The spin-$1/2$ baryon wavefunctions are slightly more complicated. There are
no spin-$1/2$ baryons in which all three quarks are the same, for then the
wavefunction would be completely symmetric in flavor, thus completely
symmetric in spin, and so spin-$3/2$. Consider a spin-$1/2$ baryon in which two
of the quarks are identical, such as $uud$. The spin wavefunction for the two 
identical quarks must be completely symmetric, so the total spin wavefunction
of the baryon in an $m=1/2$ state must have the form
\[
\uparrow \uparrow \downarrow + \lambda (\uparrow \downarrow + \downarrow 
\uparrow) \uparrow.
\]
The constant $\lambda=-1/2$ can be determined by requiring that the raising
operator $J_{+}$ annihilates the state, since it is a $j=1/2$, $m=1/2$ state.
Thus the wavefunction of the baryon can be written as
\begin{equation}\label{5.1}
{1\over \sqrt{6}}\, uud\, \left[ 2 \uparrow \uparrow \downarrow - \uparrow \downarrow 
\uparrow - \downarrow 
\uparrow \uparrow\right].
\end{equation}
Actually, one should add cyclic permutations (and divide by $1/\sqrt 3$) to
ensure that the wavefunction is completely symmetric. However, for most
calculations, one can work just as well with eq.~(\ref{5.1}). We have
determined the wavefunctions of six of the octet baryons. The remaining two
states have three different quarks, $uds$. The $\Sigma^{0}$ is constructed
using the combination $(uds+dus)/\sqrt 2$. This symmetrizes the wavefunction
in the first two flavors, and so one constructs the wavefunction as in 
eq.~(\ref{5.1}),
\[
{1\over \sqrt{12}}\, (uds+dus)\, \left[ 2 \uparrow \uparrow \downarrow - \uparrow \downarrow 
\uparrow - \downarrow 
\uparrow \uparrow\right].
\]
The $\Lambda$ state is constructed by antisymmetrizing $uds$ in the first two
flavors, $(uds-dus)/\sqrt 2$. The spin-wavefunction must also be antisymmetric
in the first two flavors, so the $\Lambda$ wavefunction is
\[
{1\over 2}\, (uds-dus)\, \left[\uparrow \downarrow \uparrow - \downarrow \uparrow 
\uparrow\right],
\]
which can be abbreviated to
\[
{1\over \sqrt 2}\, uds\, \left[\uparrow \downarrow \uparrow - \downarrow \uparrow 
\uparrow\right].
\]
The entire spin of the $\Lambda$ is carried by the $s$-quark. The spin-$1/2$
octet is shown in fig.~\ref{fig:33}. The spin-$1/2$ octet and spin-$3/2$
decuplet together make up the $\bf 56$ of $SU(6)$. The permutation symmetry
properties of the baryons under spin $SU(2)$ and flavor $SU(3)$ are:
\[
{\bf 8} = \left(\Yfund,\ \Yadjoint\right) \qquad 
{\bf 10} = \left(\Ythrees,\ \Ythrees \right).
\]

It is straightforward to compute baryon properties in the non-relativistic
quark model. The axial coupling constant $g_{A}$ of the proton is given by the
matrix element $\me{p} {\bar q \gamma^{\mu}\gamma_{5}\tau^{3}}{p}$. In the
non-relativistic limit, this reduces to the matrix element of $q^{\dagger}
\sigma^{3} \tau^{3} q$, an operator which is $+1$ on $u\uparrow$ and $d
\downarrow$, $-1$ on $u \downarrow$ and $d \uparrow$ and zero on strange
quarks. The proton matrix element is
\[
g_{A} = {4\over6}\left[1+1+1 \right] + {1\over 6}\left[1-1-1 \right] +
{1\over 6}\left[-1+1-1 \right] = \frac{5}{3},
\]
where the first term is $(2/\sqrt 6)^{2}$ times the matrix elements for
$u\uparrow$, $u\uparrow$, $d\downarrow$, etc. obtained using the wavefunction
eq.~(\ref{5.1}).

Similarly, one can compute the magnetic moments of the baryons. The magnetic
moment operator is $\mu \sigma^{3}$, where $\mu$ is the magnetic moment of the
quark, and $\sigma^3$ is the spin operator. The quarks have magnetic moments
$\mu_u$, $\mu_d$ and $\mu_s$. If the $u$ and $d$ quarks are degenerate in mass,
$\mu_u=-2\mu_d$. The computation of the baryon magnetic moments in the
non-relativistic quark model is left as an exercise.

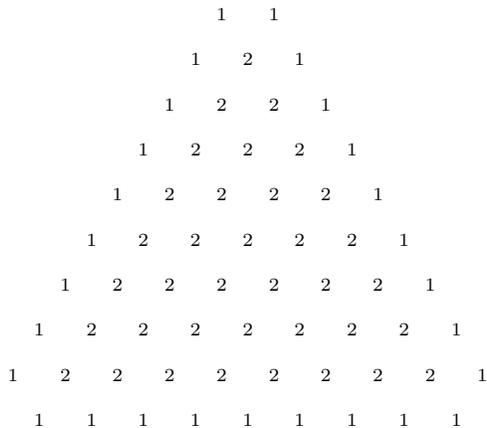
\begin{figure}[tbp]
\begin{center}
\def\onedot{\makebox(0,0){$\scriptstyle 1$}}
\def\twodot{\makebox(0,0){$\scriptstyle 2$}}
\setlength{\unitlength}{3mm}
\centerline{\hbox{
\begin{picture}(20.79,18)(-10.395,-8)
\multiput(-1.155,10)(2.31,0){2}{\onedot}
\multiput(-2.31,8)(4.62,0){2}{\onedot}
\multiput(-3.465,6)(6.93,0){2}{\onedot}
\multiput(-4.62,4)(9.24,0){2}{\onedot}
\multiput(-5.775,2)(11.55,0){2}{\onedot}
\multiput(-6.93,0)(13.86,0){2}{\onedot}
\multiput(-8.085,-2)(16.17,0){2}{\onedot}
\multiput(-9.24,-4)(18.48,0){2}{\onedot}
\multiput(-10.395,-6)(20.79,0){2}{\onedot}
\multiput(-9.24,-8)(2.31,0){9}{\onedot}
\multiput(0,8)(2.31,0){1}{\twodot}
\multiput(-1.155,6)(2.31,0){2}{\twodot}
\multiput(-2.31,4)(2.31,0){3}{\twodot}
\multiput(-3.465,2)(2.31,0){4}{\twodot}
\multiput(-4.62,0)(2.31,0){5}{\twodot}
\multiput(-5.775,-2)(2.31,0){6}{\twodot}
\multiput(-6.93,-4)(2.31,0){7}{\twodot}
\multiput(-8.085,-6)(2.31,0){8}{\twodot}
\end{picture}
}}
\caption{Flavor $SU(3)$ weight diagram for the spin-1/2 baryons for $N$ colors.
The numbers represent the degeneracy of each weight. The long edge of the
diagram has $(N+1)/2$ states. \label{fig:33a}}
\end{center}
\end{figure}

One can repeat the entire non-relativistic quark model analysis for $N$ colors.
It is convenient to choose $N=2m+1$, so that $N$ is always odd. The baryons
form the completely symmetric tensor of $SU(6)$ with $N$ indices,
\begin{equation}
\Ythrees \cdots \Ysymm
\end{equation}
This decomposes under $SU(2)_{\rm spin} \times SU(3)_{\rm flavor}$ as a tower
of representations with $J=1/2$, $J=3/2$, \ldots $J=N/2$,
\begin{equation}
\left(\frac12,\ \Yasymm\hskip-0.4pt%
\Yasymm\hskip-0.4pt%
\Yasymm \cdots \Yadjoint\right)
\qquad
\left(\frac32,\ \Yasymm\hskip-0.4pt%
\Yasymm \cdots \Yadjoint\hskip-0.4pt%
\raisebox{3pt}{\drawsquare{6.5}{0.4}}\hskip-0.4pt%
\raisebox{3pt}{\drawsquare{6.5}{0.4}}\right) \cdots 
\left(\frac N 2,\ \raisebox{3pt}{\drawsquare{6.5}{0.4}}\hskip-0.4pt%
\raisebox{3pt}{\drawsquare{6.5}{0.4}}\hskip-0.4pt%
\raisebox{3pt}{\drawsquare{6.5}{0.4}}
\cdots
\raisebox{3pt}{\drawsquare{6.5}{0.4}}\hskip-0.4pt%
\raisebox{3pt}{\drawsquare{6.5}{0.4}}\hskip-0.4pt%
\right)
\end{equation}
The $SU(3)$ representations are complicated for arbitrary values of $N$. For
example the $SU(3)$ weight diagram of the $J=1/2$ baryons is shown in
fig.~\ref{fig:33a}. The flavor representations simplify for two light flavors,
where the states are $I=J=1/2$, \ldots $I=J=N/2$.

Baryon transformation properties under spin and flavor are $N$ dependent for
$N_F \ge 3$, unlike for mesons. To apply the $1/N$ expansion to baryons it is
convenient to make some identification between the arbitrary $N$ baryons and
the $N=3$ baryons. The proton state for arbitrary $N$ will be taken to be the
strangeness zero $I=J=1/2$ state. It contains $(m+1)$ $u$ quarks and $m$ $d$
quarks. The spin $J_u$ of the $u$ quarks must be $(m+1)/2$, since the spin
wavefunction has to be completely symmetric. Similarly, the spin $J_d$ of the
$d$ quarks must be $m/2$. The proton is then the $J=1/2$ state made by
combining $J_u$ and $J_d$ to form spin-1/2. This is sufficient information to
compute many of the proton properties as a function of $N$~\cite{karl}. For
example, one can compute $g_A$, which is the matrix element of $\sigma^3 \tau^3
= 2 \left(J_u^3-J_d^3\right)$. By the Wigner-Eckart theorem
\begin{eqnarray*}
\me{p}{{\mathbf J}_u}{p} = \lambda_u \me{p}{{\mathbf J}}{p},\\
\me{p}{{\mathbf J}_d}{p} = \lambda_d \me{p}{{\mathbf J}}{p},
\end{eqnarray*}
so that
\[
\lambda_u = {{\mathbf J} \cdot {\mathbf J}_u \over {\mathbf J}^2},
\qquad \lambda_d = {{\mathbf J} \cdot {\mathbf J}_d \over {\mathbf J}^2}.
\]
Using ${\mathbf J}={\mathbf J}_u+{\mathbf J}_d$, one finds that
\[
2 {\mathbf J} \cdot {\mathbf J}_u = {\mathbf J}^2+{\mathbf J}_u^2-
{\mathbf J}_d^2,\qquad 2 {\mathbf J} \cdot {\mathbf J}_d = {\mathbf J}^2+
{\mathbf J}_d^2-{\mathbf J}_u^2,
\]
so that
\[
\lambda_u = {N+5\over6},\qquad \lambda_d = -{N-1\over6}.
\]
In the large $N$ proton, $u$-quarks tend to have spin up, and $d$-quarks tend
to have spin down. The axial coupling is
\[
g_A=\lambda_u-\lambda_d={N+2\over3},
\]
which reduces to $5/3$ when $N=3$. $g_A \sim \mathcal{O}(N)$ in the large $N$
limit, which is consistent with the $N$-counting rules for the matrix element
of a one-quark operator.

\begin{hw}[Baryon Magnetic Moments]\sl
\begin{itemize}
\end{itemize}
\noindent Show the following:
\begin{enumerate}
\item The magnetic moment of a spin-$1/2$ baryon $q_{a}q_{a}q_{b}$ with two 
identical quarks is $4 \mu_{a}/3 - \mu_{b}/3$
\item The magnetic moment 
of the $\Lambda$ is $\mu_{s}$
\item The magnetic moment of the 
$\Sigma^{0}$ is the average of the $\Sigma^{+}$ and $\Sigma^{-}$ 
magnetic moments.
\item The magnetic moments of the 
spin-$3/2$ baryons is the sum of the moments of the constituent quarks.
\item Find the $\Delta^+ \rightarrow p \gamma$ transition magnetic moment.
\end{enumerate}
\end{hw}

\section{Spin-Flavor Symmetry for Baryons}

\subsection{Consistency Conditions}\label{sec:cons}

The large $N$ counting rules for baryons imply some highly non-trivial
constraints among baryon couplings. The simplest to derive are relations among
pion-baryon couplings, or equivalently, baryon axial current matrix elements.
Related results also hold for $\rho$-baryon couplings, etc.\ and are discussed
later. To derive the axial current relations, consider pion-nucleon scattering
at fixed energy in the $N \rightarrow \infty$ limit. The argument is simplest
in the chiral limit where the pion is massless, but this assumption is not
necessary. The two assumptions required are that the baryon mass and $g_A$ are
both of order $N$. One expects the baryon mass to be proportional to $N$, since
it contains $N$ quarks. We have seen that the $N$-counting rules imply that
$g_A$ is order $N$, unless there is a cancellation among the leading terms. In
the non-relativistic quark model, $g_A=(N+2)/3$, so such a cancellation does
not occur. It is reasonable that $g_A$ is of order $N$ in QCD, even though it
need not have the value $(N+2)/3$.

The pion-nucleon vertex is
\[
{\partial_\mu \pi^a \over f} \me{B}{\bar q \gamma^\mu \gamma_5 T^a q}{B},
\]
and is of order $\sqrt N$, since $g_A \sim N$ and $f_\pi \sim \sqrt N$. Recoil
effects are of order $1/N$, since the baryon mass is order $N$ and the pion
energy is order one, and can be neglected. This allows one to simplify the
expression for the nucleon axial current. The time component of the axial
current between two nucleons at rest vanishes. The space components of the
axial current between nucleons at rest can be written as
\begin{equation}\label{5.5}
\bra{B} \bar\psi\gamma^i\gamma_5 T^a\psi \ket{B} = g N \bra{B} X^{ia}
\ket{B},
\end{equation}
where $\bra{B} X^{ia} \ket{B}$ and $g$ are of order one. The coupling constant
$g$ has been factored out so that the normalization of $X^{ia}$ can be chosen
to simplify future expressions. $X^{ia}$ is an operator (or $4\times4$ matrix)
defined on nucleon states $p\uparrow$, $p\downarrow$, $n\uparrow$,
$n\downarrow$, and $X^{ia}$ has a finite $N \rightarrow \infty$ limit.

\begin{figure}[tbp]
\begin{center}
\epsfig{file=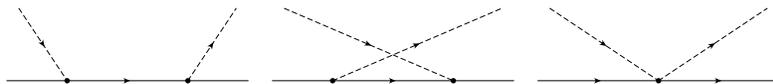,height=10mm}
\caption{Pion-nucleon scattering diagrams of order $E$. The third diagram is
$1/N^2$ suppressed in the large $N$ limit.\label{fig:37}}
\end{center}
\end{figure}

The leading contribution to pion-nucleon scattering is from the pole graphs in
fig.~\ref{fig:37}, 
which contribute at order $E$ provided the intermediate state is degenerate
with the initial and final states. Otherwise, the pole graph contribution is of
order $E^2$. In the large $N$ limit, the pole graphs are of order $N$, since
each pion-nucleon vertex is of order $\sqrt N$. There is also a direct
two-pion-nucleon coupling that contributes at order $E$, which is of order
$1/N$ in the large $N$ limit and can be neglected.

The pion-nucleon scattering amplitude for $\pi^a(q) + B (k) \rightarrow
\pi^b(q^\prime)+ B (k^\prime)$ from the pole graphs is
\begin{equation}\label{5.6}
-i\ q^i q^{\prime\,j} {N^2g^2\over f_\pi^2}
\left[{ 1\over q^0} X^{jb }X^{ia}
- {1 \over q^{\prime\,0} } X^{ia}X^{jb}\right],
\end{equation}
where the amplitude is written in matrix form, with the matrix labels denoting
the spin and flavor quantum numbers of the initial and final nucleons. Both
initial and final nucleons are on-shell, so $q^0=q^{\prime\,0}$. The product of
the $X$ matrices in eq.~(\ref{5.6}) sums over the possible spins and isospins
of the intermediate nucleon. Since $f_\pi\sim\sqrt N$, the overall amplitude is
of order $N$, which violates unitarity at fixed energy, and also contradicts
the large $N$ counting rules of Witten. Thus large $N$ QCD with a $I=J=1/2$
nucleon multiplet interacting with a pion is an inconsistent field theory.
There must be other states that cancel the order $N$ amplitude in
eq.~(\ref{5.6}) so that the total amplitude is order one, consistent with
unitarity. One can then generalize $X^{ia}$ to be an operator on this
degenerate set of baryon states, with matrix elements equal to the
corresponding axial current matrix elements. With this generalization, the form
of eq.~(\ref{5.6}) is unchanged, and we obtain the first consistency condition
for baryons \cite{dm},
\begin{equation}\label{5.7}
\left[X^{ia},X^{jb}\right]=0 \ .
\end{equation}
This consistency condition implies that the baryon axial currents are
represented by a set of operators $X^{ia}$ which commute in the large $N$
limit, a result also derived by Gervais and Sakita.~\cite{gs}. There are
additional commutation relations,
\begin{eqnarray}\label{xjcomm}
\left[J^i, X^{jb}\right] &=& i\,\epsilon_{ijk}\, X^{kb}, \\
\left[T^a, X^{jb}\right] &=& i\, f_{abc}\, X^{jc}, \nonumber
\end{eqnarray}
since $X^{ia}$ has spin one and isospin one.

The algebra in eqs.~(\ref{5.7}) and (\ref{xjcomm}) is a contracted $SU(2N_F)$
algebra, where $N_F$ is the number of quark flavors. To see this, consider the
algebra of operators in the non-relativistic quark model, which has an
$SU(2N_F)$ symmetry. The operators are
\[
J^i = q^\dagger {\sigma^i\over2}q,\ \
T^a = q^\dagger {T^a}q,\ \
G^{ia} = q^\dagger {\sigma^i \over 2} T^a q,
\]
where $J^i$ is the spin, $T^a$ is the flavor generator, and $G^{ia}$ are the
spin-flavor generators. The commutation relations involving $G^{ia}$ are
\begin{eqnarray}
\left[G^{ia},G^{jb}\right] &=& \frac i {2N_F}\, \epsilon_{ijk} \delta_{ab}\, J^k 
+\frac i 4\,f_{abc} \delta_{ij}\, T^c + \frac i 2\, \epsilon_{ijk} d_{abc}
\, G^{kc},\nonumber \\
\left[J^i,G^{jb}\right] &=& i\,\epsilon_{ijk}\, G^{kb},\nonumber \\
\left[T^a,G^{ib}\right] &=& i\,f_{abc}\, G^{jc}. \label{sucomm}
\end{eqnarray}
The algebra for large $N$ baryons in QCD is given by taking the limit
\begin{equation}\label{xlimit} 
X^{ia} \equiv \lim_{N\rightarrow\infty} {G^{ia} \over N}. 
\end{equation} 
The $SU(2N_F)$ commutation relations eq.~(\ref{sucomm}) turn into the
commutation relations eqs.~(\ref{5.7}--\ref{xjcomm}) in the large $N$ limit.
The limiting process eq.~(\ref{xlimit}) is known as a Lie algebra contraction.

We have just proved that the large $N$ limit of QCD has a contracted $SU(2N_F)$
symmetry in the baryon sector. The unitary irreducible representations of the
contracted Lie algebra can be obtained using the theory of induced
representations, and can be shown to be infinite dimensional. The simplest
irreducible representation for two flavors is a tower of states with $I=J=1/2,
3/2$, etc.\ which is the set of states of the Skyrme model, or the large $N$
non-relativistic quark model. The irreducible representations for three flavors
are more complicated. The $1/N$ expansion allows one to compute baryonic
quantities using $SU(2N_F)$ symmetry in the $N \rightarrow \infty$ limit. The
$1/N$ corrections allow one to systematically study the form of $SU(2N_F)$
symmetry breaking at finite $N$.

The pion-baryon coupling matrix $X^{ia}$ can be completely determined (up to an
overall normalization $g$), since it is a generator of the $SU(2N_F)_c$
algebra. It is easy to show that the large $N$ QCD predictions for the
pion-baryon coupling ratios are the same as those obtained in the Skyrme model
or non-relativistic quark model~\cite{am} in the $N\rightarrow \infty$ limit,
because both these models also have a contracted $SU(2N_F)$ symmetry in this
limit. In the Skyrme model, the axial current in the $N\rightarrow\infty$ limit
is $X^{ia}\propto \Tr A T^i A^{-1} T^a$, where $A$ is the Skyrmion collective
coordinate. The $X$'s commute (and so form part of an $SU(2N_F)_c$ algebra),
since $A$ is a $c$-number. We have already seen how the quark model algebra
reduces to $SU(2N_F)_c$ in the large $N$ limit. While we have shown that
$SU(2N_F)_c$ is a symmetry of QCD in the large $N$ limit, we have not shown
that $SU(2N_F)$ is a symmetry of QCD for finite $N$. There is no reason to
believe this is the case, so the $SU(2N_F)$ symmetry of the quark model is not
a symmetry of QCD. Nevertheless, many results obtained in the quark model will
be rederived in QCD using the $SU(2N_F)_c$ symmetry that is exact when $N
\rightarrow \infty$.

It is useful to have an explicit representation of the $N \rightarrow \infty$
$SU(2N_F)_c$ algebra to compute baryon properties in the $1/N$ expansion. There
are two natural choices discussed above, the quark model (acting on $N$ quark
baryons), or the Skyrme model. The two give equivalent results for physical
quantities. An operator $O$ (such as $X^{ia}$) in the quark representation can
be written as a $1/N$ expansion of Skyrme operators, and vice-versa,
\begin{equation}\label{5.8a}
O_{\rm quark} = O_{\rm Skyrme}^{(0)} + {1\over N} O_{\rm Skyrme}^{(1)} + \ldots.
\end{equation}
A
typical expression for a physical quantity is 
\[
a_0 O_0 + {a_1 \over N} O_1 + {a_2 \over N^2} O_2 + \ldots
\] 
where $a_i$ are coefficients, and $O_i$ are operators (such as $X^{ia}$ or
$J^i$, etc.) in either the quark or Skyrme representation. The coefficients
$a_i$ are not the same in the two representations, and the relation
eq.~(\ref{5.8a}) can be used to relate them. Clearly, any difference between
the representations is of higher order in $1/N$ than the terms retained in the
calculation. This is similar to the scheme dependence of physical quantities in
Feynman diagram perturbation theory.

The $1/N$ expansion provides some deep connections between the quark model,
the Skyrme model, and QCD. I do not have time in these lectures to discuss both
the quark and Skyrme models, and the connection between them. Some of these
results can be found in refs.~\cite{am,djm1,djm2}. In the rest of these
lectures, I will compare the large $N$ results mostly with the quark model.

\begin{hw}[$SU(2N_F)$ Commutation Relations]\sl\nobreak
\begin{itemize}
\end{itemize}\nobreak
\noindent Derive the $SU(2N_F)$ commutation relations eq.~(\ref{sucomm}) using
the quark operator relation $\left[q , q^\dagger \right]=1$.
\end{hw}

\subsection{$1/N$ Corrections}\label{sec:corr}

What makes the $1/N$ expansion for baryons interesting is that it is possible
to compute $1/N$ corrections. This allows one to compute results for the
physically relevant case $N=3$, rather than for the strict $N=\infty$ limit,
which is only of formal interest.

\begin{figure}[tbp]
\begin{center}
\epsfig{file=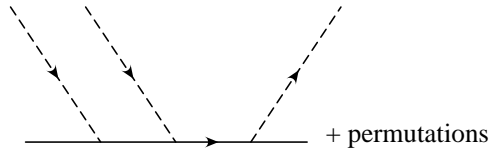,height=20mm}
\caption{The leading contribution to $\pi N \rightarrow \pi \pi N$ in the large
$N$ and chiral limits.\label{fig:38}}
\end{center}
\end{figure}

The $1/N$ corrections to the axial couplings $X^{ia}$ are determined by
considering the scattering process $\pi^a+ B\rightarrow \pi^b+\pi^c + B$ at low
energies. The baryon pole graphs that contribute in the large $N$ limit are
shown in fig.~\ref{fig:38}. 
The axial coupling $X^{ia}$ can be expanded in a series in $1/N$,
\begin{equation}\label{gexp}
X^{ia}= X^{ia}_0 +{1\over N} X^{ia}_1+\ldots,
\end{equation}
where $X^{ia}_0$ satisfies eq.~(\ref{5.7}). An explicit calculation shows that
the amplitude for pion-nucleon scattering from the diagrams in
fig.~\ref{fig:37} is proportional to
\[
N^{3/2} \left[X^{ia},\left[X^{jb},X^{kc}\right]\right]
\]
times kinematic factors, and violates unitarity unless the double commutator
vanishes at least as fast as $N^{-3/2}$, so that the amplitude is at most of
order one. In fact, one expects that the double commutator is of order $1/N^2$
from the large $N$ counting rule that the amplitude is order $1/\sqrt{N}$.
Substituting eq.~(\ref{gexp}) into the constraint gives the consistency
condition
\begin{equation}\label{acons}
\left[X^{ia}_0,\left[X_1^{jb},X^{kc}_0\right]\right] +
\left[X^{ia}_0,\left[X^{jb}_0,X_1^{kc}\right]\right] = 0,
\end{equation}
using $\left[X^{ia}_0,X^{jb}_0\right]=0$ from eq.~(\ref{5.7}). For two
flavors, the only solution to eq.~(\ref{acons}) is that $X_1^{ia}$ is
proportional to $X_0^{ia}$~\cite{dm}. This can be verified by an explicit
computation of $X_1^{ia}$ as shown in section~\ref{sec:res}~\cite{dm,djm1}. In
the rest of this section, I will state the solutions of the various large $N$
consistency conditions. The solutions can be obtained using the methods
discussed in detail in section~\ref{sec:soln}.

At order $1/N$, the baryons in an irreducible representation of the contracted
$SU(2N_F)$ Lie algebra are no longer degenerate, but are split by an order
$1/N$ mass term $\Delta M$. The intermediate baryon propagator in
eq.~(\ref{5.6}) should be replaced by $1/(E-\Delta M)$. The energy $E$ of the
pion is order one, whereas $\Delta M$ is of order $1/N$, so the propagator can
be expanded to order $1/N$ as
\begin{equation}\label{propexp}
{1\over E-\Delta M} = {1\over E} + {\Delta M\over E^2} +\ldots\ \ .
\end{equation}
Including the $1/N$ corrections to the propagator does not affect the
derivation of eq.~(\ref{5.7}), as the two terms in eq.~(\ref{propexp}) have
different energy dependences. The first term leads to the consistency condition
eq.~(\ref{5.7}) and the second gives the consistency condition on the baryon
masses \cite{j,djm1},
\begin{equation}\label{mcons}
\left[X^{ia},\left[X^{jb},\left[X^{kc},\Delta M\right]\right]\right]=0.
\end{equation}
This constraint can be used to obtain the $1/N$ corrections to the baryon
masses. The constraint eq.~(\ref{mcons}) is equivalent to a simpler constraint
obtained by Jenkins using chiral perturbation theory \cite{j}
\begin{equation}\label{mconsII}
\left[X^{ia},\left[X^{ia},\Delta M\right]\right]={\rm constant}.
\end{equation}
The solution of eq.~(\ref{mcons}) or (\ref{mconsII}) is that the baryon mass
splitting $\Delta M$ must be proportional to $J^2/N=j(j+1)/N$, where $j$ is
the spin of the baryon \cite{j}. This is precisely the form of the baryon mass
splitting in the Skyrme model~\cite{anw} or non-relativistic quark model.

\begin{figure}[tbp]
\setlength{\unitlength}{6mm}
\centerline{\hbox{
\begin{picture}(10,7.725)(-1.9,-0.525)
\def\level{\line(1,0){5}}
\thicklines
\put(0,6.4){\level}
\put(0,4.9){\level}
\put(0,3.6){\level}
\put(0,2.5){\level}
\put(0,1.6){\level}
\put(0,0.9){\level}
\put(0,0.4){\level}
\put(0,0.1){\level}
\put(0,0){\level}
\thinlines
\put(5.5,6.4){\line(1,0){1}}
\put(5.5,4.9){\line(1,0){1}}
\put(5.5,0.125){\line(1,0){1}}
\put(5.5,-0.025){\line(1,0){1}}
\put(-1.9,0){\line(1,0){1}}
\put(-1.9,6.4){\line(1,0){1}}
\put(6,5.65){\makebox(0,0){$1$}}
\put(7.5,0){\makebox(0,0){$1/N$}}
\put(-1.65,3.05){$N$}
\put(-1.4,3.6){\vector(0,1){2.7}}
\put(-1.4,2.8){\vector(0,-1){2.7}}
\put(6,7.2){\vector(0,-1){0.8}}
\put(6,4.1){\vector(0,1){0.8}}
\put(6,-0.525){\vector(0,1){0.5}}
\put(6,0.625){\vector(0,-1){0.5}}
\end{picture}
}}
\caption{The baryon mass spectrum including the $J^2/N$ term. The top of the
tower is $j=N/2$, and the bottom is $j=1/2$. \label{fig:39}}
\end{figure}
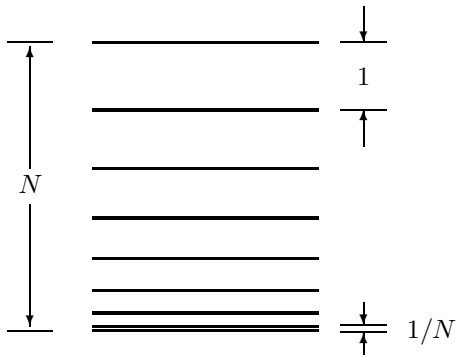

The structure of the $1/N$ corrections shows that the expansion parameter is
$J/N$, where $J$ is the spin of the baryon. For example, the baryon mass
spectrum including the $J^2/N$ mass splitting has the form shown in
fig.~\ref{fig:39}. The correction terms are only small near the bottom of the
(infinite) baryon tower. For this reason, the $1/N$ expansion will only be
considered for baryons with spin $J$ held fixed as $N \rightarrow \infty$.

\subsection{Solution of Consistency Conditions}\label{sec:soln}

\begin{figure}[tbp]
\begin{center}
\epsfig{file=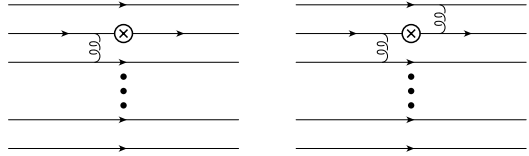,height=20mm}
\caption{Two-body and three-body operator contributions to the baryon matrix
elements of a QCD operator.\label{fig:53}}
\end{center}
\end{figure}

The solution of the large $N$ consistency conditions is the key to
understanding the structure of baryons. The answer can be given quite
simply~\cite{djm1,djm2,cgo,lm}. A quantity $Q$ that is of order $N^r$ in the
large $N$ limit can be expanded in terms of the basic operators of the quark
representation, $G^{ia}$, $J^i$ and $T^a$ as
\begin{equation}\label{5.8}
{Q \over N^r} = {\mathcal P} \left( { G^{ia} \over N}, {J^i \over N}, {T^a \over N} \right),
\end{equation}
where $\mathcal P$ is a polynomial in its arguments, with coefficients that
have an expansion in $1/N$. For example, we will soon see that the baryon mass
in the flavor limit has the expansion
\begin{equation}\label{5.10}
{M_B \over N} = \left[ a_0 + a_1 \left({J \over N}\right)^2 + 
a_2 \left({J \over N}\right)^4 + \ldots \right],
\end{equation}
where the $a_i(1/N)$ are unknown expansion coefficients. It is important to
remember that no assumption is being made about the validity of the
non-relativistic quark model. The operators $G^{ia}$, $J^i$ and $T^a$ are used
as a way to do the $SU(2N_F)$ group-theoretic computations. The expansion
eq.~(\ref{5.8}) is true irrespective of the quark mass. For very heavy quarks,
the non-relativistic Hartree picture is valid, and one can see that
eq.~(\ref{5.8}) is consistent with the diagrammatic analysis of $1/N$ factors
due to gluon exchange~\cite{cgo,lm}. Two-body operators on the right-hand side
of eq.~(\ref{5.8}) are generated by an insertion of a one-body operator plus
one-gluon exchange, three-body operators by a one-body operator plus two gluon
exchanges, and so on, as shown in fig.~\ref{fig:53}. Each gluon exchange 
brings with it a factor of $1/N$ from the two couplings, which reproduces the
$N$-counting in eq.~(\ref{5.8}). What is non-trivial is that the $N$-counting
rules also hold if the quarks are massless.

At first sight, all terms in the expansion eq.~(\ref{5.8}) are equally
important, since an $r$-body operator in the expansion of $\mathcal P$ has a
coefficient of order $1/N^r$, and matrix elements of $r$-body operators are of
order $N^r$. Including all terms in eq.~(\ref{5.8}) is equivalent to saying
that all possible $SU(2N_F)$ representations are equally important, so that
there is no predictive power. There are operator identities which allow
one to simplify the general expansion eq.~(\ref{5.8}), and drop certain terms
as subleading in $1/N$. It is these identities which allow one to use the $1/N$
expansion to make non-trivial predictions for baryons in QCD. The complete set
of operator identities was derived in~ref.~\cite{djm2}. They are listed in
Tables~\ref{tab:su6iden} and \ref{tab:su4iden} for the case of three and two
flavors, respectively.

\begin{table}[tbp]
\caption{ $SU(6)$ Identities. Some of the identities need to be projected onto
a given spin or flavor channel, which is given in parentheses.}
\label{tab:su6iden}
\setlength{\tabcolsep}{0.12em}
\renewcommand{\arraystretch}{1.5} 
\begin{tabular}{@{}cc@{}}
\hline
$2\left\{J^i,J^i\right\} + 3\left\{T^a,T^a\right\} + 12
\left\{G^{ia},G^{ia}\right\} = 5 N \left(N+6\right)$\\
\hline
$d^{abc}\left\{G^{ia}, G^{ib}\right\} + {2\over 3} \left\{J^i,G^{ic}
\right\} + {1\over 4} d^{abc}\left\{T^a, T^b\right\} = {2\over 3}
\left(N+3\right) T^c $\\
$\left\{T^a,G^{ia}\right\} = {2\over3}\left(N+3\right)\ J^i $ \\
${1\over 3}\left\{J^k,T^c\right\} + d^{abc} \left\{T^a,G^{kb}\right\}
-\epsilon^{ijk} f^{abc} \left\{G^{ia}, G^{jb}\right\}
= {4\over3}
\left(N+3\right) G^{kc} $\\
\hline
$-12\ \left\{G^{ia},G^{ia}\right\} + 27\ \left\{T^a,
T^a\right\} - 32\ \left\{J^i,J^i\right\}=0 $\\
$d^{abc}\ \left\{G^{ia}, G^{ib}\right\} + {9\over 4} \ d^{abc}\ \left\{
T^a, T^b\right\} - {10\over3}\ \left\{J^i,G^{ic}\right\} = 0$ 
\\
$4\ \left\{G^{ia},G^{ib}\right\} = \left\{T^a,T^b\right\}$ & $({\bf 27})$ \\
$\epsilon^{ijk}\ \left\{ J^i,G^{jc}\right\} = f^{abc} \ \left\{T^a,G^{kb}
\right\}$ \\
$3\ d^{abc}\ \left\{T^a,G^{kb}\right\} = \left\{J^k,T^c\right\} -
\epsilon^{ijk} f^{abc}\ \left\{G^{ia}, G^{jb}\right\} $\\
$\epsilon^{ijk}\ \left\{G^{ia},G^{jb}\right\} = f^{acg} d^{bch}\ \left\{
T^g,G^{kh}\right\}$ & $(\bar{{\bf 10}}+ {\bf 10}) $ \\
$3\ \left\{G^{ia}, G^{ja}\right\} = \left\{J^i, J^j
\right\}$ & $(J=2)$\\
$3\ d^{abc}\ \left\{G^{ia}, G^{jb}\right\} =
\left\{J^i,G^{jc}\right\}$ & $(J=2)$\\
\hline
\end{tabular}
\end{table}

\begin{table}[tbp]
\caption{$SU(4)$ Identities. Some of the identities need to be projected onto
a given spin or flavor channel, which is given in the second column. The last
column gives the transformation properties of the identities under
$SU(2)\times SU(2)$.}
\label{tab:su4iden}
\renewcommand{\arraystretch}{1.5} 
\begin{tabular}{ccc}
\hline
$\left\{J^i,J^i\right\} + \left\{I^a,I^a\right\} +
4\ \left\{G^{ia},G^{ia}\right\} ={3 \over2} N \left(N+4\right)
$ & & $(0,0)$ \\
\hline
$2 \left\{J^i,G^{ia}\right\} = \left(N+2\right)\ I^a$
& & $(0,1)$ \\
$2 \left\{I^a,G^{ia}\right\} = \left(N+2\right)\ J^i$ & $(1,0)$ \\
${1\over 2}\ \left\{J^k,I^c\right\}
-\epsilon^{ijk} \epsilon^{abc} \left\{G^{ia}, G^{jb}\right\} =
\left(N+2\right)\ G^{kc} $& &$(1,1)$ \\
\hline
$\left\{I^a,
I^a\right\} - \left\{J^i,J^i\right\}=0 $ & $(0,0)$ \\
$ 4\ \left\{G^{ia},G^{ib}\right\} = \left\{I^a,I^b\right\}$ & $(I=2)$ &
$(0,2)$ \\
$\epsilon^{ijk}\ \left\{ J^i,G^{jc}\right\} = \epsilon^{abc} \
\left\{I^a,G^{kb}
\right\}$ & & $(1,1)$ \\
$4\ \left\{G^{ia}, G^{ja}\right\} = \left\{J^i, J^j
\right\}$ & $ (J=2)$ & $(2,0)$ \\
\hline
\end{tabular}
\end{table}

The operator identities allow one to eliminate certain operator combinations in
the general expansion eq.~(\ref{5.8}). For example, for three flavors,
\[
\left\{T^a, G^{ia}\right\} = \frac 2 3\left(N+3\right) J^i
\]
can be used to rewrite the two-body operator on the left-hand side in terms of
the one-body operator on the right-hand side. The two-body operator matrix
element is of order $N^2$, and the one-body operator matrix element is of order
$N$, which is consistent with the coefficient of proportionality being of order
$N$. The operator identities have been written in terms of anticommutators,
since they are hermitian, and since commutators can be simplified using the
$SU(2N_F)$ commutation relations eq.~(\ref{sucomm}). A study of the identities
leads to the following reduction rules:
\begin{itemize}
\item[{\bf Operator Reduction Rule (three flavors):}] All operator products in
which two flavor indices are contracted using $\delta^{ab}$, $d^{abc}$ or
$f^{abc}$, or two spin indices on $G$'s are contracted using $\delta^{ij}$ or 
$\epsilon^{ijk}$ can be eliminated.
\end{itemize}
\begin{itemize}
\item[{\bf Operator Reduction Rule (two flavors):}] All operators in which two
spin or isospin indices are contracted with a $\delta$ or $\epsilon$-symbol can
be eliminated, with the exception of $J^2$. 
\end{itemize}
The inclusion of $J^2$, but not $I^2$, in the set of independent operators does
not break the symmetry between spin and isospin, because of the identity $I^2 =
J^2$.\footnote{For two flavors, the isospin $T^a$ will also be denoted by
$I^a$.}

As an application of the operator reduction rule, consider the baryon masses in
the $SU(3)$ symmetry limit. The general form of the baryon mass is given by
eq.~(\ref{5.8}),
\[
{M \over N}= \mathcal{P}\left( { G^{ia} \over N}, {J^i \over N}, 
{T^a \over N} \right),
\]
where the terms in $\mathcal{P}$ must be spin-zero flavor singlets, since those
are the quantum numbers of $M$. Thus all terms in $\mathcal{P}$ are obtained by
contracting the spin and flavor indices on $G^{ia}$, $J^i$ and $T^a$ using spin
and flavor invariant tensors, such as $\delta^{ab}$ or $f^{abc}$. The operator
reduction rule tells us that all terms involving $T^a$ or $G^{ia}$ can be
eliminated, so that one is left with an expansion in $J^i$. Rotational
invariance implies that the expansion is in $J^2$, and so has the form
eq.~(\ref{5.10}), and leads to the spectrum in fig.~\ref{fig:39}.

Equation~(\ref{5.10}) shows that the baryon expansion parameter is $J/N$. One
can compute baryon properties in a systematic expansion in $1/N$ provided one
looks at states with $J$ fixed as $N \rightarrow \infty$. Generically, $J$ is
a one-body operator and can have matrix elements of order $N$. The $1/N$ 
expansion is only valid for states in which $J$ is of order unity, i.e.\ for
states at the bottom of the baryon tower in fig.~\ref{fig:39}. In these
states, there is a cancellation between the $N$ possible insertions of the
one-body operator $J$ on the $N$ quark lines. Generically, matrix elements of
$G^{ia}$ and $T^a$ are $\ord N$, so that $G^{ia}/N$ and $T^a/N$ are $\ord 1$.
In the case of two flavors, there is an additional simplification that $I^a$
is $\ord 1$, so that $I^a/N$ is $1/N$ suppressed.

\begin{hw}\sl
\begin{itemize}
\end{itemize}
\noindent Prove the identity
\[
4 J^i G^{ia} = (N+2) I^a
\]
for two quark flavors.

\noindent {\bf Solution:}
It is convenient to write $G^{ia} = (1/4)\sum_\alpha \sigma^i_\alpha
\tau^a_\alpha$, where the sum is over the different quark lines in the baryon,
each of which has a different color. Then
\begin{eqnarray*}
8 J^i G^{ia} &=& \sum_{\alpha\beta}
\sigma^i_\alpha \sigma^i_\beta \tau^a_\beta\\
&=& \sum_{\alpha=\beta}
\sigma^i_\alpha \sigma^i_\beta \tau^a_\beta
+ \sum_{\alpha \not =\beta}
\sigma^i_\alpha \sigma^i_\beta \tau^a_\beta.
\end{eqnarray*}
The term with $\alpha=\beta$ can be written as
\begin{eqnarray*}
\sum_{\alpha=\beta}
\sigma^i_\alpha \sigma^i_\alpha \tau^a_\beta
&=& 3 \sum_\alpha \tau^a \\
&=& 6 I^a.
\end{eqnarray*} 
The terms with $\alpha\not=\beta$ can be simplified using the identity
\[
\left[\sigma^i_\alpha\right]^a_b\ \left[\sigma^i_\beta\right]^c_d = 
2 \delta^a_d \delta^c _b - \delta^a_b \delta^c_d 
\equiv 2 S_{\rm exch}(\alpha,\beta) - 1
\]
where $S_{\rm exch}(\alpha,\beta)$ is the spin-exchange operator that exchanges
the spins of the two quarks, to give
\[
\sum_{\alpha \not =\beta}
\sigma^i_\alpha \sigma^i_\beta \tau^a_\beta = 
\sum_{\alpha \not =\beta} \left[ 2 S_{\rm exch}(\alpha,\beta)-1\right]
 \tau^a_\beta
\]
The final state baryon is completely symmetric in spin $\otimes$ flavor, so one
can replace $S_{\rm exch}(\alpha,\beta)$ by the flavor exchange operator
$F_{\rm exch}(\alpha,\beta)$. The identity
\[
[\tau^g_\alpha]^a_b\ [\tau^g_\beta]^c_d = 2 \delta^a_d \delta^c _b
- \delta^a_b \delta^c_d \equiv 2 F_{\rm exch}(\alpha,\beta) - 1
\]
allows one to rewrite this as
\[
\sum_{\alpha \not =\beta} \tau^g_\alpha \tau^g_\beta
 \tau^a_\beta.
\]
Then
\[
\tau^g_\beta \tau^a_\beta = \delta^{ga}_\beta + i \epsilon_{gah} \tau^h_\beta
\]
implies that
\begin{eqnarray*}
\sum_{\alpha \not =\beta} \tau^g_\alpha \tau^g_\beta
\tau^a_\beta &=& \tau^g_\alpha \left( \delta^{ga}_\beta + 
i \epsilon_{gah} \tau^h_\beta\right) \\
&=& \sum_{\alpha \not =\beta} \tau^a_\alpha + 
\sum_{\alpha,\beta}i \epsilon_{gah} \tau^g_\alpha \tau^h_\beta
-\sum_{\alpha=\beta}i \epsilon_{gah} \tau^g_\alpha\tau^h_\beta\\
&=& (N-1) \sum_\alpha \tau^a_\alpha + i \epsilon_{gah} 
\sum_\alpha \tau^g_\alpha \sum_\beta \tau^h_\beta \nonumber \\
&& \ \ -i \sum_\alpha 
\epsilon_{gah}\left( \delta^{gh} + i \epsilon_{ghr} \tau^r_\alpha \right) \\
&=& 2 (N-1) I^a + 4 i \epsilon_{gah} I^g I^h + 2\epsilon_{gah}\epsilon_{ghr}
I^r \\
&=& 2 (N-1) I^a + 2 i \epsilon_{gah}\left[I^g,I^h \right] + 2\epsilon_{gah}
\epsilon_{ghr} I^r\\
&=& 2 (N-1) I^a
\end{eqnarray*}
Combining the pieces gives the desired identity. More examples of this kind can
be found in refs.~\cite{djm2,cgo,lm}.
\end{hw}

\section{Masses with SU(3) Breaking}

The baryon masses can be computed in a systematic expansion in powers of $1/N$
and $SU(3)$ breaking. I will use a simplified version of the analysis of
Jenkins and Lebed~\cite{jl}, by neglecting isospin breaking. This example is
discussed in detail, to show how the formalism we have developed can be
applied.

There are eight isospin-averaged baryon masses for the $N$, $\Lambda$,
$\Sigma$, $\Xi$, $\Delta$, $\Sigma^*$, $\Xi^*$ and $\Omega$. The general form
of the $SU(3)$ singlet mass term has already been worked out in
eq.~(\ref{5.8}). At first order in $SU(3)$ breaking, the baryon mass term
transforms as an $SU(3)$ octet. The most general spin-zero $SU(3)$ octet is a
polynomial in $J^i$, $T^a$ and $G^{ia}$ with one free flavor index set to $8$.
All operators with contracted flavor indices can be eliminated using the
operator reduction rule, so one is left with terms with a single $T^8$ or
$G^{i8}$, and powers of $J^i$,
\begin{eqnarray}
&&\epsilon b_1 T^8 + \epsilon b_3 {J^2 T^8 \over N^2} +\epsilon b_5 {J^4 T^8
 \over N^4} + \ldots \nonumber \\
&&+\epsilon b_2 {J^i G^{i8} \over N} +\epsilon b_4 {J^2 J^i G^{i8} \over N^3} 
+ \ldots
\label{5.11}
\end{eqnarray}
where $\epsilon$ is a measure of $SU(3)$ breaking. $SU(3)$ breaking is due to
the strange quark mass $m_s$. Chiral perturbation theory shows that there are
non-analytic contributions to the baryon mass, such as $m_s^{3/2}$ and $m_s^2
\log m_s$. The structure of $SU(3)$ breaking as a function of $m_s$ is highly
non-trivial, but goes beyond the scope of these introductory lectures. For the
purposes of the analysis here, I will treat $\epsilon$ as a small parameter of
order $SU(3)$ breaking.

At second order in $SU(3)$ breaking, one can get a tensor with two free flavor
indices set to $8$, and at third order, with three free flavor indices set to
$8$. One can work out the general form of these terms, as in eq.~(\ref{5.11}).
Eventually, the matrix elements will be taken between baryons for $N=3$. For
this reason, one can stop the expansion of the mass term at three-body
operators, since a baryon for $N=3$ contains three quarks. Operators with more
than three quarks can be written as a linear combination of operators with
three or fewer quarks. One simple way to see this is to write operators in
normal ordered form, with all $q^\dagger$'s to the left of the $q$'s. Then
$r$-body operators with $r>3$ vanish on baryons containing three quarks. 

The order $\epsilon^2$ terms up to three-body operators are
\begin{equation}\label{5.12}
\epsilon^2 c_2 {T^8 T^8 \over N} + \epsilon^2 c_3 {T^8 J^i G^{i8} \over N^2}
\end{equation}
and the $\epsilon^3$ term is
\begin{equation}\label{5.13}
\epsilon^3 d_3 {T^8 T^8 T^8\over N^2}.
\end{equation}
Combining eqs.~(\ref{5.8})--(\ref{5.13}) and stopping at three-body operators
gives the baryon mass formula
\begin{eqnarray}
M &=& a_0 + a_1 {J^2 \over N} + 
\epsilon b_1 T^8 + \epsilon b_2 {J^i G^{i8} \over N} +
\epsilon b_3 {J^2 T^8 \over N^2} + \epsilon^2 c_2 {T^8 T^8 \over N}\nonumber \\
&& + 
\epsilon^2 c_3 {T^8 J^i G^{i8} \over N^2} + 
\epsilon^3 d_3 {T^8 T^8 T^8\over N^2},
\label{5.14}
\end{eqnarray}
which gives the eight baryon masses in terms of eight parameters. This must be
the case, since the baryon masses are independent if one works to arbitrary
order in $1/N$ and $\epsilon$. The non-trivial information contained in
eq.~(\ref{5.14}) is the $\epsilon$ and $N$ dependence of the various terms. 
Equation~(\ref{5.14}) will be used to derive a hierarchy of baryon mass
relations to a given order in $\epsilon$ and $1/N$. To obtain these relations,
one needs the matrix element of eq.~(\ref{5.14}) between baryon states. The
matrix element of
\[
\sqrt{12}\ T^8 = 
\left(\begin{array}{ccc}
1\\
&1\\
&&-2\\
\end{array}\right) = 
\left(\begin{array}{ccc}
1\\
&1\\
&&1\\
\end{array}\right) - 
\left(\begin{array}{ccc}
0\\
&0\\
&&3\\
\end{array}\right)
\]
is 
\begin{equation}\label{5.15}
T^8 = {1\over \sqrt{12}}\left(N-3N_s\right),
\end{equation}
where $N_s$ is the number of strange quarks. The operator $G^{i8}$ is
\[
G^{i8} = {1\over \sqrt{12}}\left(J^i-3 J_s^i\right),
\]
where $J^i$ is the quark spin, and
\[
J_s^i = q^\dagger {\sigma^i\over 2}\left(\begin{array}{ccc}
0\\
&0\\
&&1\\
\end{array}\right) q
\]
is the strange quark spin. This gives
\[
J^i G^{i8} = {1\over \sqrt{12}}\left(J^2-3 J \cdot J_s\right).
\]
The total spin of the baryon is $J=J_{ud} + J_s$, where $J_{ud}$ is the spin of
the $u$- and $d$-quarks, and $J_s$ is the spin of the strange quarks. One can
write $J^2-3 J \cdot J_s = 3 J_{ud}^2 - J^2 - 3 J_s^2$, and then use the
identity $J_{ud}^2=I^2$, where $I$ is the isospin, to obtain
\begin{equation}\label{5.16}
J^i G^{i8} = {1\over \sqrt{12}}\left(3I^2 - J^2- 3 J_s^2\right).
\end{equation}
All the baryons are eigenstates of $N_s$, $I^2$, $J^2$ and $J_s^2$, so the
matrix element of eq.~(\ref{5.14}) can be computed simply using
eqs.~(\ref{5.15}) and (\ref{5.16}). The matrix elements are listed in
Table~\ref{tab:b}.

\begin{table}[tbp]
\caption{Matrix elements of the mass operators, eq.~(\ref{5.14}). The first
section of the table lists the matrix elements of the basic operators $N$,
$N_s$, $J^2$, $J_s^2$ and $I^2$. These are used to compute the matrix elements
of the remaining operators using eqs.~(\ref{5.15})--(\ref{5.16}).
\label{tab:b}}
\renewcommand{\arraystretch}{1.25}
\begin{tabular}{ccccccccc}
\hline
& $N$ & $\Lambda$ & $\Sigma$ & $\Xi$ & $\Delta$ & $\Sigma^*$ & $\Xi^*$ &
$\Omega$ \\
\hline
$N$ & $3$ & $3$ & $3$ & $3$ & $3$ & $3$ & $3$ & $3$ \\
$N_s$ & $0$ & $1$ & $1$ & $2$ & $0$ & $1$ & $2$ & $3$ \\
$J^2$ & $3/4$ & $3/4$ & $3/4$ & $3/4$ & $15/4$ & $15/4$ & $15/4$ & $15/4$ \\
$J_s^2$ & $0$ & $3/4$ & $3/4$ & $2$ & $0$ & $3/4$ & $2$ & $15/4$ \\
$I^2$ & $3/4$ & $0$ & $2$ & $3/4$ & $15/4$ & $2$ & $3/4$ & $0$ \\
\hline
$1$ & $1$ & $1$ & $1$ & $1$ & $1$ & $1$ & $1$ & $1$ \\
$J^2$ & $3/4$ & $3/4$ & $3/4$ & $3/4$ & $15/4$ & $15/4$ & $15/4$ & $15/4$ \\
$2\sqrt 3 T^8$ & $3$ & $0$ & $0$ & $-3$ & $3$ & $0$ & $-3$ & $-6$ \\
$4 \sqrt 3 J^i G^{i8}$ & $3/2$ & $-3$ & $3$ & $-9/2$ & $15/2$ & $0$ & $-15/2$ &
$-15$ \\
$2 \sqrt 3 J^2 T^8$ & $9/4$ & $0$ & $0$ & $-9/4$ & $45/4$ & $0$ & $-45/4$ & 
$-45/2$ \\
$\left(2\sqrt 3 T^8\right)^2$ & $9$ & $0$ & $0$ & $9$ & $9$ & $0$ & $9$ & 
$36$ \\
$24 T^8 J^i G^{i8}$ & $9/2$ & $0$ & $0$ & $27/2$ & $45/2$ & $0$ & $45/2$ &
$90$ \\
$\left(2\sqrt 3 T^8\right)^3$ & $27$ & $0$ & $0$ & $-27$ & $27$ & $0$ & $-27$ & 
$-216$ \\
 \hline
\end{tabular}
\end{table}

Combining eq.~(\ref{5.14}) and Table~\ref{tab:b} gives the baryon masses
\begin{equation}\label{20.1}
M = H A,
\end{equation}
where
\[
M = \left( 
\begin{array}{cccccccc}
m_N & m_\Lambda & m_\Sigma & m_\Xi & m_\Delta & m_{\Sigma^*} &
m_{\Xi^*} & m_\Omega \\
\end{array}
\right),
\]
is a row vector of baryon masses,
\[ 
A = \left( \begin{array}{cccccccc}
1 & 1 & 1 & 1 & 1 & 1 & 1 & 1 \\
3/4 & 3/4 & 3/4 & 3/4 & 15/4 & 15/4 & 15/4 & 15/4 \\
3 & 0 & 0 & -3 & 3 & 0 & -3 & -6 \\
3/2 & -3 & 3 & -9/2 & 15/2 & 0 & -15/2 &-15 \\
9/4 & 0 & 0 & -9/4 & 45/4 & 0 & -45/4 & -45/2 \\
9 & 0 & 0 & 9 & 9 & 0 & 9 & 36 \\
9/2 & 0 & 0 & 27/2 & 45/2 & 0 & 45/2 & 90 \\
27 & 0 & 0 & -27 & 27 & 0 & -27 & -216 \\
\end{array}\right)
\]
is the array of matrix elements from Table~\ref{tab:b}, and
\begin{eqnarray*}
H &=& \left( 
\begin{array}{cccccccc}
H_1 & H_2 & H_3 & H_4 & H_5 & H_6 & H_7 & H_8 \\
\end{array}
\right) \\
&=& \left( 
\begin{array}{cccccccc}
a_0 & {a_1\over N} & {\epsilon b_1 \over 2 \sqrt 3} & 
{\epsilon b_2 \over 4 \sqrt 3 N} &
{\epsilon b_3 \over 2 \sqrt 3 N^2} & 
{\epsilon^2 c_2 \over 12 N} & {\epsilon^2 c_3 \over 24 N^2}& 
{\epsilon^3 d_3 \over 24 \sqrt 3 N^3} \\
\end{array}
\right)
\end{eqnarray*}
is the row vector of coefficients.

The classification of the mass operators in powers of $1/N$ and $\epsilon$ is
shown in Table~\ref{tab:c}. 
\begin{table}[tbp]
\caption{The order in $1/N$ and $\epsilon$ of the eight mass operators.
\label{tab:c}}
\renewcommand{\arraystretch}{1.2}
\begin{tabular}{ccccc}
\hline
& $N$ & $1$ & $1/N$ & $1/N^2$ \\
\hline
$1$ & $1$ & & $J^2$ \\
$\epsilon$ & & $T^8$ & $J^i G^{i8}$ & $J^2 T^8$ \\ 
$\epsilon^2$ & & & $T^8 T^8$ & $T^8 J^i G^{i8}$ \\
$\epsilon^3$ & & & & $T^8 T^8 T^8$ \\
\hline
\end{tabular}
\end{table}
There are no relations if all the operators are retained in eq.~(\ref{20.1}).
The most accurate relation is obtained if one omits the operator $T^8 T^8 T^8$
which contributes at order $\epsilon^3/N^2$. This gives a baryon mass relation
that has an error of order $\epsilon^3/N^2$. The mass relation is obtained by
writing eq.~(\ref{20.1}) as
\begin{equation}\label{20.10}
H = M A^{-1}.
\end{equation}
Omitting $T^8 T^8 T^8$ means that $H_8=0$, which gives the relation $R8$ (on
multiplying by 162 to eliminate fractional coefficients)
\begin{equation}\label{r8}
(R8):\qquad\qquad\Delta - 3 \Sigma^* + 3 \Xi^* - \Omega =
\ord{\epsilon^3/N^2},
\end{equation}
since $H_8$ is $\ord{\epsilon^3/N^2}$. The next most accurate relation is
obtained by dropping $T^8 T^8 T^8$ and $T^8 J^i G^{i8}$. This is equivalent to
$H_8=0$ and $H_7=0$ in eq.~(\ref{20.10}). The relation $H_8$ is eq.~(\ref{r8}).
The relation $H_7=0$ is
\[
(R7):
14 N - 21 \Lambda - 7 \Sigma+ 14 \Xi - 8 \Delta + 10 \Sigma^* - 4 \Xi^*
-6 \Omega = \ord{\epsilon^2/N^2}.
\]
At the next step, one can drop either the $T^8 T^8$ operator, which gives a
new relation with an error $\ord{\epsilon^2/N}$, or drop the $J^2 T^8$
operator, which gives a relation with an error $\ord{\epsilon/N^2}$. Dropping
both gives no additional independent relation. One can then drop $J^i G^{i8}$
to get a $\ord{\epsilon/N}$ relation, and then drop either $T^8$ or $J^2$ to
get $\ord{\epsilon}$ or $\ord{1/N}$ relations. 

This procedure gives a hierarchy of mass relations $H_i=0$. There are no free
parameters in any of the relations, since all coefficients were fixed by group
theory. It is convenient to write down relations that have definite $SU(6)$ and
$SU(2) \times SU(3)$ transformation properties. These relations are orthogonal
to each other, since different irreducible representations of a symmetry group
are orthogonal. One can make the relations $H_i=0$ orthogonal using the
Gram-Schmidt procedure starting with $H_8=0$ and working down to  $H_1=0$, with
respect to the metric (see problem~\ref{prob:orth})
\begin{equation}\label{gmetric}
g = {\rm diag}\left(
\begin{array}{cccccccc}
\frac 1 4 & \frac 1 2 & \frac 1 6 & \frac 1 4 & \frac 1 {16} &
\frac 1 {12} & \frac 1 {8} & \frac 1 4
\end{array} \right).
\end{equation}
The entries in this matrix are the reciprocals of the number of baryon states,
e.g. there are four nucleon states, $p\uparrow$, $p\downarrow$, $n\uparrow$,
$n\downarrow$, two $\Lambda$ states $\Lambda\uparrow$, $\Lambda\downarrow$,
etc. The resulting relations are tabulated in Table~\ref{tab:d}.
\begin{table}[tbp]
\caption{Baryon mass relations with the order in $\epsilon$ and $1/N$. The
entries in the table are the coefficients of $m_N$, etc.\ in the eight
relations. The fractional error computed using the experimental values for the
masses is listed in the last column. \label{tab:d}}
\setlength{\tabcolsep}{0.6em}
\begin{tabular}{@{}crrrrrrrrcr@{\hspace{0.12em}}c@{\hspace{0.12em}}l@{}}
\hline
&$ N $&$ \Lambda $&$ \Sigma $&$ \Xi $&$ \Delta $&$ \Sigma^* $&$ \Xi^* $&$
\Omega $& Order & \multicolumn{3}{c}{{Frac. Error}} \\
\hline
$R1 $&$ 2$&$1$&$3$&$2$&$8$&$6$&$4$&$2 $&$ N $&&& \\
$R2 $&$ -10$&$-5$&$-15$&$-10$&$16$&$12$&$8$&$4 $&$ 1/N $&$ 0.18 $&$ \pm $&$
0.0004$ \\
$R3 $&$ 1$&$0$&$0$&$-1$&$4$&$0$&$-2$&$-2 $&$ \epsilon $&$ 0.27 $&$ \pm $&$
0.0007$ \\
$R4 $&$ -7$&$-2$&$6$&$3$&$4$&$0$&$-2$&$-2 $&$ \epsilon/N $&$ 0.052 $&$ \pm $&$ 
0.0003 $\\
$R5 $&$-2$&$3$&$-9$&$8$&$4$&$0$&$-2$&$-2 $&$ \epsilon/N^2 $&$ 0.011 $&$
 \pm $&$ 0.0003 $\\
$R6 $&$2$&$-3$&$-1$&$2$&$16$&$-20$&$-8$&$12 $&$ \epsilon^2/N $&$ 0.0048 $&$ 
\pm $&$ 0.0004 $\\
$R7 $&$-14$&$21$&$7$&$-14$&$8$&$-10$&$-4$&$6 $&$ \epsilon^2/N^2 $&$ 0.0017
 $&$ \pm $&$ 0.0002 $\\
$R8 $&$0$&$0$&$0$&$0$&$1$&$-3$&$3$&$-1 $&$ \epsilon^3/N^2 $&$ 0.0009 $&$ 
\pm $&$ 0.0003 $\\
\hline
\end{tabular}
\end{table}

\begin{figure}[tbp]
\begin{center}
\epsfig{file=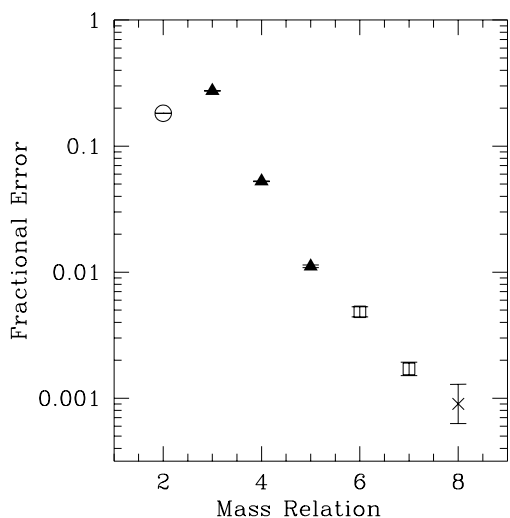,height=70mm}
\caption{Accuracy of the large $N$ baryon mass relations from
Table~\ref{tab:d}. The error bars are those on the experimentally measured
baryon masses. The shape of the points gives their order in $SU(3)$ breaking:
$\bigcirc$ is $\ord 1$, $\blacktriangle$ is $\ord \epsilon$, $\Box$ is $\ord
{\epsilon^2}$, and $\times$ is $\ord{\epsilon^3}$. The differences within the
$\ord \epsilon$ and $\ord {\epsilon^2}$ relations are explained by including
the factors of $1/N$. The accuracies of the baryon mass relations are $1/N^2$,
$\epsilon/N$, $\epsilon/N^2$, $\epsilon/N^3$, $\epsilon^2/N^2$,
$\epsilon^2/N^3$ and $\epsilon^3/N^3$, respectively. \label{fig:40}}
\end{center}
\end{figure}

The mass relations can now be compared with experiment. The relations derived
are homogeneous relations, and have no standard normalization, i.e.\ they can be
multiplied by an arbitrary overall coefficient. It is therefore best to compute
the fractional error on each relation. Relations which hold to higher order in
$1/N$ or $\epsilon$ should work better, if $1/N$ and $\epsilon$ are good
expansion parameters. The fractional error is defined using the following
procedure. Each relation is written in the form $L=R$, where both sides are
linear combinations of masses with positive coefficients. For example, the
$\ord{\epsilon^3/N^2}$ relation $R8$ is written as
\[
L = \Delta + 3 \Xi^* = 3 \Sigma^* + \Omega = R.
\]
One then computes the fractional accuracy of the relation, $\abs{L-R}/\left(
\left(L+R \right)/2 \right)$ using the experimental values for the isospin
averaged baryon masses (in MeV),
\[
\kern-1.5em\begin{array}{cccccccc}
m_N & m_\Lambda & m_\Sigma & m_\Xi \\
938.91897 \pm 0.0002 & 1115.684 \pm 0.006 & 1193.12 \pm 0.04 &
1318.11 \pm 0.31 \\ \\
m_\Delta & m_{\Sigma^*} &
m_{\Xi^*} & m_\Omega \\
1231.3 \pm 1.1 & 1384.6 \pm 0.4 & 1533.4 \pm 0.3 &
1672.45 \pm 0.29 
\end{array}.
\]
Since the baryon mass is of order $N$, the fractional accuracy of $R8$ is 
$\ord{\epsilon^3/N^3}$, since the denominator is order $N$.  The fractional
accuracy of the baryon mass relations is listed in Table~\ref{tab:d} and
plotted in fig.~\ref{fig:40}. 
The error bars on the points are from the experimental errors on the measured
baryon masses. The points in fig.~\ref{fig:40} have been plotted so that
relations of the same order in $\epsilon$ have the same symbol. The standard
$SU(3)$ analysis of baryon masses which might be familiar to some of you is
equivalent to using the relations we have derived, but ignoring the powers of
$N$. The Gell-Mann--Okubo formula for the baryon octet, and the equal spacing
rule for the baryon decuplet are linear combinations of the two $\epsilon^2$
relations and one $\epsilon^3$ relation we have obtained. Clearly, $SU(3)$
breaking alone is not the whole story, and including the factors of $1/N$
provides a much better understanding of the data than $SU(3)$ alone. It is
obvious from the figure that the $1/N$ and $\epsilon$ expansions explain the
observed data. There is also clear evidence for the validity of the $1/N$
expansion. For example, the three order $\epsilon$ relations are of the same
order in $SU(3)$ breaking, but different orders in $1/N$, and the $1/N$
suppression is obvious in the experimental data. One also gets
new relations (such as $R1$) which cannot be derived using $SU(3)$, and work
just as well as the relations derived using $SU(3)$ symmetry.

\begin{hw}[Mass Relations]\label{prob:orth}\sl
\begin{itemize}
\end{itemize}
\begin{enumerate}
\item
Show that relations which transform as different $SU(6)$ and $SU(2) \times
SU(3)$ representations must be orthogonal with respect to the metric $g$ in
eq.~(\ref{gmetric}). Hint: linear combinations of baryon states that transform
as given weights of $SU(6)$ representations are orthonormal combinations of the
${\bf 56}$, i.e.\ when $p\uparrow$, $p\downarrow$, $n\uparrow$, $n\downarrow$,
are used as the basis states.
\item
Derive the mass relations in Table~\ref{tab:d}.
\end{enumerate}
\end{hw}

\section{Other Results for Baryons}\label{sec:res}

The procedure of the previous section can be used to analyze other baryon
properties, such as magnetic moments and axial couplings~\cite{jm,lmw,ddjm}.
The $1/N$ expansion has been applied to the nucleon-nucleon potential, and
explains the origin of Wigner supermultiplet symmetry in light
nuclei~\cite{dk-ms,dk-am}. Recently, there has been some interesting work on
excited baryons using the $1/N$ expansion~\cite{cgkm,goity,pirjol}. I do not have
time to go over all these results in these lectures, and the reader is referred
to the literature for details. In this section, I will briefly discuss a few
more results obtained using the $1/N$ expansion for baryons.

The $1/N$ analysis for two light flavors is much simpler than for three light
flavors, since the baryon representations form representations of $SU(4)_c$
symmetry, rather than $SU(6)_c$. In the strangeness-zero sector, the baryon
quantum numbers are $I=J=1/2,3/2,\ldots$. These states will be referred to as a
baryon tower. The $SU(4)$ analysis can also be used in the case of more than
two flavors. One can apply $SU(4)$ spin-flavor symmetry to baryons in a given
strangeness sector, so that the $p$ and $\Delta$ are related to each other, the
$\Lambda$, $\Sigma$, and $\Sigma^*$ are related to each other, and so on. What
is more interesting is that one can relate different strangeness sectors to
each other without assuming approximate $SU(3)$ symmetry~\cite{j}. For
example, requiring that the $K+N \rightarrow \pi + \Sigma$ amplitude is unitary
relates the pion couplings of the $N$ and $\Delta$ to those of the $\Lambda$,
$\Sigma$, and $\Sigma^*$. It also constrains the form of the $KN\Sigma$
coupling~\cite{j,djm1}. Examples of relations obtained using large $N$, but not
assuming $SU(3)$ symmetry are baryon mass relations such
as~\cite{j,djm1,j:hmass}
\begin{eqnarray}
\Sigma^*-\Sigma &=& \Xi^*-\Xi + \ord{1/N^2}, \nonumber \\
{3\Lambda + \Sigma \over 4} - {N+\Xi \over 2} &=& -{1\over4}\left( \Omega - 
\Xi^* - \Sigma^* + \Delta \right) + \ord{1/N^2}, \nonumber \\
\left( \Sigma^* - \Delta \right) + \left( \Omega - \Xi^* \right) & = &
2 \left( \Xi^* - \Sigma^* \right)  + \ord{1/N^2} ,\nonumber \\
{1\over3}\left(\Sigma+3\Sigma^*\right) - \Lambda &=& {2\over3}\left(\Delta -
N\right) + \ord{1/N^2} ,\nonumber \\
\Sigma^*_Q-\Sigma_Q &=& \Xi^*_Q-\Xi_Q' + \ord{1/N^2} ,\nonumber \\
{1\over3}\left(\Sigma_Q+3\Sigma^*_Q\right) - \Lambda &=
& {2\over3}\left(\Delta -
N\right) + \ord{1/N^2}, \label{mrels}
\end{eqnarray}
and coupling constant relations such as
\begin{equation}
g_A = g_{c,b} + \ord{1/N}.
\label{greln}
\end{equation}
Equation~(\ref{mrels}) relates mass splittings of baryons of different
strangeness, and also relates mass splittings of heavy quark baryons to mass
splittings of baryons containing only light quarks. A more detailed analysis
allows one to predict heavy baryon mass differences to high accuracy using the
$1/N$ expansion, by relating them to the known mass differences of the octet
and decuplet baryons~\cite{j:hmass}. Equation~(\ref{greln}) relates pion
couplings of $c$ and $b$ baryons to the pion coupling of the nucleon, $g_A$.
This relation was originally derived using the Skyrme model in
ref.~\cite{gskyrme}. One also obtains information on the Isgur-Wise function
for heavy baryons using large $N$ QCD~\cite{chow}.

In a given strangeness sector, one can predict the ratios of pion couplings to
order $1/N^2$. This result is simple to derive. Consider the operator expansion
eq.~(\ref{5.8}) where the flavor group is now $SU(2)$. The pion couplings are
the same as the axial current matrix elements, by the Goldberger-Treiman
relation. As in eq.~(\ref{5.5}), the only non-zero matrix element is that of
the space component of the axial current. Thus the operator expansion is for
spin one and isospin one. The expansion of the axial current has the form
\begin{equation}
A^{ia} = g G^{ia} + h {J^i I^a \over N} + \ldots,
\label{6.1}
\end{equation}
using the operator reduction rule for two flavors. The $1/N$ expansion is
applied to states with $J$ of order unity. In the three-flavor case, matrix
elements of $G^{ia}$ and $T^a$ can be of order $N$. However, in the two-flavor
case, matrix elements of the isospin are of order unity, since $I^2=J^2$. The
$J^i I^a$ term is therefore a $1/N^2$ correction, so one can predict the ratios
of axial current matrix elements (such as $g_{\pi NN}/g_{\pi N \Delta}$) with
an accuracy of $1/N^2$. The Skyrme model is one particular representation of
the $SU(4)_c$ spin-flavor symmetry, so we have shown that the QCD predictions
for the {\it ratios} of pion couplings, such as $g_{\pi NN}/ g_{\pi N \Delta}$
or $g_{\pi NN}/ g_{\pi \Delta \Delta}$ are equal to the prediction in the
Skyrme model up to corrections of order $1/N^2$. The Skyrme model results for
the couplings are listed in Table~\ref{tab:skyrme}. 
\begin{table}[tbp]
\caption{Values for the pion couplings from ref.~\cite{anw}. The values of
$g_{\pi N \Delta}$ and $g_{\pi N N}$ are obtained using the Skyrme model. The
ratio $g_{\pi N \Delta}/g_{\pi N N}$ obtained using the Skyrme model is the
same as that predicted using large $N$ QCD to order $1/N^2$.}
\label{tab:skyrme}
\begin{tabular}{cccc}
\hline
& Method & Theory & Experiment \\
\hline
$g_{\pi N \Delta}$ & Skyrme Model & $13.2$ & $20.3$ \\ 
$g_{\pi N N}$ & Skyrme Model & $8.9$& $13.5$ \\
$g_{\pi N \Delta}/g_{\pi N N}$ & Large N QCD/Skyrme Model & $1.5$ & $1.48$\\
\hline
\end{tabular}
\end{table}
Clearly, the values of the individual couplings do not agree that well with the
experimental data. Nevertheless, the ratio is in excellent agreement. Only the
prediction for the ratio can be derived directly from QCD to $\ord{1/N^2}$, and
does not make any assumption about the validity of the Skyrme model. The
overall scale of the couplings depends on the details of the Skyrme model
Lagrangian, and is not a prediction of large $N$ QCD. Similarly, other results
in the Skyrme model literature that are model independent (using the
terminology of ref.~\cite{an}) can also be derived using large $N$ QCD.

The axial coupling constant $g$ in eq.~(\ref{6.1}) is the same for any baryon
tower, but it can be different for different towers, i.e.\ it can depend on the
strangeness. One can prove that $g$ is a constant at leading order, and is
linear in strangeness at order $1/N$. The value of $g$ extracted from the
different decuplet decays are given in Table~\ref{tab:5}, and clearly satisfy
this result. One can also show that the $F/D$ ratio for the baryon axial
currents and magnetic moments is $2/3$, with an error of order $1/N^2$. This is
consistent with the experimental values of $0.58 $ and $0.59$, respectively.
\begin{table}[tbp]
\caption{Axial couplings extracted from ${\bf 10} \rightarrow {\bf 8} + \pi$
decays. The last column gives the baryon strangeness. \label{tab:5}}
\setlength{\tabcolsep}{2em}
\begin{tabular}{ccc}
\hline
Decay & $g$ & S\\
\hline
$\Delta \rightarrow N \pi$ & $1.8$ & 0\\
$\Sigma^* \rightarrow \Sigma \pi$ & $1.5$ & $-1$ \\
$\Sigma^* \rightarrow \Lambda \pi$ & $1.5$ & $-1$ \\
$\Xi^* \rightarrow \Xi \pi$ & $1.3$ & $-2$ \\
\hline
\end{tabular}
\end{table}

The $I=J$ rule of Mattis and collaborators~\cite{mattis1,mattis2,mattis3} can be derived in the
$1/N$ expansion. This rule was originally derived using the Skyrme model, and
is in good agreement with the experimental data. The $I=J$ rule states that
meson-baryon couplings satisfy $I=J$, where $I$ is the isospin of the meson,
and $J$ is the spin transfer at the vertex. For example, the $\rho$ meson has
$I=1$, and its coupling to baryons is dominantly spin one, i.e.\ of magnetic
moment type, or proportional to the $F_2$ form factor. The $\omega$ has $I=0$,
and its coupling to baryons is dominantly spin zero, i.e.\ of charge type, or
proportional to the $F_1$ form factor. One can prove a slightly stronger form
of the result using the $1/N$ expansion~\cite{djm2}, the meson-baryon vertex is
of order $\left( \sqrt N \right)^{1-\abs{I-J}}$.

\begin{figure}[tbp]
\begin{center}
\epsfig{file=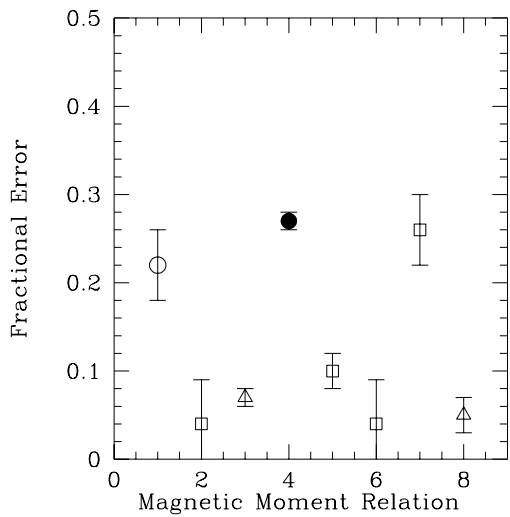,height=70mm}
\caption{Comparison of the baryon magnetic moment relations of ref.~\cite{jm}
with the experimental data. The circles are $\ord{1/N}$ relations, the squares
are $\ord{1/N^2}$ relations, and the triangles are $\ord{\epsilon/N}$
relations. All the relations except the filled circle can also be derived in
the non-relativistic quark model. The error bars are due to the experimental
errors on the measured baryon magnetic moments. \label{fig:54}}
\end{center}
\end{figure}

Finally, let me show two plots in which large $N$ predictions have been
compared with the experimental data. The first plot is fig.~\ref{fig:54}
for the baryon magnetic moment relations. The non-relativisitic quark model is
known to provide a good description of the baryon magnetic moments at the 20\%
level. All the magnetic moment relations of the non-relativistic quark model
can be derived using large $N$ QCD to some order in $1/N$. It is clear from
fig.~\ref{fig:54} that the $1/N$ expansion also explains why some relations
work better than others; something that cannot be understood solely on the
basis of the quark model. In addition, one gets one new relation (the filled
circle) that that works just as well as the relations that are also valid in
the quark model. The only relation that does not work as well as one might
expect is relation 7, which is an $\ord{1/N^2}$ relation but is violated at the
25\% level. This is a prediction for the $\Delta^+ \rightarrow p \gamma$
transition amplitude, and is a known problem for the quark model. 

\begin{figure}[tbp]
\begin{center}
\epsfig{file=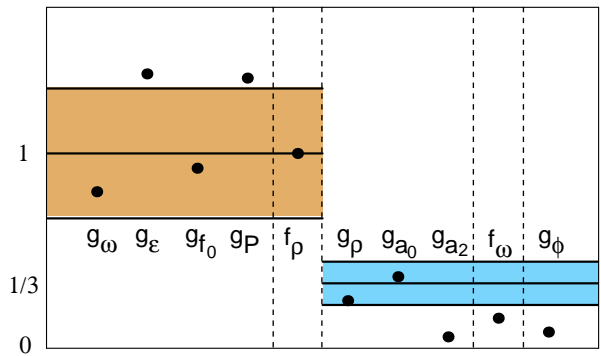,height=50mm}
\caption{Comparison of the large $N$ predictions for the strengths of the
various terms in the Nijmegen potential~\cite{nij1,nij2}. The ratios of
couplings (relative to $f_\rho$) have been plotted, and the expected size of
the ratio is shown by a horizontal line. The shaded regions are an estimate of
the size of the $\ord{1/N}$ corrections to the leading result. \label{fig:55}}
\end{center}
\end{figure}

The second plot is a comparison of the large $N$ predictions for the
nucleon-nucleon potential with the experimental data~\cite{dk-am}. It is
difficult to compare the predictions directly with nucleon scattering data.
What is actually shown in fig.~\ref{fig:55} is a comparison of the large $N$
predictions with coupling constants in the Nijmegen potential~\cite{nij1,nij2},
which provides a good description of the experimental nucleon-nucleon
scattering data. The $1/N$ expansion provides a satisfactory explanation of the
size of the various terms.

\section{Large N and Chiral Perturbation Theory}

\begin{figure}[tbp]
\begin{center}
\epsfig{file=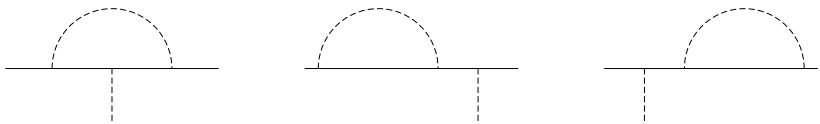,height=15mm}
\caption{One loop correction to the pion-baryon vertex. \label{fig:61}}
\end{center}
\end{figure}

The large $N$ expansion for baryons can be combined with baryon chiral
perturbation theory~\cite{dm,j:chpt}. In the large $N$ limit, the baryon is
heavy, and one can use the formalism of refs.~\cite{jm:chpt}. The pion-baryon
constant is order $\sqrt N$. The one-loop correction to the pion-baryon
coupling constant is shown in fig.~\ref{fig:61}.
The diagrams produce a correction to the pion-nucleon coupling constant of the
form
\begin{equation}\label{8.1}
\delta g \propto N g^3 m_q \log m_q/\mu
\end{equation}
Here $g \sqrt N$ is a generic pion-baryon coupling, $m$ is a light quark mass,
and $\mu$ is the scale parameter of dimensional regularization. For simplicity,
I will denote the light quark masses by $m_q$, and not distinguish between the
$u$, $d$ and $s$ quark masses. One can see from eq.~(\ref{8.1}) that the chiral
expansion naively breaks down in the large $N$ limit, since the correction
grows with $N$. This would be true if one computed the diagrams in
fig.~\ref{fig:61} by only including intermediate nucleon states. However, in
the large $N$ limit, the $\Delta$-nucleon mass difference is of order $1/N$,
and the $\Delta$ must be included as an intermediate state as well. In the
notation of section~\ref{sec:cons}, the diagrams in fig.~\ref{fig:61} are
proportional to the group-theoretic factor
\begin{equation}\label{9.1}
- 2 X^{ia} X^{jb} X^{ia} + X^{ia} X^{jb} X^{jb} + X^{jb} X^{jb} X^{ia}.
\end{equation}
The first term in eq.~(\ref{9.1}) is the vertex correction, and the last two
terms are from wavefunction renormalization. The relative factor of $-1/2$ for
the wavefunction diagrams arises since their contribution to the amplitude is
$1/\sqrt Z$. The sum over mesons is the sum on $ia$, and the sum over
intermediate baryons is the matrix multiplication of $X$. Equation~(\ref{9.1})
is the double commutator
\[
\left[ X^{ia}, \left[ X^{ia}, X^{ib} \right] \right] = \ord{1/N^2}
\]
which is of order $1/N^2$ from the consistency conditions of
section~\ref{sec:cons}. This converts eq.~(\ref{8.1}) into
\[
\delta g \propto {1\over N} g^3 m_q \log m_q/\mu,
\]
and the expansion parameter is now suppressed by one power of $N$. Including
the entire tower of large $N$ baryon states is crucial for this $1/N$
suppression; the double commutator is $\ord 1$ if only intermediate nucleon
states are included. The loop expansion in the baryon sector is now a $1/N$
expansion, as in the meson sector. Hadronic dynamics in the meson and baryon
sectors becomes semiclassical in the large $N$ limit~\cite{am:semi}. The
consistency conditions of section~\ref{sec:cons} can be derived by requiring
that the chiral expansion has a sensible large $N$ limit~\cite{dm}.

It is interesting to see what happens if one studies $\pi$-nucleon scattering
in the large $N$ limit without including the complete tower of large $N$
states. The nucleon is infinitely heavy in this limit, and the theory reduces
to a strong coupling theory of pions interacting with a static nucleon, a model
studied by Pauli and Dancoff~\cite{pauli}. They showed that the theory produces
an infinite tower of states with $I=J=1/2,3/2,\ldots$, precisely the spectrum
of the baryon tower in large $N$ QCD. The ratios of the pion-couplings are also
the same as those of large $N$ QCD. One therefore has two possible ways of
thinking about baryon chiral perturbation theory: (a) One starts with a
Lagrangian with pions and nucleons. The theory is strongly coupled, with loop
graphs of order $N$, and dynamically generates a $\Delta$ resonance. (b) One
starts with a Lagrangian with pions, nucleons and $\Delta$ fields. The theory
is weakly coupled, with loop graphs of order $1/N$. All predictions of (a) and
(b) for observable quantities are the same. The main effect of the $\ord N$
strong interactions in (a) is to produce a $\Delta$ resonance. Once the effects
of this resonance are explicitly included, as in (b), the residual 
interactions are $\ord {1/N}$ and weak.

Baryon chiral perturbation theory can be formulated including intermediate
$\Delta$ states~\cite{jm:chpt}, and has been applied to baryon axial couplings,
masses, magnetic moments, etc.~\cite{misc1,misc2,misc3}. Results including the
$\Delta$ resonance provide a good description of the experimental data. In
fact, these original calculations showed that baryon couplings were very close
to their $SU(6)$ values, that there were interesting cancellations in diagrams
when the $\Delta$ was included, and that ignoring the $\Delta$ led to
disagreement with experiment. The large $N$ approach to baryons originated in
trying to understand these results.

\section{Conclusions}

The $1/N$ expansion is an extremely useful tool for understanding the
properties of mesons and baryons. S.~Coleman, at the end of his 1979 Erice
Lectures~\cite{coleman} remarks that ``For the baryons, things are not so
good. Witten's theory is an analytical triumph but a phenomenological
disaster.'' He also concludes ``I feel future progress in this field rests
upon constructing the leading approximation.'' Since these remarks much has
been learned. The leading approximation (i.e.\ the master field) has been
constructed only for $1+1$ dimensional chromodynamics, not for $3+1$
dimensional chromodynamics. Nevertheless, a lot of progress has been made.
Many of the applications to mesons phenomenology in section~\ref{sec:mesons}
are new. Baryon phenomenology is also very successful; the new results exploit
large $N$ spin-flavor symmetry, rather than the Hartree picture of Witten.

\section{Acknowledgments}
I would like to thank the organizers R.~Gupta, A.~Morel, and E.~de Rafael for
making this a successful Les Houches summer school. I would also like to thank
R.~Lebed for providing me with his 't~Hooft model computer programs, and
R.~Flores, Z.~Ligeti, E.~Poppitz and W.~Skiba for carefully reading the
manuscript. This work was supported in part by Department of Energy Grant
DOE-FG03-97ER40546.

\end{document}